\def\m0{m^{\!\!\!\!^o}}
\def\barqslash{\FMslash{\bar{q}}}

\newcommand{\qslash}{\FMslash q}

\newcommand{\wslash}{\FMslash w}

\documentclass[12pt]{elsart}
\usepackage{feynmp,epsfig,graphics}

\begin{document}
GSI-Preprint-01-30, December 2001
\begin{frontmatter}
\title{Scattering of vector mesons off nucleons}
\author{M.F.M. Lutz$^{a,b}$, Gy. Wolf$^{\,c}$ and B. Friman$^{a,b}$}
\address{$^a$ GSI, Planckstr. 1, D-64291 Darmstadt, Germany}
\address{$^b$ Institut f\"ur Kernphysik, TU Darmstadt\\
D-64289 Darmstadt, Germany}
\address{$^c$ KFKI RMKI,
H-1525 Budapest POB. 49, Hungary}
\begin{abstract}
We construct a relativistic and unitary approach to 'high' energy
pion- and photon-nucleon reactions taking the
$\pi N, \pi \Delta $, $\rho N$, $\omega N $, $\eta N, K
\Lambda, K \Sigma $ final states into account.
Our scheme  dynamically generates the
s- and d-wave nucleon resonances $N(1535)$, $N(1650)$ and $N(1520)$
and isobar resonances $\Delta(1620)$ and
$\Delta(1700)$ in terms of quasi-local interaction vertices.
The description of photon-induced processes is based on a generalized
vector-meson dominance assumption
which directly relates the electromagnetic quasi-local 4-point interaction vertices
to the corresponding vertices involving the $\rho$ and $\omega$ fields.
We obtain a satisfactory description of the elastic and inelastic pion- and
photon-nucleon scattering data in the channels considered.
The resulting s-wave $\rho $- and $\omega $-nucleon scattering amplitudes are presented.
Using these amplitudes we compute the leading density modification of the $\rho$ and
$\omega $ energy distributions in nuclear matter. We find a repulsive energy shift for the
$\omega $ meson at small nuclear density but predict considerable strength
in resonance-hole like $\omega $-meson modes. Compared to previous calculations
our result for the $\rho$-meson spectral function shows a significantly
smaller in-medium effect. This reflects a fairly small coupling strength of the
N(1520) resonance to the $\rho N$ channel.
\end{abstract}
\end{frontmatter}

\section{Introduction}

The decay of vector mesons into $e^+e^-$ and $\mu^+\mu^-$ pairs
offers a unique tool to explore the properties of dense and hot
matter in nuclear collisions. The lepton pairs provide virtually
undistorted information on the current-current correlation
function $\langle j_\mu j_\nu\rangle$ in the medium~\cite{RW}. At
invariant masses in the range $500 - 1000$ MeV, $\langle j_\mu
j_\nu\rangle$ is sensitive to in-medium modifications of the mass
distribution of the light vector mesons $\rho$ and $\omega$.
Indeed, the observed enhancement of the di-lepton yield at small
invariant masses in ultra relativistic heavy ion collisions
~\cite{CERES} is presently interpreted in terms of in-medium
modifications of the mass and width of the $\rho$
meson~\cite{BR,Kapusta,su-huong,CS,HFN,Wolf,Wolf:2,Wambach,LPM,Leupold,Friman:Pirner,Urban:Buballa:Rapp:Wambach}.
To leading orders in the baryon density, modifications of the mass distribution are
determined by the vector-meson nucleon scattering amplitudes.
Since these amplitudes are not directly constrained by data, there
are presently large theoretical ambiguities in predictions of the
vector-meson spectral densities in nuclear matter.

This work is an attempt to overcome this problem by using the
constraints from data on pion- and photon-nucleon reactions
considering in particular the $\omega$ and $\rho$ production data
in a systematic way. We develop a coupled-channel approach for
meson-baryon and photon-baryon scattering, including the
$\gamma\,N, \pi N$, $\rho N$, $\omega N$, $\pi \Delta$, $\eta N$,
$K\Lambda$ and $K\Sigma$ channels where the interaction kernel is
approximated by quasi-local 4-point interaction vertices.
The description of photon-induced reactions is based on a vector-meson dominance type assumption
which directly relates the electromagnetic quasi-local 4-point interaction
vertices to the corresponding vertices involving the $\rho$ and $\omega$ fields.
In such a scheme the amplitudes for experimentally non-accessible processes
like $\rho N$ and $\omega N$ scattering are constrained by the data on elastic $\pi N$
scattering and inelastic reactions like the pion-and photon-induced
production of vector mesons. Our primary goal is to derive the
vector-meson nucleon scattering amplitudes close to threshold,
which in turn determine the self energy of a vector-meson at rest
in nuclear matter to leading order in density. Thus it is sufficient to consider
only s-wave scattering in the $\rho N$ and $\omega N$ channels. As a consequence
of parity conservation, this implies that in the $\pi N$ and $\pi\Delta$
channels we need only s- and d-waves. In particular, we consider
the $S_{11}, S_{31}, D_{13}$ and $D_{33}$ partial waves of $\pi N$
scattering. To derive systematically  the momentum dependence of
the vector-meson self energy, vector-meson nucleon scattering in
higher partial waves would have to be considered in addition.
We concentrate on the energy window 1.4 GeV $< \sqrt{s} <$ 1.8 GeV and base our analysis
on the conjecture that all baryon resonances except the baryon octet and decuplet ground
states are generated by coupled channel dynamics. Under these assumptions
the application of quasi-local interaction terms is justified, because
in all considered channels the interaction kernel is slowly varying in energy.

In this paper we expand the scope of existing coupled channel
calculations~\cite{Feuster:Mosel:1,Feuster:Mosel:2,Vrana:Dytman:Lee,Manley}
in several ways. First, we attempt a
comprehensive description of pion and photon induced reactions,
including for the first time the $\rho N$, $\omega N$
photon-nucleon channels on equal footing. Thus, we describe elastic
$\pi N$ scattering as well as pion and photon induced production
cross sections within the same model. Second, our approach is
consistent with analyticity and causality, i.e. the amplitudes
satisfy dispersion relations. This is not the case for models that
employ the K-matrix approximation or extensions
thereof~\cite{Feuster:Mosel:1,Feuster:Mosel:2,Manley}. Analyticity is a
fundamental property of the S-matrix and the analytic structure of the
scattering amplitude plays an important role e.g. close to thresholds (see
e.g.~\cite{Kondratyuk:Scholten}). Since we are addressing the physics in an energy
range where several production thresholds are located, this is an
important feature of our model. Finally, in our approach, which applies and
generalizes the covariant projector technique of \cite{LK},
baryon resonances can be generated by coupled channel dynamics.
This offers an appropriate framework for testing the conjecture of
dynamical generation of resonances~\cite{LK}. If this conjecture is
correct, a more economical description of the scattering amplitude
follows, since there is no need for introducing explicit baryon
resonances in the interaction kernel. We note that in the K-matrix
approximation, the lack of analyticity prohibits the dynamical
generation of resonances. Thus, in such an approach the baryon
resonances must be explicitly included in the interaction kernel,

We obtain a satisfactory description of the elastic and inelastic $\pi N$ and
$\gamma N$ reaction data in the channels considered.
The s- and d-wave nucleon resonances $N(1535)$, $N(1650)$ and $N(1520)$
and isobar resonances $\Delta(1620)$ and $\Delta(1700)$
are generated dynamically in our approach. Whereas we predict
considerable strength for the $N(1520)$ nucleon-hole component
of the $\omega$ meson in nuclear matter, our analysis suggests only a
moderate importance of the $N(1520)$ nucleon-hole states in the in-medium
spectral function of the $\rho$ meson.

In section 2 we first present the general hadronic coupled channel
approach within the framework of the covariant projector operator
technique of \cite{LK}. We first introduce the projection
operators for elastic $\pi N$ scattering and subsequently
generalize the scheme to include inelastic reactions between the
different coupled channels. In the following section the specifics
of the $\rho N$ and $\pi \Delta$ channels are presented, while in
section 4 we give the framework for the treatment of
electromagnetic interactions. There we also motivate our
choice for the vector meson dominance assumption. Readers who are
not interested in the technical details of the calculation may skip
sections 2-4 and jump to the results, which are presented in
section 5. There the basic ingredients of our approach are recalled
with appropriate references to the central equations in sections 2-4.

\section{Relativistic coupled-channel dynamics}

In this section we present the theoretical framework for our
analysis of the meson-baryon scattering process. We construct an
effective Lagrangian with quasi-local four-point meson-baryon contact
interactions. Within this framework, the Bethe-Salpeter equation for the
coupled-channel system reduces to a matrix equation. The justification
of our approximation strategy follows from two assumptions. First, we conjecture
that all baryon resonances except the baryon octet and decuplet ground states are
dynamically generated by coupled channel dynamics. And second, we aim at a description
of elastic and inelastic pion-nucleon scattering in an energy window
1400 MeV $< \sqrt{s}<$ 1800 MeV only, where the interaction kernel is slowly
varying in energy in all channels considered. This will be explained in detail later.

In order to keep the discussion transparent, we first restrict
ourselves to the pion-nucleon sector. Consider the general
pion-nucleon four-point interaction in momentum space
\begin{eqnarray}
{\mathcal L}_{\pi N} (\bar k, k; w) = {\textstyle{1\over 2}}\,\sum_{ijkl}\, \pi^\dagger_i(\bar q)\,
\bar N_j(\bar p)\, K_{ijkl}(\bar k, k; w)\, N_k(p)\, \pi_l(q) \,,
\label{piN-lag-0}
\end{eqnarray}
where $i,j,k,l$ are isospin indices, and $K$ is a matrix in spinor
space, which in general is a function of the relative momenta in
the initial and final states $k=\frac 12 (p-q)$ and $\bar k=\frac
12 (\bar p-\bar q)$ as well as the total momentum $w = p+q = \bar p
+ \bar q$. Here we use a generic two-body interaction kernel $K$ at tree-level as
to specify the conventions applied throughout this work. The relation of the
momentum-space object (\ref{piN-lag-0}) to the Lagrangian density ${\mathcal L}(x)$
is  obtained by defining the action,
\begin{eqnarray}
&& \int d^4x\,{\mathcal L}(x) = \int \frac{d^4k}{(2\,\pi)^4}\,
\frac{d^4\bar k}{(2\,\pi)^4}\,\frac{d^4w}{(2\,\pi)^4}\,
{\mathcal L}(\bar k ,k ;w)\;,
\nonumber\\
&& \pi (q) = \int d^4 x \,e^{i\,q\cdot x}\,\pi (x)\,,\quad
\pi^\dagger  (\bar q) = \int d^4 x \,e^{-i\,\bar q\cdot x}\,\pi (x) \,,
\nonumber\\
&& N(p) = \int d^4 x \,e^{i\,p\cdot x}\,N (x)\,,\quad \!
\bar N(\bar p) = \int d^4 x \,e^{-i\,\bar p\cdot x}\,\bar N (x)\,.
\label{lag-momentum}
\end{eqnarray}
Without loss of generality we may assume a crossing symmetric interaction kernel with
\begin{eqnarray}
K_{ijkl}(\bar k, k; w) = K_{ljki}((\bar p+q)/2, (p-\bar q)/2; p-\bar q)\,,
\label{}
\end{eqnarray}
simplifying the identification of any exchange term contribution.
The fact that the effective interaction ${\mathcal L}(\bar k ,k ;w)$ depends
only on three combinations of the four 4-momenta introduced in (\ref{lag-momentum})
is a consequence of the invariance of the coordinate-space interaction kernel under
time- and space-translation. Note for instance that the momenta $q$ and $\bar q$ of the
$\pi (q)$ and $\pi^\dagger (\bar q)$ fields do not represent the 4-momenta of the physical
pion states. This interpretation is legitimate only after the pion field is
decomposed into positive and negative frequency
parts\footnote{We use the normalization of Itzykson and Zuber~\cite{IZ}.}
\begin{eqnarray}
\pi^{(+)}_i(x) &=& \int\frac{d^3p}{(2\pi)^{3}2p_0}\,a_i(p) \,e^{-ipx}\,,
\nonumber\\
\pi^{(-)}_i(x) &=& \int\frac{d^3p}{(2\pi)^{3}2p_0}\, a_i^\dagger(p) \,
e^{ipx}\,.
\end{eqnarray}
In order to construct the isospin projectors, it is useful to split also
the nucleon field into positive and negative
frequency parts, defined by
\begin{eqnarray}
N^{(+)}(x) &=& \sum_s \int
\frac{d^3p}{(2\pi)^{3}}\,\frac{m_N}{p_0}\,u(p,s)\,b(p,s)
e^{-ipx}\,,\nonumber\\
N^{(-)}(x) &=&  \sum_s \int
\frac{d^3p}{(2\pi)^{3}}\,\frac{m_N}{p_0}\,v(p,s)
\,d^\dagger(p,s) e^{ipx}\, ,
\end{eqnarray}
where $u(p,s)$ and $v(p,s)$ are the positive and negative energy
spinors, respectively. Here, $a_i(p)$ and $a_i^\dagger(p)$ are the
pion annihilation and creation operators, while $b(p,s)$ is the
nucleon annihilation operator and $d^\dagger(p,s)$ is the
anti-nucleon creation operator.

For the construction of the Bethe-Salpeter kernel, it is convenient to
rearrange the interaction Lagrangian, ${\mathcal L}_{\pi N}(\bar k ,k
;w)$, which is considered in
momentum space, as a sum of terms that project
onto states of good isospin\footnote{We assume perfect isospin
symmetry.}
\begin{eqnarray}
\label{piN-lag}
&& {\mathcal L}_{\pi N}(\bar k ,k ;w)= {\textstyle{1\over 3}}\,
\bar N^{(+)}(\bar p)\,[\vec\tau\cdot\vec\pi^{(-)}(\bar q)]
\,K^{(\frac 12)}(\bar k ,k ;w )\,[\vec\pi^{(+)}(q)
\cdot\vec\tau]\,N^{(+)}(p)
\nonumber\\
&& \qquad +\bar N^{(+)}(\bar p)\, [\vec T\cdot\vec\pi^{(-)}(\bar
q)]
\,K^{(\frac 32)}(\bar k ,k ;w )\,[\vec\pi^{(+)}(q)
\cdot\vec T]\,N^{(+)}(p)\,+ \dots\,,
\label{dots}
\end{eqnarray}
where $\tau_i$ is a Pauli matrix in isospin space and $T_i$ an
isospin $1/2$ to $3/2$ transition matrix and $K^{(I)}$ the
interaction kernel for states with isospin $I$. The normalization
of the isospin transition matrices is such that
$T^\dagger_i\,T_j=\delta_{ij}-\tau_i\,\tau_j/3$. In
(\ref{piN-lag}) we have only written the terms that explicitly
contribute to pion-nucleon scattering in the Born approximation.
These are sufficient to specify the tree-level Bethe-Salpeter
kernel. The additional terms, indicated by the dots in (\ref{dots}),
contribute to the production and absorption of two pions and are
uniquely determined by covariance.

It is straightforward to generalize this scheme to the
coupled-channel problem at hand. The interaction Lagrangian can
thus be written in the form
\begin{eqnarray}
&&{\mathcal L}(\bar k ,k ;w)=
\sum_{I}\,R^{(I)\,\dagger }(\bar q,\bar p)\,\gamma_0
\,K^{(I)}(\bar k ,k ;w )\,R^{(I)}(q,p)\,+ \dots\,\;\;\;\;
\label{eff-lagr}
\end{eqnarray}
where again we show only those terms that contribute to
meson-baryon scattering in the Born approximation. The state
vectors $R^{(I)}(q,p)$ are given by
\begin{eqnarray}
&&R^{(\frac{1}{2})} =
\left(
\begin{array}{c}
\textstyle{1\over\sqrt{3}}\,\big(\vec \pi^{(+)} \cdot \vec \tau
\big)\,N^{(+)} \\
\textstyle{1\over\sqrt{2}}\,\big(\vec \pi^{(+)}\cdot \vec T^\dagger
\big)\, \Delta^{(+)}_\mu \\
\textstyle{1\over\sqrt{3}}\,\big(\vec \rho_\mu^{(+)} \cdot \vec\tau
\big) \,N^{(+)} \\
\omega_\mu^{(+)} \,N^{(+)} \\
\eta^{(+)} \,N^{(+)} \\
K^{(+)} \,\Lambda^{(+)}\\
\textstyle{1\over\sqrt{3}}\,\big( \vec \Sigma^{(+)} \cdot \vec \tau
\,\big) \, K^{(+)}
\end{array}
\right) \;,\;\;\;\;
R^{(\frac{3}{2})}= \left(
\begin{array}{c}
\big(\pi^{(+)} \cdot T\big)\,N^{(+)} \\
\sqrt{\textstyle{3\over 5}}\,T_i \,\big(\vec\pi^{(+)}\cdot \vec\tau
\big)\,
T_i^\dagger\,\Delta_\mu^{(+)}  \\
\big(\vec\rho_\mu^{(+)} \cdot \vec T\big)\,N^{(+)}  \\
\big(\vec\Sigma^{(+)}  \cdot \vec T\big) \,K^{(+)}
\end{array}
\right) \;.
\label{r-def}
\end{eqnarray}
Note that the spin and isospin structure of the states is implicit
in (\ref{r-def}). Thus, $\rho_\mu $ is an isovector and $\omega_\mu $
an isoscalar vector field, while e.g. the nucleon is an isospin
doublet Dirac spinor field. Similarly, the $\Delta_\mu $ is a
Rarita-Schwinger isospin 3/2 field. These fields
combine with the corresponding elements of the kernel
$K^{(I)}(\bar k,k,w)$ and the isospin operators in (\ref{r-def}).
For instance, the vector index of the $\rho_\mu $ and $\omega_\mu $
fields
are contracted with a vector index of the corresponding term
in the kernel.

The coupled-channel Bethe-Salpeter equation \footnote{We apply here the
conventions of
Itzykson and Zuber \cite{IZ} where the 'amputated' Greens function
$G^{\rm amp}_{X\to Y}$ defines the scattering amplitudes with
$G^{\rm amp}_{\pi N\to \pi N}=i\,T_{\pi N\to \pi N}$, $G^{\rm
amp}_{V N\to V N}=-i\,T_{V N\to V N}$ but $G^{\rm amp}_{\pi N\to V
N}= T_{\pi N\to V N}$ and $G^{\rm amp}_{V N\to \pi N}= T_{V N\to
\pi N}$. The analogous convention is chosen for the $\pi \Delta $
channel.} projected onto a given
isospin then reads
\begin{eqnarray}
T^{(I)}_{ab}(\bar k ,k ;w ) &=& K^{(I)}_{ab}(\bar k ,k ;w )
+\sum_{c,d}\int\!\! \frac{d^4l}{(2\pi)^4}\,K^{(I)}_{ac}(\bar k , l;w )\,
G^{(I)}_{cd}(l;w)\,T^{(I)}_{db}(l,k;w )\;,
\nonumber\\
G^{(I)}_{cd}(l;w)&=&-i\,D_{\Phi(I,c)\Phi(I,d)}(\half\,w-l)\,S_{B(I,c)B(I,d)}(
\half\,w+l)\,,
\label{BS-coupled}
\end{eqnarray}
where $D_{\Phi(I,c)\Phi(I,d)}(q)$ and $S_{B(I,c)B(I,d)}(p)$ denote
the meson and baryon propagators in a given channel. Evidently the interaction
kernel $K^{(I)}_{ab}(\bar k ,k ;w )$ in the Bethe-Salpter equation
(\ref{BS-coupled}) is not necessarily restricted to the tree-level
expressions discussed in detail above. It encompasses all
Feynman diagrams which are two-particle irreducible in the s-channel.
Off-diagonal components of $G^{(I)}$ are due to the matrix structure of the
vector-meson and isobar propagators. A convenient labelling of
the channels is obtained by defining
\begin{eqnarray}
&&\Phi({\textstyle{1\over 2}},a)=(\pi,\pi,\rho_\mu ,\omega_\mu,\eta ,K,
K)_a\;,\;\;\;\;\;\;\
B({\textstyle{1\over 2}},a)=(N,\Delta_\mu,N, N,N,\Lambda , \Sigma )_a\;,
\nonumber\\
&&\Phi({\textstyle{3\over 2}},a)=(\pi,\pi, \rho_\mu, K)_a\;,\;\;\qquad
\quad \quad  \;\,\;\;
B({\textstyle{3\over 2}},a)=(N, \Delta_\mu,N,\Sigma )_a\;.
\label{def-channel}
\end{eqnarray}
The meson and baryon propagators $D_{\Phi(I,c)\Phi(I,d)}(q)$ and
$S_{B(I,c)B(I,d)}(p)$ in (\ref{BS-coupled}) are
\begin{eqnarray}
&& D_{\pi\pi}(q)= \frac{1}{q^2-m_\pi^2+i\,\epsilon}\;, \qquad
S_{NN}(q)
=
\frac{1}{\qslash-m_N +i\,\epsilon}  \;,
\nonumber\\
&& D_{\rho^\mu \rho^\nu}(q) =
\int_{4\,m_\pi^2}^{\infty } \,\frac{d\,m^2}{\pi}\,
\left(g^{\mu \nu}-\frac{q^\mu\,q^\nu}{m^2}\right)
\,\frac{\rho_\rho(m^2)}{q^2-m^2+i\,\epsilon } \;,
\nonumber\\
&& S_{\Delta^\mu \Delta^\nu}(q) =
\int_{(m_N+m_\pi)^2}^{\infty } \,\frac{d\,m^2}{\pi}\,
\Bigg( g^{\mu \nu}\left(\qslash +\sqrt{m^2}\right)
+\frac{1}{3}\,\gamma^\mu\,\left(\qslash -\sqrt{m^2}
\right)\,\gamma^\nu
\nonumber\\
&& \qquad \qquad \quad + \frac{1}{3\,\sqrt{m^2}}\left( q^\mu
\left(\qslash -\sqrt{m^2} \right)\,\gamma^\nu +\gamma^\mu
\left(\qslash -\sqrt{m^2} \right)\,q^\nu
\right)
\nonumber\\
&& \qquad \qquad \quad
-\frac{2}{3}\,\frac{q^\mu\,q^\nu}{m^2}\left(\qslash +2\,\sqrt{m^2}
\right)
\Bigg) \, \frac{\rho_\Delta(m^2)}{q^2-m^2+i\,\epsilon} \;.
\label{def-prop}
\end{eqnarray}
The spectral distributions of the vector mesons, $\rho_{\rho}$ and
$\rho_\omega$, and that of the isobar-resonance, $\rho_\Delta$,
will be specified in the subsequent section, where the pion-induced
vector-meson and isobar production processes are
discussed. The propagators of the $\eta$ and $K$ mesons as well as of
the
$\Lambda$ and $\Sigma$ baryons are, except for the mass, identical
to the pion and nucleon propagators, respectively.

A convenient way to perform the partial-wave expansion is offered
by the covariant projector formalism introduced in ~\cite{LK}.
A detailed discussion of the underlying strategy leading to the
projector formulation can be found in that paper. Here we generalize
this approach in order to include the $\rho_\mu
\,N$, $\omega_\mu \,N$ and $\pi
\,\Delta_\mu$ channels. The interaction kernel $K_{ab}^{(I)}$ is
expanded in a series of channel-dependent projectors
$\big[Y^{(\pm)}_{n}(\bar q,q;w)\big]_{ab}$,
\begin{eqnarray}
K^{(I)}_{ab}(\bar q, q;w)&=&
\sum_{n=0}^\infty\,V^{(I+)}_{ab}(\sqrt{s};n)\,\big[{Y}_{n}^{(+)}(\bar q,
q;w)\big]_{ab}
\nonumber\\
&+&\sum_{n=0}^\infty\,V^{(I-)}_{ab}(\sqrt{s};n)\,\big[{Y}_{n}^{(-)}(\bar
q, q;w)\big]_{ab} \;,
\label{not-def}
\end{eqnarray}
with $s= w^2$, $n= J-{\textstyle{1\over 2}}$ and a set of
energy-dependent effective interaction kernels
$V^{(I\pm)}_{ab}(\sqrt{s};n)$. The projectors are constructed in
a way that states  corresponding to different projectors with
given total angular momentum $J=n+{\textstyle{1\over 2}}$ do not
couple in the Bethe-Salpeter equation (\ref{BS-coupled}). Any
given projector ${Y}_{n}^{(+)}(\bar q, q;w)$ is a finite
polynomial in the Dirac matrices $\gamma_5$ and $\gamma_\mu $ and
the four-momenta $q_\mu, \bar q_\mu, w_\mu $. A further crucial
property of the projectors is that they are free of any
kinematical singularities in $q \cdot w$, $\bar q \cdot q$, $q^2$
and $\bar q^2$. Only then the set of projectors decouples in the
Bethe-Salpeter equation (\ref{BS-coupled}) in any frame.

In general the Bethe-Salpeter interaction kernel $K^{(I)}_{ab}(\bar
q, q;w)$ may have additional contributions not of the form
(\ref{not-def}). Such terms do not lead to s-channel unitarity cuts
if iterated in the Bethe-Salpeter equation, since, by construction
of the projectors, they vanish for on-shell kinematics. In
\cite{LK} it is shown explicitly that these terms may be considered as
part of
the irreducible loop corrections of an effective interaction kernel
$\bar K^{(I)}_{ab}(\bar q, q;w)$. Note that the form of the
projectors implies a particular off-shell behavior of the
interaction kernel. From a field theoretical point of view the
off-shell structure of the interaction kernel and also that of the
scattering amplitude is not defined because it may be changed at
will by redefining the hadronic interpolating fields
\cite{off-shell,Fearing}. However, as was emphasized in \cite{LK} the
proper treatment of the Bethe-Salpeter equation requires a systematic on-shell
reduction as to avoid an unphysical and uncontrolled dependence on the
choice of chiral coordinates or the choice of interpolating fields. A covariant
on-shell reduction can be defined unambiguously with respect to the
covariant set of projectors that solve the Bethe-Salpeter equation for any choice
of quasi-local interaction terms. The systematic on-shell reduction of the
Bethe-Salpeter interaction kernel then leads to an effective interaction kernel
of the form imposed in (\ref{not-def}).

Given an interaction kernel of the form (\ref{not-def}), the
coupled-channel scattering amplitudes $T^{(I)}_{ab}(\bar k,k;w)$
can be decomposed accordingly:
\begin{eqnarray}
T^{(I)}_{ab}(\bar q, q;w)&=&
\sum_{n=0}^\infty\,M^{(I+)}_{ab}(\sqrt{s};n)\,
\big[{Y}_{n}^{(+)}(\bar q, q;w)\big]_{ab}
\nonumber\\
&+&\sum_{n=0}^\infty\,M^{(I-)}_{ab}(\sqrt{s};n)\,
\big[{Y}_{n}^{(-)}(\bar q, q;w)\big]_{ab} \;.
\label{t-sum}
\end{eqnarray}
The reduced amplitudes $M^{(I\pm)}_{ab}(\sqrt{s};n)$ are given by
\begin{eqnarray}
M^{(I\pm)}_{ab}(\sqrt{s};n)
&=& \Bigg[\left( 1- V^{(I\pm )}_{}(\sqrt{s};n)\,J^{(I\pm
)}_{}(\sqrt{s};n)\right)^{-1} V^{(I\pm
)}_{}(\sqrt{s};n)\Bigg]_{ab} \,,
\label{result-loop:ab}
\end{eqnarray}
where the loop matrix $J^{(I\pm)}_{ab}(\sqrt{s};n)$ is defined by
\begin{eqnarray}
&& J^{(\pm)}_{ab}(\sqrt{s};n)\,\Big[{Y}^{(\pm)}_n(\bar q,q;w)\Big]_{ab}
= -i\int \frac{d^4l}{(2\pi)^4}\,\Big[{Y}^{(\pm)}_{n}(\bar
q,l;w)\Big]_{ac}\,
\nonumber\\
&& \qquad \qquad  \quad \;
\times \, S_{B(c)
B(d)}(w-l)\,D_{\Phi(c)\,\Phi(d)}(l)\,\Big[{Y}^{(\pm)}_n(l,q;w)]_{db}\;.
\label{j-n-def}
\end{eqnarray}
We stress that the definition of the loop functions in
(\ref{j-n-def}) is consistent only because the projectors are constructed to
carry good total angular momentum $J$. A priori it is unclear
that the left hand side of (\ref{j-n-def}) does not involve
additional terms not proportional to $Y^{(\pm)}_n$.
Furthermore, since the integral in (\ref{j-n-def}) diverges, the loop functions are
defined only in conjunction with a suitable renormalization scheme, to be discussed below. The
consistency of (\ref{j-n-def}) is verified by explicit
calculations.

Before we proceed with the presentation of the projectors and loop
functions, we first discuss the general form of the interaction
kernel used in this work. It may be obtained from a meson exchange
model or more systematically from the chiral Lagrangian. Here we
use the concept of an effective field theory in a sense that
differs somewhat from common practice. Usually this concept is used
to describe the physics close to threshold, like e.g. in chiral
perturbation theory, which is based on a systematic expansion in
powers of the momentum \cite{Weinberg}. In this paper we
construct an effective field theory for meson-nucleon scattering in
the resonance region. The philosophy of this approach is to
approximate the slowly varying interaction kernels $K_{ab}$ with
simple coupling functions, and to treat rescattering in the
s-channel explicitly. By allowing an arbitrary energy dependence of
$K_{ab}$, one could certainly describe the phase shifts at any
energy, as long as all open channels are included. However, such a
model has essentially no predictive power. In our scheme, the
predictive power results from the assumption that the interaction
kernel is slowly varying in energy in the relevant window 1.4 GeV$
<\sqrt{s}< 1.8$ GeV.

This assumption is based on the following observation.
A small scale, which implies a strong energy dependence of the
scattering amplitude, is a reflection of nearby singularities.
In general the partial-wave
amplitudes have an analytic structure which is more complicated
than the s-channel singularities generated by rescattering. There
are left-hand cuts generated by $t$- and $u$-channel exchanges of
mesons and baryons, which are not treated explicitly in our approach.
These cuts are located in an unphysical region, but the exchange of
light mesons, e.g. the pion, may give rise to a rapid energy
variation close to threshold because the branch point is located
relatively close to the physical region. However, far above
threshold, the energy variation due to the left-hand cuts is
unimportant, and the interaction kernel can be approximated by
slowly varying function. We have checked by explicit calculation
that, for the partial waves considered, this is the case in the
energy region of interest here. Consequently, we employ the
following simple ansatz for the interaction kernel:
\begin{eqnarray}
V^{(\pm)}_{ab}(\sqrt{s};n) \simeq \frac{s}{m_N^2} \,g^{(\pm)}_{n,ab} \;,
\label{ansatz}
\end{eqnarray}
where the linear dependence in (\ref{ansatz}) on $s$ is imposed in
order to guarantee a regular behaviour of the scattering amplitude
at $\sqrt{s}=0$ (see (\ref{cov-proj})). The linear dependence in
$s$ removes an unphysical kinematical singularity of the scattering
amplitudes at $s=0$, which is due to the form of the
projectors\footnote{An alternative, equally viable, way to
eliminate this singularity is to include a factor proportional to
$s$ in the definition of the projection operators and take the
interaction kernel independent of $s$.} \cite{LK}. Although this
simple interpolation perhaps does not yield an accurate
representation of the scattering amplitude outside the window of
applicability, because it does not account for e.g. the $t$- and
$u$-channel singularities, we expect at least a qualitative
description also there.

In this context, one may worry that the s-channel contributions of
baryon ground states and resonance to the interaction kernel may
lead to a rapid energy variation in our energy window 1.4 GeV$
<\sqrt{s}< 1.8$ GeV. Our ansatz (\ref{ansatz}) is justified by the
following observations. In this work we consider only s-wave
interactions in the vector-meson nucleon channels. Because of
parity conservation these couple only to the s- and d-wave
pion-nucleon channels. In these channels the large $N_c$ baryon
ground-state states, the positive parity baryon octet and decuplet
states, do not lead to pole structures in the corresponding
kernels\footnote{The partial wave content of the various projectors
will be identified below.}, $V^{(+)}_{ab}(\sqrt{s};0)$ and
$V^{(-)}_{ab}(\sqrt{s};1)$. Furthermore, the large $N_c$ excited
states are not explicitly included in (\ref{ansatz}) since we aim
to generate those as resonances of the ground-state degrees of
freedom \cite{LK}. The basis for this strategy is the hypothesis
that large $N_c$ counting rules can be applied to the interaction
kernel but not to the scattering amplitude. This is
supported by the following argument. The reducible diagrams summed
by the Bethe-Salpeter equation are typically enhanced by a factor
$2\pi$ compared to irreducible diagrams. Thus, a reducible diagram,
which in the large $N_c$ expansion is suppressed by a factor
$1/N_c$, acquires a total factor $2\pi/N_c$. For $N_c=3$ this
factor is about $2$, which indicates that the large $N_c$
expansion is not useful for reducible diagrams. Consequently, it is
unclear whether baryon resonances are already present in the
interaction kernel, i.e. included in the irreducible diagrams, or
generated by coupled-channel dynamics, i.e. by including reducible
diagrams. The fact that baryon resonances typically have large
decay widths we take as an indication that indeed the
coupled-channel dynamics is the driving mechanism for the creation
of baryon resonances.

We return to the set of projectors and recall those that are
relevant for elastic pion-nucleon scattering \cite{LK}
\begin{eqnarray}
&&\big[{Y}^{(\pm )}_{n}(\bar
q,q;w)\big]_{11}=\frac{1}{2}\,\Bigg(\frac{\wslash}{\sqrt{w^2}}\pm
1\Bigg)\,
\bar Y_{n+1}(\bar q,q;w)
\nonumber\\
&&\;\;\;\;\;\;\;\;\;\;-\frac{1}{2}\,\Bigg(  \barqslash -\frac{w\cdot
\bar q}{w^2}\,\wslash \Bigg)
\Bigg(\frac{\wslash}{\sqrt{w^2}} \mp 1\Bigg)\,
\Bigg( \qslash -\frac{w\cdot q}{w^2}\,\wslash \Bigg)
\bar Y_{n}(\bar q,q;w)\;,
\nonumber\\
&&\bar Y_{n}(\bar q,q;w)=
\sum_{k=0}^{[(n-1)/2]}\,\frac{(-)^k\,(2\,n-2\,k) !}{2^n\,k !\,(n-k)
!\,(n-2\,k -1) !}\,Y_{\bar q \bar q}^{k}\,Y_{\bar q q}^{n-2\,k-1}\,Y_{q
q}^{k}\;,
\nonumber\\
&&Y_{\bar q \bar q}=\frac{(w\cdot \bar q )\,(\bar q\cdot w)}{w^2}
-\bar q \cdot \bar q
\;,\;\;\;
Y_{q  q}=\frac{(w\cdot  q )\,( q\cdot w)}{w^2} -q \cdot  q \;,
\nonumber\\
&&Y_{\bar q q}=\frac{(w\cdot \bar q )\,(q\cdot w)}{w^2} -\bar q \cdot q
\;.
\label{cov-proj}
\end{eqnarray}
We list the leading order projectors $[{Y}_{n}^{(\pm)}]_{11}$ for
s-,p- and d-wave scattering with total angular momentum
$J=\frac{1}{2},\frac{3}{2}$ explicitly
\begin{eqnarray}
\big[{Y}_{0}^{(\pm )}(\bar q,q;w)\big]_{11} &=& \frac{1}{2}\,\left(
\frac{\wslash}{\sqrt{w^2}}\pm 1 \right)\;,
\nonumber\\
\big[{Y}_{1}^{(\pm)}(\bar q,q;w)\big]_{11} &=& \frac{3}{2}\,\left(
\frac{\wslash}{\sqrt{w^2}}\pm 1 \right)
\left(\frac{(\bar q \cdot w )\,(w \cdot q)}{w^2} -\big( \bar q\cdot
q\big)\right)
\nonumber\\
&-&\frac{1}{2}\,\Bigg(  \barqslash -\frac{w\cdot \bar q}{w^2}\,\wslash
\Bigg)
\Bigg(\frac{\wslash}{\sqrt{w^2}}\mp 1\Bigg)\,
\Bigg( \qslash -\frac{w\cdot q}{w^2}\,\wslash \Bigg)\;.
\label{}
\end{eqnarray}
As demonstrated in \cite{LK} these projectors indeed lead to
reduced scattering amplitudes $M^{(I\pm)}_{ab}(\sqrt{s};n) $ of
the form (\ref{result-loop:ab}). The loop functions are
\begin{eqnarray}
J^{(I\pm)}_{11}(\sqrt{s}; n) &=&
p_{\pi N}^{2\,n}\,I_{\pi N}(\sqrt{s}\,)
\left( \frac{\sqrt{s}}{2}+ \frac{m_N^2-m_\pi^2}{2\,\sqrt{s}}\pm m_N
\right)\; ,
\label{result-loop}
\end{eqnarray}
with the master function $I_{\pi N}(\sqrt{s}\,)$ given by
\begin{eqnarray}
\!\!\! \!\!\! I_{\pi N}(\sqrt{s}\,)&=& -i\,\int \frac{d^4
l}{(2\pi)^4}\,\frac{1}{l^2-m_\pi^2}\,\frac{1}{(l-w)^2-m_N^2}
\nonumber\\
&=&\frac{1}{16\,\pi^2}
\left( \frac{p_{\pi N}}{\sqrt{s}}\,
\left( \ln \left(1-\frac{s-2\,p_{\pi N}\,\sqrt{s}}{m_\pi^2+m_N^2}
\right)
-\ln \left(1-\frac{s+2\,p_{\pi N}\sqrt{s}}{m_\pi^2+m_N^2} \right)\right)
\right.
\nonumber\\
&+&\left.
\left(\frac{1}{2}\,\frac{m_\pi^2+m_N^2}{m_\pi^2-m_N^2}
-\frac{m_\pi^2-m_N^2}{2\,s}
\right)
\,\ln \left( \frac{m_\pi^2}{m_N^2}\right) +1 \right)+I_{\pi N}(0)\;,
\nonumber\\
p_{\pi N}^2 &=&
\frac{s}{4}-\frac{m_N^2+m_\pi^2}{2}+\frac{(m_N^2-m_\pi^2)^2}{4\,s}  \;.
\label{ipin-analytic}
\end{eqnarray}
We recall \cite{LK} that the projectors $Y_{n}^{(\pm)}$ with $J=n+\frac{1}{2}$ require
a physical mass dependent renormalization procedure with vanishing renormalized tadpole
contributions, $I_\pi \to 0$ and $I_N \to 0$. The loop-orthogonality of projectors
with different total angular momenta $J$ and $J'$ follows only in a
renormalization scheme where tadpole contributions vanish effectively. Moreover, we observe
that the pion and nucleon tadpole diagrams $I_\pi$ and $I_N$
determine the master-loop function $I_{\pi N}(\sqrt{s}\,)$ at $s=0$ via the algebraic identity
\begin{eqnarray}
I_{\pi N}(\sqrt{s}=0) = \frac{-1}{m_N^2-m_\pi^2}\left(I_N-I_\pi\right)
\; .
\label{mast-id}
\end{eqnarray}
The identity (\ref{mast-id}), which can be verified in dimensional
regularization, then leads to $I_{\pi N}(0) =0$ in (\ref{ipin-analytic}).
We conclude that the renormalization prescription, to drop tadpole contributions
in reducible diagrams, leads to finite expressions for
all loop functions defined by the set of covariant projectors (\ref{j-n-def}).

The reduced amplitudes\footnote{For simplicity we suppress the
isospin index in this paragraph.}$M^{(\pm)}_{11}(\sqrt{s};n)$ of
(\ref{t-sum}) have a unique interpretation in terms of the
pion-nucleon scattering phase shifts $\delta^{(l)}_{J=l\pm
\frac{1}{2}}$ and inelasticity parameter $\eta^{(l)}_{J=l\pm
\frac{1}{2}} $ \cite{LK}. The matching can be carried out in a
straightforward manner by identifying the s-channel unitarity cut
of $M^{(\pm)}_{11}(\sqrt{s};n)$ and the partial-wave amplitudes
$f^{(J)}_{\pm }(s)$
\begin{eqnarray}
&&p_{\pi N}\,f^{(J)}_{\pm}(s) =
\frac{1}{2\,i}\left(\eta^{(l)}_{J=l\pm \frac{1}{2}}(s)\,
e^{2\,i\,\delta^{(l)}_{J=l\pm \frac{1}{2}} (s)}-1 \right) \;,
\nonumber\\
&&f^{(J)}_{\pm }(s) = \frac{(p_{\pi N})^{2\,J-1}}{8\,\pi\,\sqrt{s}}
\left( \frac{\sqrt{s}}{2}+\frac{m_N^2-m_\pi^2}{2\,\sqrt{s}} \pm
m_N\right)
M^{(\pm )}(\sqrt{s};J-{\textstyle{1\over 2}}) \;.
\label{match}
\end{eqnarray}
The angular momentum $l$ of a given projector can be inferred from
the threshold behavior of the associated loop function. From
(\ref{result-loop}) it follows that $\Im \,J_{\pi
N}^{(+)}(\sqrt{s};n)
\sim p_{\pi N}^{2\,n+1}$ but $\Im \,J_{\pi N}^{(-)} (\sqrt{s};n)
\sim p_{\pi N}^{2\,n+3}$. This confirms the angular momentum
assignment implicit in (\ref{match}). In the presence of inelastic
channels the inelasticity parameter is smaller than the value for
elastic scattering, $\eta =1$.

The projectors of the $\eta N$, $K \Lambda$ and $K \Sigma $ channels
are identical to the ones of the pion-nucleon channels. Therefore they
need not to
be specified separately. For example in the isospin ${\textstyle {1\over
2}}$ channel
one has $\big[{Y}^{(\pm )}_{n}\big]_{11}= \big[{Y}^{(\pm
)}_{n}\big]_{xy}$
for $x,y\in \{1,5,6,7\}$. In order to arrive at a convenient notation
for the
pion-induced pseudo-scalar meson production cross sections it is useful
to generalize the
notation of (\ref{match}). Take as an example the pion-induced $\eta $
production amplitude for which
we write
\begin{eqnarray}
&& N^{(\pm)}_{\pi N} = \frac{\sqrt{s}}{2}+
\frac{m_N^2-m_\pi^2}{2\,\sqrt{s}}\pm m_N
\,, \qquad
N^{(\pm)}_{\eta \,N} = \frac{\sqrt{s}}{2}+
\frac{m_N^2-m_\eta^2}{2\,\sqrt{s}}\pm m_N\,,
\nonumber\\
&& f^{({\textstyle{1\over2}}J)}_{\pm ,\pi N\,\to \eta N}(\sqrt{s}\,) =
\frac{\sqrt{N^{(\pm)}_{\pi N} \, N^{(\pm)}_{\eta N}}}{8\,\pi\,\sqrt{s}}
\,
\Big(p_{\eta N}\,p_{\pi N}\Big)^{J-{\textstyle{1\over2}}}\,
M^{({\textstyle{1\over2}}\pm)}_{15}(\sqrt{s},J-{\textstyle{1\over2}})\;,
\label{eta-notation}
\end{eqnarray}
with  $\sqrt{s}=\sqrt{m_N^2+p_{\eta N}^2}+\sqrt{m_\eta^2+p_{\eta N}^2}$.
The expressions for the pion-induced pseudo-scalar meson production
cross sections
have the simple form
\begin{eqnarray}
&& \!\!\!\! \!\!\!\!\sigma_{\pi^-\,p\,\to\,\,\eta \,\,n \;} \;=
4\,\pi \,\frac{p_{\eta N}}{p_{\pi N}}\,
 \frac{2}{3}\,\Big| f^{(\frac{1}{2}\frac{1}{2} )}_{+,\pi N \rightarrow
\eta N}\Big|^2  \;,
\\
&& \!\!\!\!\!\!\!\!\sigma_{\pi^-p\,\to\,K^0 \,\Sigma^0 } =
4\,\pi \,\frac{p_{K \Sigma}}{p_{\pi N}}\,\frac{2}{9}
 \Big| f^{(\frac{1}{2}\frac{1}{2} )}_{+,\pi N \rightarrow K \Sigma }
- f^{(\frac{3}{2}\frac{1}{2} )}_{+,\pi N \rightarrow K \Sigma }\Big|^2
\;,
\nonumber\\
&& \!\!\!\!\!\!\!\!\sigma_{\pi^-p\,\to\,K^+ \,\Sigma^- } =
4\,\pi \,\frac{p_{K \Sigma}}{p_{\pi N}}\,\frac{1}{9}
 \Big|2\, f^{(\frac{1}{2}\frac{1}{2} )}_{+,\pi N \rightarrow K \Sigma }
+ f^{(\frac{3}{2}\frac{1}{2} )}_{+,\pi N \rightarrow K \Sigma }\Big|^2
\;,
\nonumber\\
&& \!\!\!\!\!\!\!\! \sigma_{\pi^+p\,\to\,K^+ \,\Sigma^+ } =
4\,\pi \,\frac{p_{K \Sigma}}{p_{\pi N}}\,\Big|
 f^{(\frac{3}{2}\frac{1}{2} )}_{+,\pi N \rightarrow K \Sigma }\Big|^2
\;,
\quad \!\!
\sigma_{\pi^-p\,\to\,K^0\,\Lambda } =
4\,\pi \,\frac{p_{K \Lambda}}{p_{\pi N}}\,
 \frac{2}{3}\,\Big| f^{(\frac{1}{2}\frac{1}{2} )}_{+,\pi N \rightarrow
K \Lambda}\Big|^2  \;, \nonumber
\label{eta-prod}
\end{eqnarray}
where we have adopted a notation  for the $K\Lambda$ and $K \Sigma
$ production amplitudes, which is analogous to
(\ref{eta-notation}). We do not consider d-wave contribution in the
$\eta N$, $K \Lambda$ or $K
\Sigma $ channels. Only with improved quality of the differential
cross section data it would be possible to determine those channels
reliably in our present approach. Moreover we do not address effects from
isospin violation. We use the approximate
mass parameters $m_N = 939$ MeV, $m_\Lambda =1116$ MeV, $m_\Sigma = 1195$ MeV,
$m_\pi = 139$ MeV, $m_\eta = 547$ MeV and $m_K = 494$ MeV throughout this work.

\section{Pion-induced vector meson and isobar production}

Pion-nucleon scattering at energies larger than $\sqrt{s}>
m_N+2\,m_\pi$ is complicated by the coupling to inelastic 2-pion
channels. In our approach we approximate the three-body final
state by including the $\rho N$ and $\Delta \pi $ channels \cite{Manley}.
The decay of the $\rho$ meson and $\Delta$ isobar are taken into
account by incorporating their energy-dependent widths. We do not consider
the $\sigma N$ channel since it would be difficult to discriminate its effects from
the $\pi \Delta$ channel which in our scheme is constrained by data only rather
indirectly. The latter channel is viewed as an effective channel that takes into account
some residual effects from the $\sigma N$ channel. We, however, consider the $\omega N$
channel because its mass is almost degenerate with that of the $\rho$ meson and
the $\omega$ meson is expected to couple strongly to baryons.

Before constructing the appropriate projectors required in
(\ref{not-def}) it is useful to describe the general framework
within which our expressions should be considered. We assume that
the vector-meson fields satisfy the Proca equation which follows
from the Lagrangian density
\begin{eqnarray}
{\mathcal L} &=& -{\textstyle{\zeta^{(0)}_\rho \over 4}} \,
\Big( \partial_\mu \,\vec \rho_\nu \,-\partial_\nu\,\vec\rho_\mu \Big)\,
 \Big(\partial^\mu \,\vec\rho^{\,\nu} \,-\partial^\nu\,\vec \rho^{\,\mu}
\Big)
\,+{\textstyle{1\over
2}}\,{\m0}^2_\rho\,\vec\rho_\mu\,\vec\rho^{\,\mu }
\nonumber\\
&&-{\textstyle{\zeta^{(0)}_\omega \over 4}} \, \Big( \partial_\mu
\,\omega_\nu -\partial_\nu\,\omega_\mu \Big)\, \Big(\partial^\mu
\,\omega^\nu -\partial^\nu\,\omega^\mu \Big) +{\textstyle{1\over
2}}\,{\m0}^2_\omega \,\omega_\mu\,\omega^\mu \,, \label{proca}
\end{eqnarray}
where we, for later convenience, allow for a wave-function
renormalization. As is well known, the corresponding Feynman propagator
sustains only (4-dimensionally) transverse modes. When interactions of the
$\rho $ meson with other fields are taken into account, its propagator is
modified by the polarization tensor $\Pi_{\mu \nu }^{(\rho )}(q)$, which
in vacuum has two independent components, $\Pi_{L}^{(\rho )}(q)$ and
$\Pi_{T}^{(\rho )}(q)$. The $\rho $-meson propagator is then given by
\begin{eqnarray}
\!\!\!D^{(\rho)}_{\mu \nu} (q) =  \left(g_{\mu
\nu}-\frac{q_\mu\,q_\nu}{q^2}\right)
\frac{1}{\zeta^{(0)}_\rho\,q^2-{\m0}_\rho^2 -\Pi_T^{(\rho)}(q^2)}
+ \frac{q_\mu\,q_\nu}{q^2}\,\frac{1}{{\m0}_\rho^2-\Pi_L^{(\rho
)}(q^2)} \;. \label{def-LT}
\end{eqnarray}
The first term is the transverse component of the propagator, while the
second one is the longitudinal one. In a scheme, where the
vector meson couples to a conserved current, the vector-meson
nucleon scattering amplitude is transverse, i.e. $\bar q_\mu\,T_{\rho N \to
\rho N}^{\mu \nu} (\bar q, q;w)= T_{\rho N \to \rho N}^{\mu \nu} (\bar q, q;w)\,q_\nu= 0 $
vanish \cite{Sakurai,KLZ}. However, in an approximation scheme, where the corresponding
current is not conserved, the longitudinal self energy differs from zero and the
longitudinal propagator  may exhibit singularities, which correspond to the propagation
of unphysical degrees of freedom. Thus it is preferable to work in a scheme with conserved
currents, when possible. Nevertheless, in non-perturbative  approximations, like the
present one, there is no systematic way of implementing current conservation. We
therefore employ an alternative scheme, which deals with this in a pragmatic way.

From a practical point of view, the minimal
requirement a model should satisfy is that a possible longitudinal
mode should have a mass large enough so that it does not lead to
cut structures within the domain of applicability of the
model. The idea is to start with the most general interaction
terms which then allows one to push the mass of the unphysical mode
to a sufficiently high value by a suitable choice of parameters.
In such an approach the vector-meson scattering amplitudes need
not be transverse. Once the propagation of longitudinal modes is
suppressed, the longitudinal parts of the scattering amplitude are
not relevant.

We illustrate this discussion with a simple calculation. To
leading order with no interactions the vector-meson fields satisfy
the Proca equation. We add the typical interaction terms required for the modelling of the
vector-meson decay widths
\begin{eqnarray}
{\mathcal L}_{int } &=&  g_{\rho \,\pi \pi}\, {\vec \rho}\,^\mu
\left((\partial_\mu \vec \pi ) \times \vec \pi \right)
+2\,\frac{g_{\rho \,\omega \,\pi}}{m_\omega } \, \epsilon^{\alpha
\beta \mu \nu }\, \left( \partial_\nu \,\omega_\beta \right)
\left( \partial_\mu \,\vec \rho_\alpha \right) \vec \pi \, .
\label{rho-width-model}
\end{eqnarray}
Of course there may be more terms, which however are not relevant
for the present discussion. Consider the lowest-order contribution
to the $\rho$-meson polarization tensor $\Pi^{(\rho)}_{\mu
\nu}(q)$, which in $d$ dimensions is given by
\begin{eqnarray}
\Pi^{(\rho)}_{\mu \nu} (q)
&=&-i\,\frac{g_{\rho \pi \pi}^2}{\mu^{d-4}}\,\int \frac{d^d\,l}{(2\,\pi
)^d}
\left(2\,l-q \right)_\mu \,\left(2\,l-q\right)_\nu\,
D_\pi (l)\,D_\pi (l-q)
\nonumber\\
&=& \Pi^{(\rho)}_{T} (q^2) \left( g_{\mu
\nu}-\frac{q_\mu\,q_\nu}{q^2} \right) + \Pi^{(\rho)}_{L}
(q^2)\,\frac{q_\mu\,q_\nu}{q^2}  \;.
\label{two-pi-loop}
\end{eqnarray}
The contribution of the second term in (\ref{rho-width-model}) to
the $\rho$-meson self energy is small, and we neglect it. However,
for the self energy of the $\omega$ meson, which we discuss below,
it plays a crucial role. The transverse and longitudinal components
can be expressed in terms of the logarithmically divergent master
loop function $I_{\pi \pi}(q^2)$ and the tadpole $I_\pi$
\begin{eqnarray}
&& \Pi^{(\rho)}_T (q^2) =\frac{g_{\rho \pi \pi}^2}{1-d}\,\Bigg(
\Big( q^2-4\,m_\pi^2 \Big)\, \Big( I_{\pi \pi}(\sqrt{q^2}\,)-I_{\pi
\pi}(0 ) \Big)
\nonumber\\
&& \qquad \qquad \qquad \quad
+ \left( \frac{q^2}{2\,m_\pi^2}\,(2-d) - 2\,(1-d) \right) \,I_\pi
\Bigg) \,,\quad
\nonumber\\
&& \Pi^{(\rho)}_L (q^2) =
-2\,g_{\rho \pi \pi}^2\,I_\pi \,,
\label{example:rho}
\end{eqnarray}
where
\begin{eqnarray}
&& I_{\pi \pi }(q^2) =-\frac{i}{\mu^{d-4}}\, \int \frac{d^d\,l}{(2\,\pi
)^d}\,D_\pi (l)\,D_\pi(q-l) \,,\quad
\nonumber\\
&& I_{\pi } =\frac{i}{\mu^{d-4}}\, \int \frac{d^d\,l}{(2\,\pi
)^d}\,D_\pi (l) \,, \qquad
I_{\pi \pi}(0) = \frac{2-d}{2\,m_\pi^2}\,I_\pi  \,.
\label{ex-def}
\end{eqnarray}
In our model, the $\rho$-meson polarization tensor does, according
to (\ref{example:rho}), exhibit a longitudinal component. In
contrast to that, the longitudinal polarization would be
identically zero in a formulation, in which the $\rho$ meson couples to
a conserved current. We observe, however, that in our scheme the
$\rho$-meson propagator is also well behaved supporting exclusively
the propagation of transverse modes for any choice of parameters.
This is a consequence of the fact that the longitudinal polarization
$\Pi_L^{(\rho)}(q^2)$ is a constant (see (\ref{def-LT})). It is
clear that if additional interaction terms would lead to a
contribution of the form $\Pi_L^{(\rho)}(q^2) = c_\rho\,q^2$, which
gave rise to unphysical propagation of longitudinal modes with mass
squared of $m^2_\rho/c_\rho $,  an appropriate counter term
$-c_\rho\,(\partial^\mu \rho_\mu )\,(\partial^\nu \,\rho_\nu )$
could be designed to exactly cancel this troublesome contribution.
This has some similarity to the St\"uckelberg mechanism
\cite{Stueckelberg}.

The renormalization of the simple model (\ref{rho-width-model}) is
performed by absorbing the tadpole
terms in (\ref{example:rho}) into the bare parameters $\m0_\rho $ and
$\zeta_\rho^{(0)}$
\begin{eqnarray}
\tilde m_\rho^2 = {\m0}_\rho^2 + 2\,g_{\rho \pi \pi}^2\,I_\pi \,,
\qquad \tilde \zeta_\rho = \zeta_\rho^{(0)} - \frac{g_{\rho \pi
\pi}^2}{2\,m_\pi^2}\,\frac{2-d}{1-d} \,. \label{ren-rho}
\end{eqnarray}
After the renormalization (\ref{ren-rho}) is implied the limit $d
\to 4$ can be performed smoothly. All together we arrive at the
renormalized propagator
\begin{eqnarray}
&&\!\!\!\!D_{\mu \nu}^{(\rho)} (q) =  \left( g_{\mu \nu}-
\frac{q_\mu\,q_\nu }{q^2 } \right) D_T^{(\rho)}(q^2)
+ \frac{q_\mu\,q_\nu }{q^2 } \, D_L^{(\rho)}(q^2) \,,\quad
D_L^{(\rho)}(q^2)= \frac{1}{\tilde m_\rho^2}\,,
\nonumber\\
&&\!\!\!\! D_T^{(\rho)}(q^2) =  \frac{1}{ \zeta_\rho \,(q^2- m_\rho^2)
+ \frac{1}{3}\,g_{\rho \pi \pi}^2\,
\Big( q^2-4\,m_\pi^2 \Big)\, \Big( I_{\pi \pi}(\sqrt{q^2}\,)-\Re I_{\pi
\pi}(m_\rho ) \Big) } \,,
\label{rho-explicit}
\end{eqnarray}
where we performed an additional finite renormalization introducing the
physical $\rho$-mass parameter
$m_\rho \simeq 779$ MeV and the wave-function renormalization parameter
$\zeta_\rho$. The explicit form of the
master-loop function $I_{\pi \pi} (q^2)$ follows from the expression
$I_{\pi N}(\sqrt{q^2}\,)$ given already in
(\ref{ipin-analytic}) if the formal limit $m_N \to m_\pi$ is applied.
The coupling constant
$g_{\rho \pi \pi} \simeq 5.79$ and $\zeta_\rho \simeq 1$ follow if the
model (\ref{rho-width-model}) is used
to reproduce the p-wave $\pi \pi$ scattering phase shift
\cite{rho-model}.

For later convenience it is useful to represent the $\rho$-meson
propagator in terms of
its spectral function $\rho_\rho (q^2)$,
\begin{eqnarray}
\rho_\rho (q^2) = \Im D_T^{(\rho)}(q^2) \,\Theta (\Lambda^2_\rho-q^2)
\;, \qquad
\int_{4\,m_\pi^2}^{\Lambda_\rho^2 } \,\frac{dm^2}{\pi}\,\rho_\rho (m^2)
= 1\,,
\label{def-rho-norm}
\end{eqnarray}
where the cutoff  $\Lambda_\rho $ is required if one demands a
normalized spectral function.
The spectral density $\rho_\rho (m^2)$ is normalized to $\pi $  by the
choice $\Lambda_\rho \simeq 1.34 $ GeV.
It is convenient to work with a normalized spectral function because
that will avoid a particular class
of divergent terms which would otherwise arise if the dressed
$\rho$-meson propagator is used inside
loop integrals. From a physical point of view this is well justified
because the simple model (\ref{rho-width-model})
is certainly incorrect at invariant masses $q^2$ much larger than the
squared $\rho$-meson mass.

We turn to the $\omega$ meson. The interaction vertex given in
(\ref{rho-width-model})
leads to the following expression for the transverse and longitudinal
polarization functions
$\Pi_{L,T}^{(\omega )} (q^2)$
\begin{eqnarray}
&&
\Pi_T^{(\omega )} (q^2) = \frac{i\,4\,g_{\omega \rho
\pi}^2}{m_\omega^2\,\mu^{4-d}}
\frac{3\,d-6}{1-d} \int \frac{d^d\,l}{(2\,\pi )^d}\, \Big(
q^2\,l^2-(l\cdot q)^2\Big)
D^{(\rho)}_{T} (l)\,D_\pi (l-q) \,,
\nonumber\\
&& \Pi_L^{(\omega )} (q^2) = 0\,,
\label{}
\end{eqnarray}
where we consider the effect of a dressed $\rho$-meson propagator in the
loop integral \cite{GSW}. Since for the $\omega$ meson the interaction terms
(\ref{rho-width-model}) lead to a vanishing longitudinal polarization tensor,
the $\omega$-meson propagator is well behaved supporting transverse
modes only. The loop integration is most economically performed applying
the spectral representation of the $\rho$-meson propagator. We write
\begin{eqnarray}
\Pi_T^{(\omega )} (q^2) = \int_{4\,m_\pi^2}^{\Lambda_\rho^2} \,d
m^2\,\rho_\rho (m^2)\,\Pi_T^{(\omega )} (q^2,m^2) \,,
\label{omega-kernel}
\end{eqnarray}
where we use the normalized spectral function introduced in
(\ref{rho-explicit}, \ref{def-rho-norm}). The integral
kernel in (\ref{omega-kernel}) is evaluated in space-time dimension $d$,
\begin{eqnarray}
&& \!\!\!
\Pi_T^{(\omega )} (q^2, m_\rho^2) = \frac{4\,g_{\omega \rho
\pi}^2}{m_\omega^2\,\mu^{4-d}}
\frac{3\,d-6}{1-d} \Bigg( {\textstyle{1\over 2}}\,q^2\, I_\pi +
{\textstyle{1\over 4}}\,q^2\,\big(q^2-m_\pi^2-3\,m_\rho^2 \big)\,I_{\pi
\rho}(0,m_\rho^2 )
\nonumber\\
&& \qquad   +
\left({\textstyle{1\over 4}}\,\Big(
q^2+m_\rho^2-m_\pi^2\Big)^2-q^2\,m_\rho^2 \right)
\Big( I_{\pi \rho}(q^2, m_\rho^2) -  I_{\pi \rho}(0, m_\rho^2) \Big)
\Bigg) \,,
\label{}
\end{eqnarray}
where we encounter exclusively the master-loop function $I_{\pi
\rho}(q^2,m_\rho^2)$
and the tadpole terms $I_\pi$ and $I_\rho$. At $d=4$ the loop function
$I_{\pi \rho}(q^2,m_\rho^2)$
can be inferred from the analogous expression $I_{\pi N}(\sqrt{q^2})$
given in (\ref{ipin-analytic}).
Note that here we make explicit the dependence of the master-loop
function
$I_{ab}(\sqrt{q^2}\,) \to  I_{ab}(q^2, m_b^2)$ on the mass parameter
$m_b$.
The expression for the $\omega$-meson polarization is renormalized by
absorbing the diverging terms
proportional to $q^2$ and $(q^2)^2$ into appropriate wave-function
renormalization terms. For the
renormalized polarization function $\Pi_{T,R}^{(\omega )} (q^2,
m_\rho^2)$ we use
\begin{eqnarray}
&& \!\!\!
\Pi_{T,R}^{(\omega )} (q^2, m_\rho^2) = -4\,\frac{g_{\omega \rho
\pi}^2}{m_\omega^2}
\left(\Big( q^2+m_\rho^2-m_\pi^2\Big)^2-4\,q^2\,m_\rho^2 \right)
\nonumber\\
&& \qquad \qquad \qquad \times
\Big( I_{\pi \rho}(q^2, m_\rho^2) - \Re I_{\pi \rho}(m_\omega^2,
m_\rho^2) \Big)  \,,
\label{omega-explicit}
\end{eqnarray}
where we performed a subtraction at the physical mass $m_\omega \simeq
782$ MeV for convenience. This
permits the identification $\m0_\omega = m_\omega $ in (\ref{proca}).
The parameter
$g_{\omega \rho \pi } \simeq 7.57 $ is adjusted to reproduce the
hadronic decay width of the
$\omega$ meson with $\Gamma_\omega \simeq  8.4$ MeV where we assume here
$\zeta_\omega =1$ for simplicity.
By analogy with the $\rho$-meson spectral function
(see (\ref{def-rho-norm})) we introduce a cutoff $\Lambda_\omega \simeq
821$ MeV. This leads to the
correct normalization of the $\omega$-meson spectral function
$\rho_\omega (m^2) $, which is defined by
(\ref{omega-explicit},\ref{def-LT}).

\subsection{Projector approach for vector mesons}

We construct the relativistic vector-meson production amplitudes with
total angular momentum $J={\textstyle{1\over 2}}$ and
$J={\textstyle{3\over 2}}$
by specifying the appropriate projectors introduced in (\ref{t-sum}). It
is sufficient to present the projectors
$\big[Y_{n,\,\mu}^{(\pm )}\big]_{13}$, $\big[Y_{n,\,\mu }^{(\pm
)}\big]_{31}$
and $\big[Y_{n,\,\mu \nu}^{(\pm)}\big]_{33}$ of the inelastic $\rho N$
channel
for the isospin $I={\textstyle{1\over 2}}$ states. The remaining
projectors follow
then by analogy. For example in the isospin $1/2$ channel it holds
$\big[Y_{n}^{(\pm )}\big]_{x3}=\big[Y_{n}^{(\pm )}\big]_{x4}$.
The $J={\textstyle{1\over 2}}$ projectors are:
\begin{eqnarray}
&& \big[Y_{0,\,\mu}^{(+)}(\bar q, q;w) \big]_{31} =
-\frac{1}{2\,\sqrt{3}}\left(\gamma_\mu -\frac{w_\mu}{w^2}\,\wslash
\right) \left(1-\frac{\wslash }{\sqrt{w^2}}\right) i\,\gamma_5 \;,
\nonumber\\
&& \big[Y_{0,\,\mu}^{(+)}(\bar q, q;w) \big]_{13} =
+\frac{1}{2\,\sqrt{3}}\,i\, \gamma_5\left(1-\frac{\wslash
}{\sqrt{w^2}}\right) \left(\gamma_\mu -\frac{w_\mu}{w^2}\,\wslash
\right) \;,
\nonumber\\
&& \big[Y_{0,\,\mu \nu }^{(+)}(\bar q, q;w) \big]_{33} =
\frac{1}{6}\left(\gamma_\mu -\frac{w_\mu}{w^2}\,\wslash \right)
\left(1-\frac{\wslash }{\sqrt{w^2}}\right)
\left(\gamma_\nu -\frac{w_\nu}{w^2}\,\wslash \right) \;.
\label{proj-v1}
\end{eqnarray}
The $J={\textstyle{3\over 2}}$ projectors are:
\begin{eqnarray}
&& \big[Y_{1,\,\mu}^{(-)}(\bar q, q;w) \big]_{31} =
-\frac{\sqrt{3}}{2}\left(1+\frac{\wslash }{\sqrt{w^2}}\right)
 \left(q_\mu-\frac{w \cdot q}{w^2}\,w_\mu \right) i\,\gamma_5
\nonumber\\
&& \qquad \qquad \qquad  \quad +\;
\frac{1}{2\,\sqrt{3}}\left(\gamma_\mu -\frac{w_\mu}{w^2}\,\wslash
\right)
\left(1-\frac{\wslash }{\sqrt{w^2}}\right)
\left(\qslash -\frac{w \cdot q}{w^2}\,\wslash \right) i\,\gamma_5 \;,
\nonumber\\
&& \big[Y_{1,\,\mu}^{(-)}(\bar q, q;w) \big]_{13} =
+\frac{\sqrt{3}}{2}\,i\,\gamma_5 \left(1+\frac{\wslash
}{\sqrt{w^2}}\right)\left(\bar q_\mu-\frac{\bar q\cdot w}{w^2}
\,w_\mu\right)
\nonumber\\
&& \qquad \qquad \qquad  \quad -\;
\frac{1}{2\,\sqrt{3}}\,i\,\gamma_5 \left(\barqslash -\frac{\bar q \cdot
w}{w^2}\,\wslash \right)
\left(1-\frac{\wslash }{\sqrt{w^2}}\right)
\left(\gamma_\mu -\frac{w_\mu}{w^2}\,\wslash \right)  \;,
\nonumber\\
&& \big[Y_{1,\,\mu \nu }^{(-)}(\bar q, q;w) \big]_{33} =
\frac{1}{2}\left(1+\frac{\wslash }{\sqrt{w^2}}\right)  \left(g_{\mu
\nu}-\frac{w_\mu\,w_\nu}{w^2} \right)
\nonumber\\
&& \qquad \qquad \qquad  \quad -\;
\frac{1}{6}\left(\gamma_\mu -\frac{w_\mu}{w^2}\,\wslash \right)
\left(1-\frac{\wslash }{\sqrt{w^2}}\right)
\left(\gamma_\nu -\frac{w_\nu}{w^2}\,\wslash \right) \;.
\label{proj-v3}
\end{eqnarray}
It is a straightforward exercise to verify that the projectors
(\ref{proj-v1},\ref{proj-v3}) have indeed the acclaimed property
(\ref{result-loop:ab},\ref{j-n-def}), provided that tadpole
contributions are dropped systematically. We emphasize that the
property (\ref{j-n-def}) is sufficient as to prove that the
projectors must carry good total angular momentum $J$. This is
evident because only states with good total angular momentum $J$
may decouple the Bethe-Salpeter equation. Note that the
$J=\frac{3}{2}$ projectors couple to additional projectors, which
probe the vector-meson nucleon state with angular momentum $L >0$.
In general the projectors for the vector mesons acquire an
additional 3 $\times$ 3 matrix structure which simply reflects the
various angular momenta $L$ accessible at given $J$. For the case
$J={\textstyle{1\over 2 }} $ that matrix collapses to a $2\times
2$ matrix describing the coupling of the two states $ | \rho^{(T)}
N, \lambda\!= \!{\textstyle{1\over 2 }} , J\!=\!{\textstyle{1\over
2 }}\rangle$ and $ | \rho^{(L)} N, \lambda \! =
\!{\textstyle{1\over 2 }} , J\!=\!{\textstyle{1\over 2 }}\rangle $
or alternatively the coupling of a s-wave and d-wave state. A
third state $ | \rho^{(T)} N,\lambda = {\textstyle{1\over 2 }},
J\!=\!{\textstyle{3\over 2 }} \rangle $ existed only for $J>
{\textstyle{1\over 2 }} $. That follows because a transverse vector
meson, $\rho^{(T)}$, cannot couple to a nucleon
forming a helicity $\lambda = {\textstyle{3\over 2 }} $ state if
the vector meson and nucleon sit in a relative s-wave. In the
$J={\textstyle{1\over 2 }} $ channel the matrix structure of the
projectors is defined with respect to specific linear combinations
of helicity states $| J\!=\!{\textstyle{1\over 2 }};1 \rangle$ and
$| J\!=\!{\textstyle{1\over 2 }};2 \rangle$ defined as follows,
\begin{eqnarray}
| J\!=\!{\textstyle{1\over 2 }};1 \rangle &=& | \rho^{(T)} N,
\lambda\!= \!{\textstyle{1\over 2 }} , J\!=\!{\textstyle{1\over 2
}}\rangle + {\textstyle{1\over \sqrt{2} }}\sqrt{1+\vec
q\,^2/m_\rho^2}\,| \rho^{(L)} N, \lambda\!= \!{\textstyle{1\over 2
}} ,
J\!=\!{\textstyle{1\over 2 }}\rangle \,, \nonumber\\
| J\!=\!{\textstyle{1\over 2 }};2 \rangle &=& | \rho^{(L)} N,
\lambda\!= \!{\textstyle{1\over 2 }} , J\!=\!{\textstyle{1\over 2
}}\rangle \,,\label{j12-trafo}
\end{eqnarray}
where $\vec q$ is the vector meson's three momentum. We point out
that the particular form of the linear combination is unique in the
sense that only with respect to the $| J\!=\!{\textstyle{1\over 2
}};1 \rangle$ and $| J\!=\!{\textstyle{1\over 2 }};2 \rangle$
states it is possible to construct an associated covariant
projector algebra $P_{ij}$ with $i,j=1,2$. For other linear
combinations there are kinematic singularities which prohibit a
covariant interpretation of the helicity states defined in the
center of mass frame only. We observe that the coupling of the
$P_{11}$ to the $P_{22}$ projector is kinematically suppressed by
the phase space factor $p_{\rho N}^2$. In other words the
associated loop functions $J_{ij}$ with $i,j=1,2$ have the
threshold property $\Im J_{11} \sim p_{\rho N} $, $\Im J_{12} =
\Im J_{21} \sim p_{\rho N}^3 $ and $\Im J_{22} \sim p_{\rho N}^5
$. It is evident that the transformation (\ref{j12-trafo}) leads
to non-diagonal loop functions because it is not a unitary
transformation. In more physical terms the transformation
(\ref{j12-trafo}) is required because both helicity states $|
\rho^{(T)} N, {\textstyle{1\over 2 }} , {\textstyle{1\over 2
}}\rangle $ and $ | \rho^{(L)} N, {\textstyle{1\over 2 }},
{\textstyle{1\over 2 }}\rangle $ carry non-zero s-wave components.
The projectors given in (\ref{proj-v1}) correspond to the leading
$| J\!=\!{\textstyle{1\over 2 }};1 \rangle$ state.

Similarly the implicit matrix structure of the projectors of the
$J={\textstyle{3\over 2 }}$ channel are defined with respect to
the three states $| J\!=\!{\textstyle{3\over 2 }};i \rangle$ with
$i=1,2,3$ reflecting the presence of one s-wave but two d-wave
states. We find the unique transformation
\begin{eqnarray}
| J\!=\!{\textstyle{3\over 2 }};1 \rangle &=& | \rho^{(T)} N,
 \!{\textstyle{3\over 2 }} , {\textstyle{3\over 2
}}\rangle + \sqrt{{\textstyle{2\over 3} }}\,\sqrt{1+\vec
q\,^2/m_\rho^2}\,| \rho^{(L)} N,  \!{\textstyle{1\over 2 }} ,
{\textstyle{3\over 2 }}\rangle +{\textstyle{1\over \sqrt{3} }}\,|
\rho^{(T)} N,
 \!{\textstyle{1\over 2 }} , {\textstyle{3\over 2
}}\rangle\,, \nonumber\\
| J\!=\!{\textstyle{3\over 2 }};2 \rangle &=& | \rho^{(L)} N,
\!{\textstyle{1\over 2 }} , {\textstyle{1\over 2
}}\rangle+{\textstyle{1\over \sqrt{2} }}\sqrt{1+\vec q\,
^2/m_\rho^2}\,| \rho^{(T)} N,
 \!{\textstyle{1\over 2 }} , {\textstyle{3\over 2
}}\rangle \,, \nonumber\\
| J\!=\!{\textstyle{3\over 2 }};3 \rangle &=& | \rho^{(T)} N,
 \!{\textstyle{1\over 2 }} , {\textstyle{3\over 2
}}\rangle\,,
 \label{j33-trafo}
\end{eqnarray}
which leads to well defined associated covariant projectors
$P_{ij}$ with $i,j=1,2,3$. The projectors given in (\ref{proj-v3})
correspond to the leading $| J\!=\!{\textstyle{3\over 2 }};1
\rangle$ state. In this work we neglect the contribution of the
$P_{ij}$ projectors with $i,j \neq 1 $ since their contributions
are suppressed close to the vector-meson nucleon threshold by at
least the phase space factor $p_{\rho N}^2$. To reliably describe
the behavior away from threshold to order $p_{\rho N }^2$ would
require the systematic inclusion of p-wave projectors also.

The leading angular momentum $L$ of a given projector
$Y^{(\pm)}_n$ follows in a straightforward manner by inspecting
the threshold behavior of its associated loop function (see
(\ref{j-n-def})). The required loop functions are first specified
in the limit of zero-width vector mesons. In this approximation
the loop functions of the $\rho N$ channels are
\begin{eqnarray}
&& J^{(I+)}_{33}( \sqrt{s},0)
= N_{\rho N}(\sqrt{s}\,)\,I_{\rho N}(\sqrt{s}\,)\, , \qquad
J^{(I-)}_{33}( \sqrt{s},1) =J^{(I+)}_{33}( \sqrt{s},0)\;,
\nonumber\\ \nonumber\\
&& N_{\rho N} =\left(1+\frac{p_{\rho N}^2}{3\,m_\rho^2}\right)\left(
\frac{\sqrt{s}}{2}+ \frac{m_N^2-m_\rho^2}{2\,\sqrt{s}}+ m_N \right)\,,
\label{om-spec}
\end{eqnarray}
where $\sqrt{s}=\sqrt{m_N^2+p_{\rho N}^2}+\sqrt{m_\rho^2+p_{\rho N}^2}$.
The master-loop function $I_{\rho N}(\sqrt{s}\,)$ is defined at hand of
the pion-nucleon loop
function $I_{\pi N}(\sqrt{s}\,)$ in (\ref{ipin-analytic}). The form of
the loop functions
in (\ref{om-spec}) at threshold indeed confirm that the considered
vector-meson nucleon
channels are s-wave like with $L=0$. It is understood that $I_{\rho
N}(\sqrt{s}=0)=0$ holds.
Our final vector-meson loop functions are obtained by folding the result
(\ref{om-spec})
with the normalized vector-meson spectral functions $\rho_{\rho}(q^2)$
and
$\rho_{\omega}(q^2)$ as specified in
(\ref{rho-explicit},\ref{def-rho-norm},\ref{omega-explicit}).
This is analogous to the folding performed when evaluating the
$\omega$-meson polarization tensor (\ref{omega-kernel}).
Our final loop functions follow with
\begin{eqnarray}
J^{(I+)}_{33}( \sqrt{s},0) \to
\int_{4\,m_\pi^2}^{\infty } \,\frac{d\,q^2}{\pi}\,\rho_\rho (q^2)\,
J^{(I+)}_{33}( \sqrt{s},0) \;,
\label{rho-loop}
\end{eqnarray}
where the integral in (\ref{rho-loop}) extends over the implicit
dependence of
$J^{(I+)}_{33}( \sqrt{s},0)$ on the vector-meson mass $m_\rho^2\to q^2$.
The analogous
folding is assumed for the $J={\textstyle{3\over 2}}$ loop function. The
loop functions of the $\omega N $ channels are given by complete analogy
with
(\ref{om-spec}) and (\ref{rho-loop}).

In order to have a convenient notation for the pion-induced
vector-meson production cross sections it is useful to generalize
(\ref{match}) and introduce corresponding partial-wave amplitudes
$f^{(IJ\pm)}(\sqrt{s}\,)$. For example in the isospin
${\textstyle{1\over 2}}$ channel we write:
\begin{eqnarray}
&& f^{({\textstyle{1\over 2}} {\textstyle{1\over 2}}+)}_{\pi
N\,\to \rho N}(\sqrt{s}\,) = \frac{\sqrt{N_{\pi N}^{(+)}\,N_{\rho
N} }}{8\,\pi\,\sqrt{s}}\, \,M^{({\textstyle{1\over
2}}+)}_{31}(\sqrt{s},0)\;,
\nonumber\\
&& f^{({\textstyle{1\over 2}} {\textstyle{3\over 2}}-)}_{\pi
N\,\to \rho N}(\sqrt{s}\,) = \frac{\sqrt{N_{\pi N}^{(-)}\,N_{\rho
N} }}{8\,\pi\,\sqrt{s}}\,p_{\pi N} \,M^{({\textstyle{1\over 2}}
-)}_{31}(\sqrt{s},1)\;. \label{f-pn:vn}
\end{eqnarray}

We derive the vector-meson production cross section in terms
of the partial-wave amplitudes in (\ref{f-pn:vn}). In the zero-width
approximation one finds
\begin{eqnarray}
\sigma_{\pi^-p\,\to\,\rho^0 \,n \;} \;&=& 4\,\pi \,\frac{p_{\rho
N}}{p_{\pi N}}\, \frac{2}{9}\, \Bigg(\Big|
f^{(\frac{1}{2}\frac{1}{2}+ )}_{\pi N \rightarrow \rho N}-
f^{(\frac{3}{2}\frac{1}{2}+ )}_{\pi N \rightarrow \rho N}\Big|^2
\nonumber\\
&&\qquad \quad  +2\,\Big| f^{(\frac{1}{2}\frac{3}{2}- )}_{\pi N
\rightarrow \rho N}-f^{(\frac{3}{2}\frac{3}{2}- )}_{\pi N
\rightarrow \rho N}\Big|^2 \Bigg) \;,
\nonumber\\
\sigma_{\pi^+p\,\to\,\rho^+ \,p \;} \;&=& 4\,\pi \,\frac{p_{\rho
N}}{p_{\pi N}}\, \Bigg(\Big|f^{(\frac{3}{2}\frac{1}{2}+ )}_{\pi N
\rightarrow \rho N}\Big|^2 +2\,\Big| f^{(\frac{3}{2}\frac{3}{2}
-)}_{\pi N \rightarrow \rho N}\Big|^2 \Bigg) \;,
\nonumber\\
\sigma_{\pi^-p\,\to\,\omega \,n \;} \;&=& 4\,\pi \,\frac{p_{\omega
N}}{p_{\pi N}}\,\frac{2}{3}\, \left( \Big|
f^{(\frac{1}{2}\frac{1}{2}+ )}_{\pi N \rightarrow \omega N}\Big|^2
+2\,\Big| f^{(\frac{1}{2}\frac{3}{2} -)}_{\pi N \rightarrow \omega
N}\Big|^2\right) \;. \label{vec-prod}
\end{eqnarray}
We include the energy-dependent width of the vector mesons by folding
the
cross section with the appropriate spectral functions.
For example the $\rho$-meson width is included by the replacement,
\begin{eqnarray}
\sigma_{\pi N \to  \rho N} \to
\int_{4\,m_\pi^2}^{\infty } \,\frac{d\,q^2}{\pi}\,\rho_\rho (q^2)\,
\sigma_{\pi N \to  \rho N} \;,
\label{rho-cross}
\end{eqnarray}
where the integrals in (\ref{rho-cross}) extents over the implicit
dependence of the bare cross section
$\sigma_{\pi N \to  \rho N}$ on the vector-meson mass $m_\rho^2\to q^2$.
We observe that the folding
integral in (\ref{rho-cross}) and (\ref{rho-loop}) are closely related,
which basically reflects the unitarity condition. This is clearly illustrated if the cross
sections are written in terms of the
original amplitudes $M$ rather than the reduced amplitudes $f$. For
instance one may rewrite the $\rho^+$
production cross section in (\ref{vec-prod}) as follows:
\begin{eqnarray}
\sigma_{\pi^+p\,\to\,\rho^+ \,p \;} \;&=&
4\,\pi \,\frac{\Im \,J_{\pi N}^{(1/2)}\,
\Im \,J_{\rho N }^{(1/2)}}{p_{\pi N}^2} \,
\Big|M_{\pi N \rightarrow \rho N}^{(1/2)}\Big|^2
\nonumber\\
&+&8\,\pi \,\frac{\Im \,J_{\pi N}^{(3/2)}\,
\Im \,J_{\rho N }^{(3/2)}}{p_{\pi N}^2} \,
\Big|M_{\pi N \rightarrow \rho N}^{(3/2)}\Big|^2 \,,
\label{}
\end{eqnarray}
where $M_{\pi N \rightarrow \rho N}^{(1/2)}(\sqrt{s})=
M_{31}^{(3/2,+)}(\sqrt{s},0)$ and
$M_{\pi N \rightarrow \rho N}^{(3/2)}(\sqrt{s})=
M_{31}^{(3/2,-)}(\sqrt{s},0)$. Similarly we write identify the loop functions
$J_{\rho N }^{(1/2)}(\sqrt{s}\,)= J_{\rho N }^{(3/2)}(\sqrt{s}\,)=
J_{33}^{(3/2,+)}(\sqrt{s},0)$  (see (\ref{j-n-def})).

\section{Photon-induced meson production}

We wish to generalize our model to describe electromagnetic processes as
to
constrain our model parameters by data from photon-induced meson
production off
the nucleon. Provided that all hadronic amplitudes of the previous
sections are well
determined by the available hadronic data set a generalized vector-meson
dominance assumption would predict a huge amount of electromagnetic
processes.
Conversely, given a vector-meson dominance assumption the photon-induced
reaction data may be used as a consistency check of the hadronic model.
In this work we follow a mixed strategy in the sense that we use the
ample
data on the photon-induced meson production off the nucleon to further
constrain
our hadronic reaction amplitudes. This is advantageous because we found
out that
the present day hadronic data set does not completely determine our
hadronic model
parameters\footnote{In view of this we question our preliminary results, which did
not consider the photon induced reactions systematically, presented in
\cite{Hirschegg}.}.
Moreover, following this strategy avoids the fitting of some data
over-precisely. One should not forget that the hadronic model
constructed in the previous sections is rather crude and approximative.
Thus it would be misleading to fix part of the model parameters by fitting
some precise data point with small error bars that are much smaller than those
expected from the accuracy level of the model. We emphasize that there
still remains a strong predictive power of our scheme to the extent that many channels
for which good data are not available so far, or which are unlikely to be
measured, will be predicted.

It is useful to describe the framework within which we consider
electromagnetic interactions and construct a generalized vector-meson dominance in our
scheme. We will generalize Sakurai's vector meson dominance conjecture by
assuming that the electromagnetic quasi-local 4-point interaction vertices
of our model are directly proportional to corresponding vertices involving the
$\rho$- and $\omega$-meson fields. The photon field $A_\mu$ must be included
in a gauge invariant manner. For instance the kinetic term of the photon field,
\begin{eqnarray}
{\mathcal L}_{kin}= -{\textstyle{1 \over 4}}\,F_{\mu \nu}\,F^{\mu \nu}
\,,
\label{}
\end{eqnarray}
is expressed in terms of the field strength tensor $F_{\mu \nu}=
\partial_\mu \,A_\nu -\partial_\nu \,A_\mu $. Since
in our approach the vector-meson fields do not couple exclusively to
conserved vector currents, an interaction vertex
$A^\mu \,\rho_\mu^{(0)}$ or $A^\mu \,\omega_\mu^{}$ suggested by Sakurai
\cite{Sakurai} would be in conflict with
electromagnetic gauge invariance. We introduce the gauge invariant
interaction terms instead
\begin{eqnarray}
{\mathcal L}_{int} &=&
i\,e\,A_\mu \,\rho_\nu^{(-)}
\,\Big(\partial_\mu\,\rho_\nu^{(+)}-\partial_\nu\,\rho_\mu^{(+)} \Big)
-i\,e\,A_\mu \,\rho_\nu^{(+)}
\,\Big(\partial_\mu\,\rho_\nu^{(-)}-\partial_\nu\,\rho_\mu^{(-)} \Big)
\nonumber\\
&-& e^2 \,\Big( A\cdot A \Big) \,\Big(\rho^{(+)} \cdot \rho^{(-)}\Big)
+ e^2 \,\Big(A \cdot \rho^{(+)} \Big)\,\Big(A \cdot \rho^{(-)} \Big)
\nonumber\\
&+&\frac{f_\rho }{2\,m_\rho^2} \,F^{\mu \nu}\,\Big(\partial_\mu
\,\rho^{(0)}_\nu-\partial_\nu \,\rho^{(0)}_\mu \Big)
+\frac{f_\omega }{2\,m_\omega^2} \,F^{\mu \nu}\,\Big(\partial_\mu
\,\omega_\nu -\partial_\nu \,\omega_\mu\Big) +\cdots  \,,
\label{gam-def}
\end{eqnarray}
where we use
\begin{eqnarray}
&& \rho^{(0)}_\mu = \rho^{(3)}_\mu \,, \quad \rho^{(\pm)}_\mu =
{\textstyle{1\over{\sqrt{2}}}}\,(\rho^{(1)}_\mu \pm \,i\,\rho^{(2)}_\mu) \;.
\label{}
\end{eqnarray}
One observes that the terms in (\ref{gam-def}) proportional to $f_\rho $
and $f_\omega $ induce a vector-meson dominance
type behavior for virtual photons giving the photon hadronic
$\rho^{(0)}_\mu$ and $\omega_\mu $ components as
suggested by phenomenology. The parameters $f_\omega$ and $f_\rho $ are
determined by the $e^+\,e^-$ decay widths
of the vector mesons
\begin{eqnarray}
m_\rho\,\Gamma_{\rho  \to e^-\,e^+ } &=&
\frac{\alpha\,f_\rho^2}{3\,m_\rho^4}\,
\sqrt{1-\frac{4\,m_e^2}{m_\rho^2}}\left(2\,m_e^2+m_\rho^2\right)\;,
\nonumber\\
m_\omega\,\Gamma_{\omega \to e^-\,e^+ } &=&
\frac{\alpha\,f_\omega^2}{3\,m_\omega^4}\,
\sqrt{1-\frac{4\,m_e^2}{m_\omega^2}}\left(2\,m_e^2+m_\omega^2\right)\;,
\label{}
\end{eqnarray}
where $\alpha= e^2/(4\pi) \simeq 1/137.04 $.
The empirical decay widths $\Gamma_{\omega\rightarrow e^+e^-} = 0.60\pm
0.02 $ keV and
$\Gamma_{\rho_0\rightarrow e^+e^-} = 6.77\pm 0.32 $ keV lead to the
values
\begin{eqnarray}
f_\rho \simeq 0.0357\,{\rm GeV}^2
\,, \qquad
f_\omega \simeq 0.0109\,{\rm GeV}^2 \;.
\label{}
\end{eqnarray}

On the other hand, it is evident that the terms proportional to $f_\rho $
or $f_\omega $ do not contribute to reactions involving real photons. Therefore
in our present approach, in which we do not consider explicit hadronic 3-point
vertices but only quasi-local two-body interaction terms, we need to introduce
effective 4-point vertices which involve the electromagnetic field tensor
$F_{\mu \nu }$ explicitly. Only then the scheme is capable to describe
the process of photon-induced meson production. We show a few typical
examples we have in mind
\begin{eqnarray}
F^{\mu \nu }\,\bar N  \,\gamma_\mu \,\vec \tau\, \partial_\nu \,(N
\, \,\vec \pi )\,, \quad F^{\mu \nu }\,\bar N  \,\gamma_\mu \,\vec
\tau\, N \,\vec \rho_\nu \,, \quad \cdots \,.
\label{quasi-local-a}
\end{eqnarray}
It is instructive to discuss such quasi-local two-body interaction terms
(\ref{quasi-local-a}) in view of the
vector-meson dominance picture. In the vector-meson dominance approach
one insists that the photon couples to
hadronic states exclusively via its hadronic components. Interaction
terms as given in (\ref{quasi-local-a})
are typically not considered. Rather they would effectively be generated
by corresponding hadronic
vertices of the generic form
\begin{eqnarray}
\omega^\mu\,\bar N  \,\gamma_\mu \,\vec \tau\,  N \, \,\vec \pi
\,, \quad \omega^\mu\,\bar N  \,\vec \tau\, N \,\vec \rho_\mu \,,
\quad \cdots \,. \label{quasi-local-b}
\end{eqnarray}
According to the original conjecture of Sakurai \cite{Sakurai} the
vector-meson converts into a real photon
via the interaction terms $A^\mu\,\rho_\mu^{(0)} $ and
$A^\mu\,\omega_\mu $. In our scheme the terms
$A^\mu\,\rho_\mu^{(0)} $ and $A^\mu\,\omega_\mu $ are not allowed. In
order to implement the successful vector-meson dominance assumption it is
therefore natural to directly relate the strength of the
$\gamma N \to X $ and $\rho^{(0)}\,N \to X$, $\omega\,N \to X$ vertices,
where $X$ is any hadronic final state.
Note that our construction has similarities with the construction of
Kroll, Lee and Zumino \cite{KLZ} only that
here we refrain from insisting that the vector mesons couple to
conserved currents only.

We specify the generalized form of the vector-meson dominance conjecture
as applied for the direct
photon-induced production vertices $K^{(I),\mu}_{\gamma N \rightarrow
X}(\bar q,q;w)$:
\begin{eqnarray}
&& {\textstyle{1\over 2}}\,\Big( K^{(\frac{1}{2}),\,\mu}_{\gamma
\,p \rightarrow X}+K^{(\frac{1}{2}),\,\mu}_{\gamma \,n \rightarrow
X} \Big) (\bar q,q;w)=  e\,K^{(\frac{1}{2})}_{\nu, \,\omega N
\rightarrow X}(\bar q,q;w) \,\Gamma^{\nu \mu}_S (q;w)\,,
\nonumber\\
&& {\textstyle{1\over 2}}\,\Big( K^{(\frac{1}{2}),\,\mu}_{\gamma
\,p \rightarrow X}-K^{(\frac{1}{2}),\,\mu}_{\gamma \,n \rightarrow
X} \Big) (\bar q,q;w)=
{\textstyle{e\over\sqrt{3}}}\,K^{(\frac{1}{2})}_{\nu, \,\rho N
\rightarrow X}(\bar q,q;w) \,\Gamma^{\nu \mu}_V (q;w)\,,
\nonumber\\
&& K^{(\frac{3}{2}),\,\mu}_{\gamma \,p \rightarrow X} (\bar
q,q;w)= e\,\sqrt{{\textstyle{2\over 3}}}\, \,
K^{(\frac{3}{2})}_{\nu,\, \rho N\rightarrow X}(\bar
q,q;w)\,\Gamma^{\nu \mu}_V (q;w)\,,
\nonumber\\
&& K^{(\frac{3}{2}),\,\mu}_{\gamma \,n \rightarrow X}(\bar q,q;w)= e\,
\sqrt{{\textstyle{2\over 3}}}\, K^{(\frac{3}{2})}_{\nu,\, \rho
N\rightarrow X}(\bar q,q;w)\,\Gamma^{\nu \mu}_V (q;w)\,,
\label{gamma-ansatz:k}
\end{eqnarray}
where $X$ stands for any hadronic two-body final state with
isospin $I$. The transverse isoscalar  $\Gamma_{S}^{\mu \nu}(q;w)$
and isovector $\Gamma_{V}^{\mu \nu}(q;w)$ objects are to be specified
transition
tensors. Note that the factors $1/\sqrt{3}$ and $\sqrt{2/3}$ in
the $\rho$-meson contribution follow from the definition of the
isospin state in (\ref{r-def}). We emphasize that our generalized
vector-meson
dominance assumption is defined with respect to the unique projector
algebra introduced in the previous section. As is evident from
(\ref{gamma-ansatz:k}) we rely on a particular off-shell interpretation
of the hadronic production vertex, which is given by the form
of the projectors.
For the transition tensor $\Gamma^{\nu
\mu}_{S(V)}(q;w )$ we make the most general ansatz,
\begin{eqnarray}
\Gamma^{\mu \nu}_{S(V)} (q;w)&=& \Gamma^{\mu \nu,+}_{S(V)}(q;w)
+\Gamma^{\mu \nu,-}_{S(V)}(q;w) \,, \nonumber\\ \Gamma^{\mu \nu,
\pm }_{S(V)}(q;w) &=& \frac{g^{(\pm)}_{S(V),1}}{m_\omega}
\,\frac{1}{2}\,\Big( 1\pm\frac{\wslash }{\sqrt{w^2}} \Big)\,\Big(
\Big(\qslash- \frac{w \cdot q}{w^2}\,\wslash \Big) \,g^{\mu \nu} -
q^\mu \,\Big(\gamma^\nu - \frac{w^\nu}{w^2}\,\wslash \Big)\Big)
\nonumber\\
&+&\frac{g^{(\pm)}_{S(V),2}}{m_\omega} \,\frac{1}{2}\,\Big(
1\pm\frac{\wslash }{\sqrt{w^2}} \Big)\,\Big( \frac{w \cdot q
}{\sqrt{w^2}}  \,g^{\mu \nu} -\frac{q^\mu \,w^\nu}{\sqrt{w^2}}
\Big)
\nonumber\\
&+&\frac{g^{(\pm)}_{S(V),3}}{m_\omega^2} \,\frac{1}{2}\,\Big(
1\pm\frac{\wslash }{\sqrt{w^2}} \Big)\,\Big( q^2 \,g^{\mu \nu}
-q^\mu \,q^\nu \Big)\,, \label{tensor-ansatz}
\end{eqnarray}
that is compatible with a vertex involving the field strength
tensor $F_{\mu \nu }$ (see (\ref{quasi-local-a})). We note that we
can not exclude the possibility that the coupling constants
$g_{S,i}^{(\pm)}$ and $g_{V,i}^{(\pm)}$ are weakly
dependent on the total energy $\sqrt{s}$ in some polynomial
fashion. We emphasize, however, that any further possible term in
(\ref{tensor-ansatz}) would either renormalize the already
existing terms or be in conflict with gauge invariance. Here we
exploit the fact that our model does not consider any  of the
hadronic 3-point interaction vertices explicitly. Such interaction
terms are supposed to be integrated out with their effect being
absorbed into the quasi-local 2-body interaction terms. As a
direct consequence the photon can couple to hadrons in $K_{\gamma N \to
X}$
exclusively via the electromagnetic field strength
tensor $F_{\mu \nu}$. The parameters $g^{(\pm)}_{S,i}$ and
$g^{(\pm)}_{V,i}$ will be determined by the photon-induced meson
production data. In fact the coupling constants
$g^{(\pm)}_{S(V),3}$ do not contribute in processes with real
photons obviously. The ratios $\sqrt{3}\,g^{(\pm)}_{S,i}/g^{(\pm)}_{V,i}$
may be identified with $f_\omega /f_\rho \simeq 0.3 $ naively. In
our scheme we consider all parameters $g^{(\pm)}_{S(V),i}$ as
free, because we would like to avoid the poorly controlled
extrapolation from virtual photon kinematics with $q^2 \simeq
m_\omega^2$ down to the real photon point at $q^2 =0$.

The generalized vector-meson dominance conjecture (\ref{gamma-ansatz:k})
leads to the $\gamma$-induced production amplitudes,
$T^{(I),\,\mu}_{\gamma \,N \rightarrow X}(\bar q,q;w)$,
\begin{eqnarray}
&& {\textstyle{1\over 2}}\,\Big( T^{(\frac{1}{2}),\,\mu}_{\gamma
\,p \rightarrow X}+T^{(\frac{1}{2}),\,\mu}_{\gamma \,n \rightarrow
X} \Big) (\bar q,q;w)=  e\,T^{(\frac{1}{2})}_{\nu, \,\omega N
\rightarrow X}(\bar q,q;w) \,\Gamma^{\nu \mu}_S (q;w)\,,
\nonumber\\
&& {\textstyle{1\over 2}}\,\Big( T^{(\frac{1}{2}),\,\mu}_{\gamma
\,p \rightarrow X}-T^{(\frac{1}{2}),\,\mu}_{\gamma \,n \rightarrow
X} \Big) (\bar q,q;w)=
{\textstyle{e\over\sqrt{3}}}\,T^{(\frac{1}{2})}_{\nu, \,\rho N
\rightarrow X}(\bar q,q;w) \,\Gamma^{\nu \mu}_V (q;w)\,,
\nonumber\\
&& T^{(\frac{3}{2}),\,\mu}_{\gamma \,p \rightarrow X} (\bar
q,q;w)= e\,\sqrt{{\textstyle{2\over 3}}}\, \,
T^{(\frac{3}{2})}_{\nu,\, \rho N\rightarrow X}(\bar
q,q;w)\,\Gamma^{\nu \mu}_V (q;w)\,,
\nonumber\\
&& T^{(\frac{3}{2}),\,\mu}_{\gamma \,n \rightarrow X}(\bar q,q;w)= e\,
\sqrt{{\textstyle{2\over 3}}}\, T^{(\frac{3}{2})}_{\nu,\, \rho
N\rightarrow X}(\bar q,q;w)\,\Gamma^{\nu \mu}_V (q;w)\,,
\label{gamma-ansatz:t}
\end{eqnarray}
that are directly proportional to the corresponding vector-meson
induced production amplitudes. This is a consequence of the
generalized vector-meson dominance conjecture
(\ref{gamma-ansatz:k}) and the particular form of the interaction
kernel (\ref{not-def}).

Given the generalized vector-meson dominance assumption
(\ref{gamma-ansatz:k})
we can now derive photon-induced meson production cross sections in
terms of the hadronic reaction amplitudes of the previous sections.
In analogy to (\ref{f-pn:vn}, \ref{vec-prod}) we express the
production cross sections in terms of reduced reaction amplitudes,
$f^{(I J\pm)}_{h,\gamma \,N \rightarrow X}$, where the lower index $h$
specifies their helicity projection and the signature $\pm$ their parity.
We provide explicit expressions of meson-production cross sections for
reactions
where multipole amplitudes are not established so far,
\begin{eqnarray}
&& \sigma_{\gamma \,p\to \, \eta \,p} = 4\,\pi \, \frac{p_{\eta
N}}{p_{\gamma N}}\, \Big|
f^{(\frac{1}{2}\frac{1}{2}+)}_{\frac{1}{2},\gamma \,p \rightarrow
\eta N}\Big|^2  \,, \qquad \sigma_{\gamma \,p \to \,K^+ \Lambda} =
4\,\pi \, \frac{p_{K \Lambda}}{p_{\gamma N}}\, \Big|
f^{(\frac{1}{2}\frac{1}{2} +)}_{\frac{1}{2},\gamma \,p \rightarrow
K \Lambda}\Big|^2  \,,
\nonumber\\
&&\sigma_{\gamma \,p \to \,K^+ \Sigma^0} = 4\,\pi \, \frac{p_{K
\Sigma}}{p_{\gamma N}}\, \frac{1}{3} \Big|
f^{(\frac{1}{2}\frac{1}{2} +)}_{\frac{1}{2},\gamma \,p \rightarrow
K \Sigma} +\sqrt{2}\,f^{(\frac{3}{2}\frac{1}{2}
+)}_{\frac{1}{2},\gamma \,p \rightarrow K \Sigma} \Big|^2 \,,
\nonumber\\
&&\sigma_{\gamma \,p \to \,K^0 \Sigma^+} = 4\,\pi \, \frac{p_{K
\Sigma}}{p_{\gamma N}}\, \frac{1}{3} \Big|
\sqrt{2}\,f^{(\frac{1}{2}\frac{1}{2}+ )}_{\frac{1}{2},\gamma \,p
\rightarrow K \Sigma} -f^{(\frac{3}{2}\frac{1}{2}
+)}_{\frac{1}{2},\gamma \,p \rightarrow K \Sigma}\Big|^2  \,.
\label{}
\end{eqnarray}
The reduced amplitudes, $f^{(I J\pm)}_{h,\gamma \,N \rightarrow X}$, are
determined
by the invariant amplitudes
$M^{(I\pm)}_{i3}(\sqrt{s};J-{\textstyle{1\over 2}})$ in
the isovector and $M^{(I\pm)}_{i4}(\sqrt{s};J-{\textstyle{1\over 2}})$
in the isoscalar channel. The helicity index $h$ is required for the
$J={\textstyle{3\over 2}}$ channel because the invariant amplitudes
introduced
in (\ref{t-sum}) couples to two distinct multipole amplitudes.
The helicity projection is $h={\textstyle{1\over 2}}$ for
$J={\textstyle{1\over 2}}$ but $h={\textstyle{1\over 2}},
{\textstyle{3\over 2}}$ for $J={\textstyle{3\over 2}}$. Note that, if we
were to consider the
effect of the kinematical suppressed states in (\ref{j12-trafo},
\ref{j33-trafo})
an analogous index would be needed for the hadronic amplitudes. We
illustrate
our notation at hand of the $\pi N$ channels,
\begin{eqnarray}
&& \!\!\!\! {\textstyle{1\over 2}}\,\Big( f^{(\frac{1}{2}
\frac{1}{2}+) }_{\frac{1}{2} ,\gamma \,p\,\to \pi N}(\sqrt{s}\,)
+f^{(\frac{1}{2} \frac{1}{2} +) }_{\frac{1}{2} ,\gamma \,n\,\to
\pi N}(\sqrt{s}\,)\Big)  = \frac{\sqrt{N_{\pi
N}^{(+)}\,N_{\frac{1}{2},\gamma N}^{(S+)} }}{8\,\pi\,\sqrt{s}} \,
M^{(\frac{1}{2} +)}_{14}(\sqrt{s};0)  \;,
\nonumber\\
&& \!\!\!\!f^{(\frac{3}{2} \frac{1}{2} +)}_{\frac{1}{2} ,\gamma
\,p\,\to \pi N}(\sqrt{s}\,) = \frac{\sqrt{N_{\pi
N}^{(+)}\,N_{\frac{1}{2},\gamma N}^{(V+)} }}{8\,\pi\,\sqrt{s} }\,
\sqrt{{\textstyle{2\over 3}}}\,M^{(\frac{3}{2} +)}_{13
}(\sqrt{s};0)\;,
\nonumber\\
&& \!\!\!\!{\textstyle{1\over 2}}\,\Big( f^{(\frac{1}{2}
\frac{3}{2}-)}_{h ,\gamma \,p\,\to \pi
N}(\sqrt{s}\,)+f^{(\frac{1}{2} \frac{3}{2}-)}_{h ,\gamma \,n\,\to
\pi N}(\sqrt{s}\,) \Big)= \frac{\sqrt{N_{\pi N}^{(-)}\,N_{h,\gamma
N}^{(S-)} }}{8\,\pi\,\sqrt{s} }\,p_{\pi N} \, M^{(\frac{1}{2}
-)}_{14}(\sqrt{s};1)  \,,
\nonumber\\
&& \!\!\!\! f^{(\frac{3}{2} \frac{3}{2}-)}_{h ,\gamma \,p\,\to \pi
N}(\sqrt{s}\,) = \frac{\sqrt{N_{\pi N}^{(-)}\,N_{h,\gamma
N}^{(V-)} }}{8\,\pi\,\sqrt{s} }\,p_{\pi N}
\,\sqrt{{\textstyle{2\over 3}}}\,M^{(\frac{3}{2}
-)}_{13}(\sqrt{s};1) \;. \label{f-gammap}
\end{eqnarray}
The difference of proton and neutron production amplitudes
not specified in (\ref{f-gammap}) follows by analogy with
(\ref{gamma-ansatz:t}).
The normalization factor $N_{\pi N}^{(\pm)}$ for the pseudo-scalar
channel was introduced already in (\ref{eta-notation}). It remains to
specify the
normalization factors $N_{h,\gamma N}^{(S(V)\pm)}$ for the initial
state,
\begin{eqnarray}
&&\sqrt{2\,\sqrt{s}\,N_{\frac{1}{2},\gamma N}^{(S(V)+)}} =
e\,\frac{p_{\gamma
N}}{\sqrt{3}}\, \left( 2\,\frac{\sqrt{s}-m_N }{m_\omega }
\,g^{(+)}_{S(V),1} +
\frac{\sqrt{s}+m_N }{m_\omega }\,g^{(+)}_{S(V),2}  \right) \,, \quad
\nonumber\\
&& \sqrt{2\,\sqrt{s}\,N_{\frac{1}{2},\gamma N}^{(S(V)-)}} =
e\,\frac{p_{\gamma
N}}{2\,\sqrt{3}}\, \left( -\frac{ \sqrt{s}-m_N}{m_\omega
}\,g^{(+)}_{S(V),1} + \frac{\sqrt{s}+m_N }{m_\omega }
\,g^{(+)}_{S(V),2} \right) \,,
\nonumber\\
&&\sqrt{2\,\sqrt{s}\, N_{\frac{3}{2},\gamma N}^{(S(V)-)}} =
e\,\frac{p_{\gamma
N}}{2}\, \left( \frac{ \sqrt{s}-m_N}{m_\omega
}\,g^{(+)}_{S(V),1} + \frac{ \sqrt{s}+m_N }{m_\omega }
\,g^{(+)}_{S(V),2} \right) \,. \label{def-norm}
\end{eqnarray}
Note that the normalization factors (\ref{def-norm})
do not involve the vector-meson dominance parameters $g_{S(V),i}^{(-)}$.
The reliable determination of the latter parameters requires the study
of further multipole amplitudes, which couple to $\pi N$ states not
considered
in this work. We imply a notation analogous to (\ref{f-gammap},
\ref{def-norm})
for the $K \Lambda$, $K \Sigma $ and  $\eta N$ channels.

For photon-induced pion production it is more appropriate to compare
with the multipole
analysis \cite{Arndt:gamma}. We identify the electric and magnetic
multipole
amplitudes $E_{l\pm}^{(p,n)}$ and $M_{l\pm}^{(p,n)}$ (see e.g.
\cite{Walker,Moorhouse}) by
\begin{eqnarray}
&&E_{0+}^{(p)}(S_{11}) \,= f^{(\frac{1}{2}
\frac{1}{2}+)}_{\frac{1}{2} ,\gamma \,p\,\to \pi N}(\sqrt{s}\,)
\,, \quad E_{0+}^{(n)}(S_{11}) \,= f^{(\frac{1}{2}
\frac{1}{2}+)}_{\frac{1}{2} ,\gamma \,n\,\to \pi N}(\sqrt{s}\,)\,
,
\nonumber\\
&& E_{2-}^{(p)}(D_{13}) -3\,M_{2-}^{(p)}(D_{13}) =
2\,f^{(\frac{1}{2} \frac{3}{2}-)}_{\frac{1}{2} ,\gamma \,p\,\to
\pi N}(\sqrt{s}\,) \,,
\nonumber\\
&& E_{2-}^{(n)}(D_{13})-3\,M_{2-}^{(n)}(D_{13})  =
2\,f^{(\frac{1}{2} \frac{3}{2}-)}_{\frac{1}{2} ,\gamma \,n\,\to
\pi N}(\sqrt{s}\,)\, ,
\nonumber\\
&& E_{2-}^{(p)}(D_{13})+M_{2-}^{(p)}(D_{13}) =
\frac{2}{\sqrt{3}}\,f^{(\frac{1}{2} \frac{3}{2}-)}_{\frac{3}{2} ,\gamma
\,p\,\to \pi N}(\sqrt{s}\,) \,,
\nonumber\\
&& E_{2-}^{(n)}(D_{13})+M_{2-}^{(n)}(D_{13})  =
\frac{2}{\sqrt{3}}\,f^{(\frac{1}{2} \frac{3}{2}-)}_{\frac{3}{2} ,\gamma
\,n\,\to \pi N}(\sqrt{s}\,)\, ,
\nonumber\\
&& E_{0+}^{(p)}(S_{31}) \,= f^{(\frac{3}{2}
\frac{1}{2}+)}_{\frac{1}{2} ,\gamma \,p\,\to \pi N}(\sqrt{s}\,)
\,,
\nonumber\\
&& E_{2-}^{(p)}(D_{33})-3\,M_{2-}^{(p)}(D_{33}) =
2\,f^{(\frac{3}{2} \frac{3}{2}-)}_{\frac{1}{2} ,\gamma \,p\,\to
\pi N}(\sqrt{s}\,)\,,
\nonumber\\
&& E_{2-}^{(p)}(D_{33})+M_{2-}^{(p)}(D_{33}) =
\frac{2}{\sqrt{3}}\,f^{(\frac{3}{2} \frac{3}{2}-)}_{\frac{3}{2} ,\gamma
\,p\,\to \pi N}(\sqrt{s}\,)
 \,. \label{match-multipole}
\end{eqnarray}
The matching of the multipole amplitudes in
(\ref{match-multipole}) is determined only up to an overall phase
reflecting a particular convention for the production amplitude. The
phase ambiguity reduces to a sign ambiguity if we insist on the
reality of the interaction kernel $V^{(\pm )}(\sqrt{s};\, n)$ in
(\ref{ansatz}). In our present scheme the overall sign of a given
off-diagonal hadronic amplitude with fixed isospin and total angular
momentum
may be flipped by a simultaneous sign change of all off-diagonal
coupling constants in that channel without affecting any hadronic
reaction
cross section considered in this work. It is evident that precise
differential
hadronic production cross sections and polarization data, not available
in most cases,
could eliminate any phase ambiguity at least in principle.
We emphasize that in our scheme all phases of the hadronic amplitudes
are
determined unambiguously once we incorporate the constraints from the
photon-induced
pion production data as parameterized conveniently in terms of the
multipole amplitudes.

Finally we present the cross sections for photon-induced vector-meson
production.
We emphasize that these production processes, once they are studied
experimentally
in sufficient detail, offer the most stringent consistency test of the
hadronic $VN \to VN$
scattering amplitudes. It would be desirable to establish a multipole
analysis of the photon-induced vector-meson production data. This offers
a powerful means to
verify or disprove any type of vector-meson dominance assumption and
permits in the former
case a detailed consistency check of the various hadronic amplitudes
with different angular
momenta. At present there are basically only total cross sections
available, for
which we present explicit expressions,
\begin{eqnarray}
&& \sigma_{\gamma \,p\rightarrow  \rho^0 \,p} = 4\,\pi \,
\frac{p_{\rho N}}{p_{\gamma N}}\, \frac{1}{3} \Bigg(\Big|
f^{(\frac{1}{2}\frac{1}{2}+ )}_{\frac{1}{2},\gamma \,p \rightarrow
\rho N}+\sqrt{2}\,f^{(\frac{3}{2}\frac{1}{2}+
)}_{\frac{1}{2},\gamma \,p \rightarrow \rho N}\Big|^2
\nonumber\\
&& \qquad \qquad \qquad \qquad +2\,\Big|
f^{(\frac{1}{2}\frac{3}{2} -)}_{\frac{1}{2},\gamma \,p \rightarrow
\rho N} +\sqrt{2}\,f^{(\frac{3}{2}\frac{3}{2}
-)}_{\frac{1}{2},\gamma \,p \rightarrow \rho N} \Big|^2\Bigg)
\nonumber\\
&& \qquad \qquad \qquad \qquad +2\,\Big|
f^{(\frac{1}{2}\frac{3}{2} -)}_{\frac{3}{2},\gamma \,p \rightarrow
\rho N} +\sqrt{2}\,f^{(\frac{3}{2}\frac{3}{2}
-)}_{\frac{3}{2},\gamma \,p \rightarrow \rho N} \Big|^2\Bigg) \,,
\nonumber\\
&& \sigma_{\gamma \,p\rightarrow  \rho^+ \,n} = 4\,\pi \,
\frac{p_{\rho N}}{p_{\gamma N}}\, \frac{1}{3} \Bigg(\Big|
\sqrt{2}\,f^{(\frac{1}{2}\frac{1}{2}+ )}_{\frac{1}{2},\gamma \,p
\rightarrow \rho N}-f^{(\frac{3}{2}\frac{1}{2}+
)}_{\frac{1}{2},\gamma \,p \rightarrow \rho N}\Big|^2
\nonumber\\
&& \qquad \qquad \qquad \qquad +2\,\Big|
\sqrt{2}\,f^{(\frac{1}{2}\frac{3}{2} -)}_{\frac{1}{2},\gamma \,p
\rightarrow \rho N} -f^{(\frac{3}{2}\frac{3}{2}-
)}_{\frac{1}{2},\gamma \,p \rightarrow \rho N} \Big|^2\Bigg)
\nonumber\\
&& \qquad \qquad \qquad \qquad +2\,\Big|
\sqrt{2}\,f^{(\frac{1}{2}\frac{3}{2} -)}_{\frac{3}{2},\gamma \,p
\rightarrow \rho N} -f^{(\frac{3}{2}\frac{3}{2}-
)}_{\frac{3}{2},\gamma \,p \rightarrow \rho N} \Big|^2\Bigg) \,,
\nonumber\\
&& \sigma_{\gamma \,p\rightarrow  \omega \,n} = 4\,\pi \,
\frac{p_{\omega N}}{p_{\gamma N}}\, \Bigg(\Big|
f^{(\frac{1}{2}\frac{1}{2}+ )}_{\frac{1}{2},\gamma \,p \rightarrow
\omega N}\Big|^2 +2\,\Big| f^{(\frac{1}{2}\frac{3}{2}
-)}_{\frac{1}{2},\gamma \,p \rightarrow \omega N}\Big|^2
 +2\,\Big|
f^{(\frac{1}{2}\frac{3}{2} -)}_{\frac{3}{2},\gamma \,p \rightarrow
\omega N}\Big|^2 \Bigg)\,, \label{def:fVred}
\end{eqnarray}
in terms of the reduced amplitudes. As is implicit in (\ref{def:fVred})
the description of the $\gamma N \to VN$ process requires various
multipole amplitudes even at energies close to the production threshold.
In order to determine the amplitudes separately experiments with a
polarized
beam and target are required. We emphasize that our model predicts all
multipole amplitudes of the $\gamma N \to VN$ process required to
describe
the vector-meson production process in any detail for energies close to
the production threshold,
\begin{eqnarray}
&& {\textstyle{1\over 2}}\,\Big( f^{(\frac{1}{2}
\frac{1}{2}+)}_{\frac{1}{2} ,\gamma \,p\,\to \rho
N}(\sqrt{s}\,)+f^{(\frac{1}{2} \frac{1}{2}+)}_{\frac{1}{2} ,\gamma
\,n\,\to \rho N}(\sqrt{s}\,)\Big) = \frac{\sqrt{N_{\rho
N}^{}\,N_{\frac{1}{2},\gamma N}^{(S+)} }}{8\,\pi\,\sqrt{s}}\,
M^{(\frac{1}{2} +)}_{43}(\sqrt{s},0)  \;,
\nonumber\\
&& f^{(\frac{3}{2} \frac{1}{2}+)}_{ \frac{1}{2},\gamma \,p\,\to
\rho N}(\sqrt{s}\,) = \frac{\sqrt{N_{\rho
N}\,N_{\frac{1}{2},\gamma N}^{(V+)} }}{8\,\pi\,\sqrt{s}}\,
\sqrt{\frac{2}{3}} \,M^{(\frac{3}{2} +)}_{33 }(\sqrt{s};0)\;,
\nonumber\\
&& {\textstyle{1\over 2}}\,\Big( f^{(\frac{1}{2} \frac{3}{2}-)}_{h
,\gamma \,p\,\to \rho N}(\sqrt{s}\,)+f^{(\frac{1}{2}
\frac{3}{2}-)}_{h ,\gamma \,n\,\to \rho N}(\sqrt{s}\,) \Big) =
\frac{\sqrt{N_{\rho N}^{}\,N_{h,\gamma N}^{(S-)}
}}{8\,\pi\,\sqrt{s} } M^{(\frac{1}{2} -)}_{43}(\sqrt{s},1)  \,,
\nonumber\\
&& f^{(\frac{3}{2} \frac{3}{2}-)}_{h ,\gamma \,p\,\to \rho
N}(\sqrt{s}\,) = \frac{\sqrt{N_{\rho N}^{}\,N_{h,\gamma N}^{(V-)}
}}{8\,\pi\,\sqrt{s} } \,\sqrt{\frac{2}{3}} \,M^{(\frac{3}{2}
-)}_{33}(\sqrt{s};1) \;. \label{f-gammap:v}
\end{eqnarray}

\newpage

\section{Results}

In this section we present the results of our fits to the data set.
In accordance with the effective field theory approach introduced
in the previous sections only data in an appropriate kinematical
window is used in the analysis. The threshold for elastic
scattering is at  $\sqrt{s} \simeq 1.1$ GeV, while that for
vector-meson production off a nucleon is at $\sqrt{s} \simeq 1.7$
GeV. Because we do not expect our scheme to be efficient close to
the elastic pion-nucleon threshold, we fit only data in the energy
range $1.4$ GeV $\leq\sqrt{s} \leq 1.8$ GeV. We use the ansatz
(\ref{ansatz}) with energy independent interaction matrices
$g_{0,1}^{(\pm)}$ that parameterize quasi-local two-body interaction vertices
of defined isospin and total angular momentum. In the model we include
four spin and isospin channels with $I=\frac{1}{2}\,,\frac{3}{2}$ and
$J=\frac{1}{2}\,,\frac{3}{2}$. For each value of spin and isospin,
channels with different mesons and baryons are coupled by the
interaction matrices. In our scheme, an acceptable fit to the data in
these channels requires that the s- and d-wave baryon resonances
$N(1535)$, $N(1650)$, $N(1520)$, $\Delta(1620)$ and
$\Delta(1700)$ are generated dynamically. This is in contrast to the more
conventional approaches \cite{Sauerman,Feuster:Mosel:1,Feuster:Mosel:2,Vrana:Dytman:Lee,Krehl}
where such resonances are already part of the interaction kernel. Our
approach leads to loop functions (see e.g. (\ref{result-loop})) and
scattering amplitudes that are analytic and
satisfy the expected dispersion-integral representation implied by micro causality
of local quantum field theories.

The set of parameters is adjusted to describe the partial-wave
pion-nucleon phase shifts including their inelasticity parameters.
Furthermore the pion- and photon-induced production cross sections
of pseudo-scalar meson and vector mesons are included in the fit.
The production cross sections provide crucial information on the
coupling to specific channels, which obviously cannot be extracted
from the inelasticity parameters of the pion-nucleon phase shifts alone.
Ideally one should try to fit the two-pion data directly rather than possibly
model dependent production cross sections for the different sub channels.
However, in the present scheme, which does
not include e.g. the p-wave channels, this would not make sense because
potentially important contributions to the total cross sections are missing.
We point out
that the relative distribution of the various inelastic channels is
constrained by the $\rho N$ and $\pi \Delta$
production data because these data clearly prevent an overestimate of
the inelasticity into the two-pion channel. Here we exploit the fact that the inelasticity
parameters of the $\pi N$ phase shifts define the total inelastic
strength. We checked that the sum of the
partial isobar-production cross sections of our fit does not exceed the
measured total $\pi \Delta$ cross sections. Since for this channel
differential cross sections are not available and since there is no
reason to expect the production cross section to be dominated by s-
and d-waves  at the energies considered, we do not fit the isobar
production
cross sections directly. Moreover, we note that the angular
distribution of the
two-pion production data is to a large extend already encoded in the
inelasticity parameters of
the pion-nucleon phase shifts, provided that the relative importance of
the dominant inelastic two-pion channels $\rho N$ and $\pi \Delta $ are
treated correctly.

In some of the data sets, like e.g. the total cross section for the
reaction $\pi^- p \rightarrow \rho^0 n$, only a few points are
included in the fit, while e.g. for the $\pi N$ phase shifts about
25 data points per channel, with fairly small error bars, are
considered. Consequently, if one naively sums up the chi-squared
errors of all data sets, a poor overall agreement with a small data
set would be punished by only a small increase of the total
chi-squared. This would lead to an undesired imbalance between the
large and small data sets. In order to obtain a uniform
reproduction of all data sets, we therefore introduce weighting
factors, which enhance the weight of small data sets, when
otherwise a reasonable overall description could not be obtained.
Moreover, for the production cross sections we reduce the upper
limit of the fitting window down to values where we expect the cross section
to be dominated by s-wave dynamics. In view of this it does
not make too much sense to provide the final chi-squared of our fit.
The parameter search is performed with the FORTRAN routine
MINUIT \cite{Minuit}. The statistical error in the coupling constants of the fit are
estimated in the standard way by using the MINUIT error analysis.
For our final solution we find that one standard deviation corresponds to
an error of typically one percent in the coupling constants. This
is certainly an underestimate of the error for two reasons. First,
the sample is not statistical due to the weighting procedure.
Second, the systematic errors, due
e.g. to approximations and model assumptions, are not accounted
for. However, it is very difficult to estimate these errors. The
expected errors obviously propagate into the resonance coupling
constants discussed below. Thus, the relative statistical errors of these
parameters are expected to be of similar magnitude.
The reader may judge the quality of the fit by
the detailed comparison of our model results with the data presented below.

\begin{figure}[t]
\begin{center}
\includegraphics[width=11.cm,clip=true]{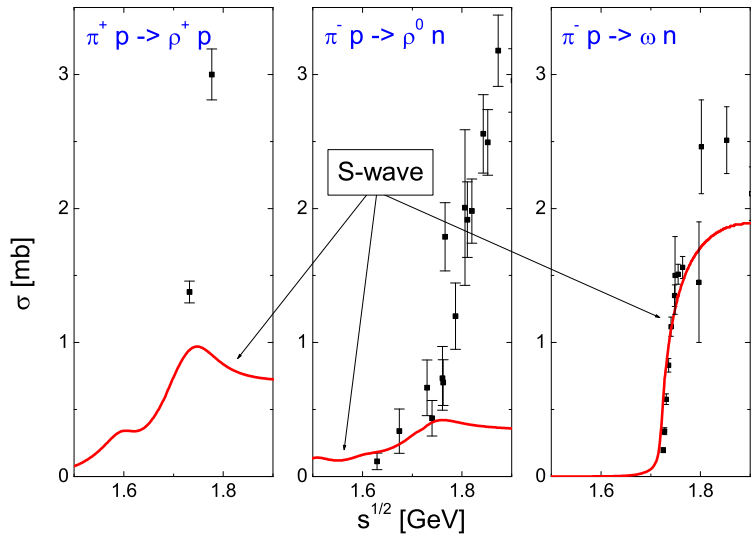}
\end{center}
\caption{Pion-induced vector-meson production. The data are taken
from \cite{Brody:et:al,keyne-karami,Landolt:Boernstein}.}
\label{fig:1}
\end{figure}

The empirical pion- and photon-induced vector-meson production data
are of crucial importance to our fit. Only when they are included
it is possible to determine the vector-meson nucleon scattering
amplitudes, the primary goal of this work.
In Fig. \ref{fig:1} we compare the results of our model for the
$\rho$- and $\omega$-meson production cross sections with data. Only
the first few data points are included in the fit, since one can
expect s-wave dominance only close to threshold. Note that a
reliable separation of the partial wave would require differential
cross sections, which in many cases are not available. The
deviation of our model from the empirical cross section at energies
$\sqrt{s} \simeq 1.8$ GeV and above leaves room for the contribution
from higher partial waves, in particular the p-waves, not included in our
model.

The bumps in the $\rho$-meson production cross section at
$\sqrt{s}$ below 1.7 GeV are due to the coupling to resonances
below the nominal threshold, like the $N(1535)$ and $\Delta(1620)$. This
illustrates the important role these resonances may play in the
$\rho$-nucleon dynamics, in qualitative agreement with the findings
of Manley and Saleski~\cite{Manley}. As compared to our previous
analysis \cite{Hirschegg}, where we did not consider the photon induced
reactions systematically, we find a much reduced importance of the $N(1520)$
resonance for the subthreshold $\rho^0$ production cross section.
Note that the very first data point which was
questioned in \cite{Post} was not included in the fit.

The large discrepancy between the analysis of Manley {\em et
al.}~\cite{Manley} and that of Brody {\em et al.}~\cite{Brody:et:al}, as shown
in \cite{Post}, illustrates the ambiguity inherent in the extraction
of the $\rho$-meson production cross section, in particular below
the nominal threshold. Within our model we are able to find a
consistent description of the $\rho$ production cross section of
Brody {\em et al.} and the $E_{2-}$ and $M_{2-}$ multipole
amplitudes. Given the vector-meson dominance assumption, we find it
difficult to accommodate a coupling of the $N(1520)$ to the $\rho
N$ channel of the strength found by Manley {\em et al.}~\cite{Manley}.

We find that the $\omega$ meson couples strongly to the $N(1520)$,
$N(1535)$ and $N(1650)$ resonances. This is manifest in the $\omega N$ scattering
amplitudes, presented in the subsequent sections. Due to the small width of the $\omega$
meson, the subthreshold resonances are not visible in the
$\omega$-meson production cross section shown in Fig. \ref{fig:1}.
Note that the production cross section was measured at fixed
relative  momentum in the final state\cite{keyne-karami} while the
calculation is done at fixed energy. This leads  to an uncertainty
in the energy resolution close to the production threshold which is
on the order of the $\omega$-meson width. For a recent K-matrix analysis
of this cross sections see \cite{Penner:Mosel}.

\begin{figure}[t]
\begin{center}
\includegraphics[width=11.cm,clip=true]{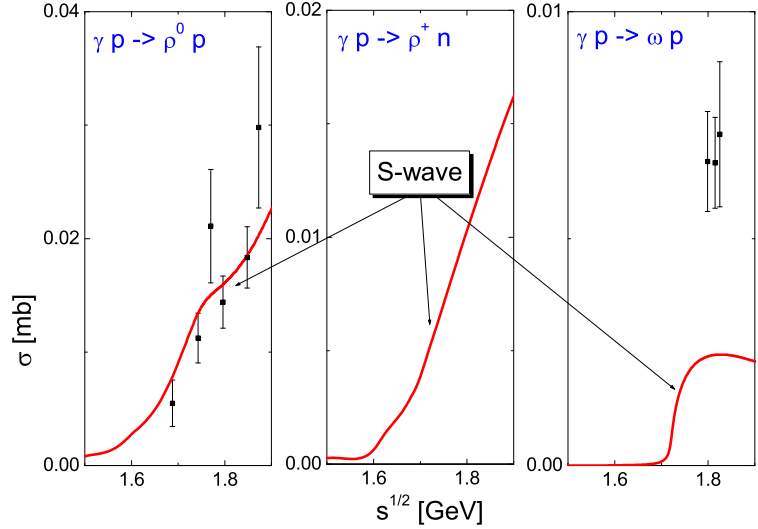}
\end{center}
\caption{Photon-induced vector-meson production. The data points are taken from \cite{ABBHH}.}
\label{fig:2}
\end{figure}

In Fig. \ref{fig:2} we compare our photon-induced $\rho$- and
$\omega$-meson production cross sections with the data.
Using the generalized vector-meson dominance
conjecture (\ref{gamma-ansatz:k},\ref{tensor-ansatz}), the
photon-induced meson-production amplitudes are given in terms of
the amplitudes for the corresponding vector-meson induced reaction
and the four coupling constants, $g_{S,1}^{(+)}, g_{V,1}^{(+)}$
and $g_{S,2}^{(+)}, g_{V,2}^{(+)}$ specifying the vector-meson
dominance assumption. Our best fit is obtained for
\begin{equation}
g_{S,1}^{(+)} = 0.083 \;, \qquad g_{V,1}^{(+)} = 0.469 \,, \qquad
g_{S,2}^{(+)} = 0.000 \;, \qquad g_{V,2}^{(+)} = 0.241
 \;. \label{VD:g}
\end{equation}
We recall that the deviation of $\sqrt{3}\,g^{(+)}_{S,i}/
g^{(+)}_{V,i} $ from the ratio $f_\omega /f_\rho \simeq 0.3$ is a
measure for the importance of an off-shell extrapolation from
virtual photon kinematics at $q^2 \simeq m_\omega^2$ down to the
photon point $q^2=0$. The parameters $f_\omega $ and $f_\rho$ are
determined by the decay of the vector mesons into lepton pairs and
therefore correspond to virtual photons with $q^2 = m_\omega^2$ or
$q^2 = m_\rho^2$. The fact that we find $ \sqrt{3}\,g^{(+)}_{S,1}/
g^{(+)}_{V,1} \simeq  f_\omega /f_\rho $ but
$ \sqrt{3}\,g^{(+)}_{S,2}/
g^{(+)}_{V,2} <  f_\omega /f_\rho $ demonstrates that this off-shell
extrapolation is quite non-trivial. Although the deviations from the naive
vector-meson dominance expectation are strong for the $g_{S(V),2}$
parameters,
the generic picture with $g^{(+)}_{S,2} < g^{(+)}_{V,2}$ is confirmed.
In the fit we again include only the first few data points close to
threshold, where one  can expect the s-wave contribution to
dominate. While our model describes the $\rho^0$-meson production
cross section reasonably well, it deviates strongly from the
empirical $\omega$-meson production cross section at $\sqrt{s}
\simeq 1.8$ GeV where the data points start. On a qualitative
level, the discrepancy may be understood if the $\omega$-meson
production cross section is dominated by the one-pion-exchange
contribution, as suggested by one-boson-exchange models (see e.g.
\cite{Friman:Soyeur}). Since the one-pion-exchange interaction
is long ranged, this would imply that there are important
contributions to the angle-averaged cross section from higher
partial waves, which are not considered in our scheme at present.
Note that the same argument applies for the pion-induced $\rho $-production
cross sections of Fig. 1 which also allow for a t-channel
one-pion-exchange
contribution. On the other hand, it is evident that there is no
one-pion-exchange contribution in the $\pi^- p \to \omega n$ reaction.
The same holds for the  $\gamma p \to  \rho^0 p$ reaction provided that
it is dominated by its isovector amplitude as suggested by (\ref{VD:g}).
Consequently one expects the interaction in these channels to be short ranged and
higher partial waves to be less important. The corresponding
cross sections are described by our model fairly well up
to higher energies
suggesting that higher partial wave contributions are indeed small.
In Fig. \ref{fig:2} we also show our prediction for the
$\gamma$-induced $\rho^+$-production cross section. At present
there are no data available for this reaction close to the
production threshold.

In the remainder of this section we discuss the results that are
specific to the four spin- and isospin-channels. For a channel with
given
isospin $I$ and angular momentum $J$ we list the coupling
constants $g^{(I\pm )}_{J-\frac{1}{2}}$, confront our model with
the appropriate pion-nucleon phase shift and if available present
further relevant data on inelastic reactions. The resulting
scattering amplitudes are compared with a schematic resonance
exchange model. The discussions are most detailed for the $S_{11}$
and $D_{13}$ channels and arguments which apply to all channels are not
repeated.

\subsection{$I \,J^P=\frac{1}{2}\,\frac{1}{2}^-$ channel}

\tabcolsep=1.8mm
\begin{table}[b]
\begin{tabular}{|r||c|c|c|c|c|c|c|c|c|c||c|c|c||c|c|} \hline
$ij $ &  11 & 15 & 16 & 17 & 55& 56 & 57 & 66 & 67 & 77 & 11 & 14 & 44 &
& [$1/m_\pi$] \\ \hline \hline

$c_{ij}^{(1)}$  & 2 & 0& -${\textstyle{3\over 2}}$
&${\textstyle{1\over 2}}$ & 0                       &
-${\textstyle{3\over 2}}$ & -${\textstyle{3\over 2}}$ & 0 & 0 & 2
& -1 & -1& -1& $h_1$& 2.62  \\ \hline $c_{ij}^{(2)}$  & 2 & 0& 0 &0
& 2 & 0                         & 0 & 2 & 0 & 2 & 2 & 0 & 2 &
$h_2$& 0.20 \\ \hline $c_{ij}^{(3)}$  & 0 & 0& 0 &1 & 0 & 1 & 0 & 0
& 0 & 0 & 0 & 1 & 0 & $h_3$& 8.79
\\ \hline $c_{ij}^{(4)}$  & 1 & 1& -${\textstyle{1\over 2}}$
&-${\textstyle{1\over 2}}$& ${\textstyle{5\over 3}}$&
${\textstyle{1\over 6}}$  & -${\textstyle{1\over 2}}$  &
${\textstyle{5\over 3}}$& 1 & 1 & 1 & 1 & 1 & $h_4$& 1.76 \\ \hline
$c_{ij}^{(5)}$  & 1 & 1& -${\textstyle{3\over 2}}$
&${\textstyle{1\over 2}}$ & -1 & ${\textstyle{1\over 2}}$  &
${\textstyle{1\over 2}}$   & 0 & 0 & -2 & 1 & -1 & 1 & $h_5$& -4.03 \\
\hline $c_{ij}^{(6)}$  & 0 & 0& 0                         &-1 & 0
& -1                        & 0                          & 0 & 0 &
0 & 0 & -1 & 0 &$h_6$& 9.88\\ \hline

\end{tabular}
\vspace*{2mm} \caption{Expansion coefficients $c_{ij}^{(k)}$ as defined
in (\ref{def-cs}). The first 10 columns
are the coefficients for the isospin one half channel, the proceeding
columns provide the coefficients for the isospin three half channel. For the
labelling of the channels see (\ref{r-def}).}
\label{tab:cs}
\end{table}

In the $I\,J^P=\frac{1}{2}\,\frac{1}{2}^-$ sector we include the $\pi
N$, $\pi \Delta $, $\rho N$, $\omega N$, $\eta N$, $K
\Lambda$ and $K \Sigma $ channels. In order to reduce the number of
parameters  we impose a partial SU(3) constraint on
the coupling constants. The interaction strengths of the four
pseudo-scalar channels $\pi N, \eta N, K \Lambda $ and $K \Sigma $
are given in terms of 6 parameters only, corresponding to the number of
available SU(3) invariant tensors. This leaves us with all together
24 parameters in this channel. Note that in our present scheme it
would not be justified to insist on SU(3) relations in the
vector-meson nucleon channels, because some of the channels
required in a SU(3) symmetric description like $K_\mu \,
\Lambda $ and $\phi_\mu \,N$ are integrated out. In order to implement
the SU(3) relations we decompose the coupling matrix for
$i,j=1,5,6,7$ in SU(3) invariants
\begin{eqnarray}
g^{(\frac{1}{2},+)}_{0,ij} = \sum_{k=1}^6 \,c^{(k)}_{ij} \,h_k \,.
\label{def-cs}
\end{eqnarray}
The expansion coefficients $c^{(k)}_{ij}$ and the best-fit values
of the parameters $h_i$ are listed in Tab. \ref{tab:cs}. We note
that the $k=1$ term in (\ref{def-cs}) has the SU(3) tensor structure of
the Weinberg-Tomozawa term. As shown by the sizable values of $h_k$
for $k>1$, we find that the flavor structure of the
Weinberg-Tomozawa term is not sufficient to describe the data in
the pseudo-scalar channels. This is not surprising
because in our scheme the interaction kernel represents the
scattering amplitude at $\sqrt{s}= 0$ where there is no reason to
expect that the Weinberg-Tomozawa interaction dominates the
amplitude. The complete coupling matrix $g^{(\frac{1}{2},+)}_{0}$ is
given in Tab. \ref{tab:s11}. Since there are important
components in basically all channels, it is difficult to select a
dominating channel. In particular we find no reason to single out
the pseudo-scalar channels.

\tabcolsep=0.6mm
\renewcommand{\arraystretch}{1.2}
\begin{table}[t]
\begin{tabular}{|r||c|c|c|c|c|c|c|} \hline
$g^{(\frac{1}{2},+)}_{0}$\,$\Big[m_\pi^{-\frac{n+m}{2}}\Big]$  & $\!\pi
N [n\!=\!1]\!$
& $\! \pi \Delta [n\!=\!3]\!$ & $\!\rho N [n\!=\!1]\!$ &$\!\omega
N[n\!=\!1]\!$ &
$\!\eta N [n\!=\!1]\!$& $\!K \Lambda [n\!=\!1]\!$ & $\!K \Sigma
[n\!=\!1]\!$\\ \hline \hline

$\pi N[m=1]$  & 3.35 & -30.55 & 6.36 & 9.78 &-2.29 & 1.24 & -2.66\\
\hline

$\pi \Delta [m=3]$  & -30.55 & 0.00 & -10.98 & -14.02& 15.48 & -19.36 &
14.35\\ \hline

$\rho N [m=1]$   & 6.36 & -10.98 & 10.66 & 3.13 & -5.50 & 3.61 & 1.86\\
\hline

$\omega N[m=1] $  & 9.78 & -14.02 & 3.13 & 16.36 & -9.35 &-9.16 & 8.56\\
\hline

$\eta N [m=1]$ & -2.29 & 15.48 & -5.50 &-9.35 & 7.33 &-6.75 & -6.83\\
\hline

$ K \Lambda [m=1] $ & 1.24 & -19.36 & 3.61&-9.16 & -6.75 & 3.30 & 1.75\\
\hline

$ K \Sigma [m=1] $ & -2.66 & 14.35 & 1.86&8.56 & -6.83 & 1.75 & 15.45 \\
\hline
\end{tabular}
\vspace*{2mm} \caption{Coupling constants in the $I J^P
=\frac{1}{2}\,\frac{1}{2}^-$ channel.
We use $m_\pi=139$ MeV.}
\label{tab:s11}
\end{table}

\begin{figure}[b]
\begin{center}
\includegraphics[width=10.5cm,clip=true]{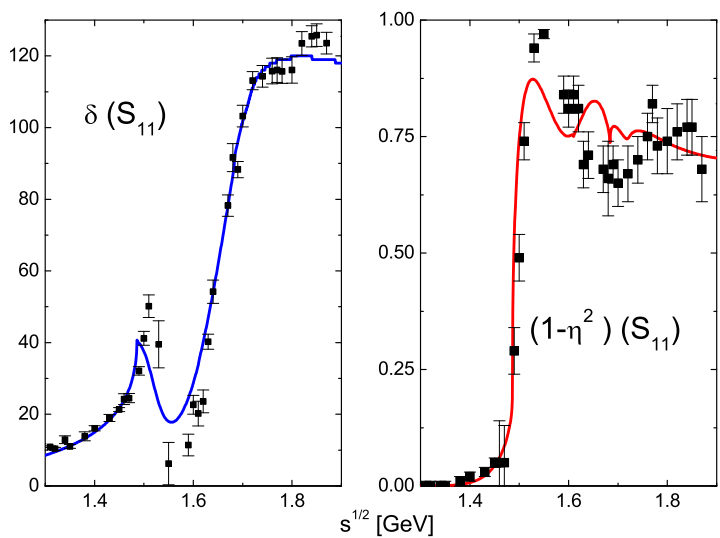}
\end{center}
\caption{Pion-nucleon scattering phase shift $\delta_{S_{11}}$ and
inelasticity parameter $\eta_{S_{11}}$ of the $S_{11}$ channel. Our results shown by
solid lines are compared to the single energy solution SP98 of \cite{Arndt:piN,SAID}.}
\label{fig:3}
\end{figure}

The pion-nucleon phase shift and the inelasticity parameter are
well reproduced by our model as demonstrated in Fig. \ref{fig:3}.
The phase shift and inelasticity parameter show characteristic
structures at $\sqrt{s}\simeq 1535$ MeV and $\sqrt{s}\simeq $ 1650
MeV, which correspond to the $I\,J^P = \frac{1}{2}\,\frac{1}{2}^-$
baryon resonances $N(1535)$ and $N(1650)$. In contrast to the work
of \cite{Weise,chi-1520}, where only the pseudoscalar
channels were included, we find that both resonances can be
generated by coupled-channel dynamics. This illustrates the
importance of the additional channels that are included in our
model, in particular the $\rho N$ and $\omega N$ channels. Clearly
a description of the scattering data at energies beyond $\sqrt{s}
\simeq 1550$ MeV is impossible in a scheme where the second resonance,
$N(1650)$, is missing.

\begin{figure}[b]
\begin{center}
\includegraphics[width=13.5cm,clip=true]{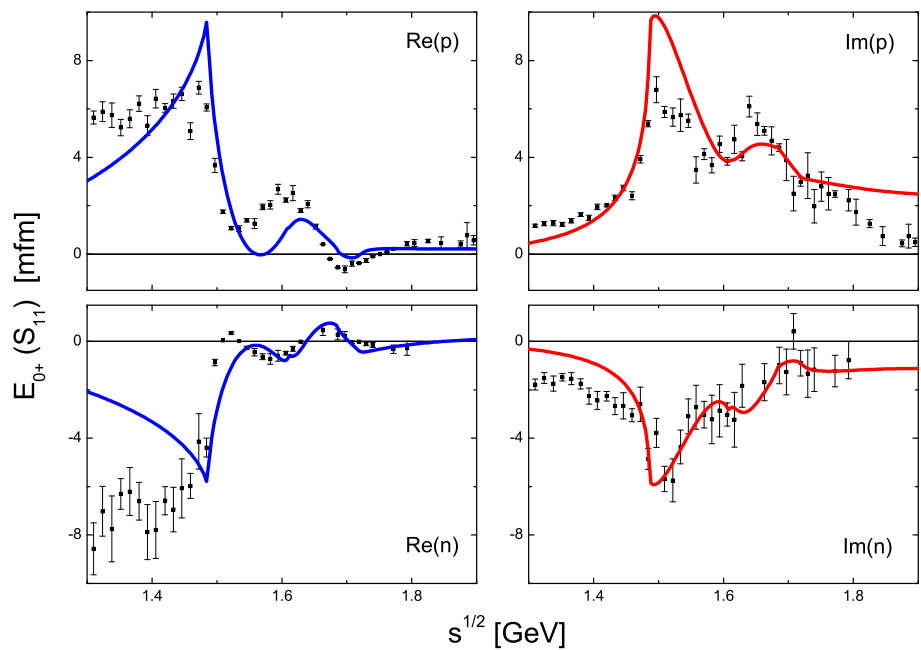}
\end{center}
\caption{Multipole amplitudes $E_{0+}^{(p)}(S_{11})$ and
$E_{0+}^{(n)}(S_{11})$ of $\gamma $-induced pion production.
Our amplitudes are confronted to those of the single energy analysis SM00
of \cite{Arndt:gamma,SAID}}
\label{fig:4}
\end{figure}

We now turn to the multipole amplitudes of the $\gamma$-induced
pion production process. Given the generalized vector-meson
dominance conjecture, the multipole amplitudes
$E_{0+}^{(p)}(S_{11})$ and $E_{0+}^{(n)}(S_{11})$, shown in Fig.
\ref{fig:4}, are related to the pion-induced vector-meson
production amplitudes. Thus, the constraints provided by the
multipole amplitudes are similar to those of the pion-induced
vector-meson production data, however, the amplitudes are tested
also at energies below the vector-meson production threshold.
Furthermore, the multipole amplitudes correspond to combinations of the
$\pi\, N
\to \rho/\omega\,N$ amplitudes that  differ from those probed in the
pion-induced vector-meson production processes. In particular,
there is interference between the isoscalar and isovector
amplitudes which gives a handle on the relative phases of the
amplitudes.
The multipole amplitudes clearly reflect the presence
of the s-wave nucleon resonances $N(1535)$ and $N(1650)$. Within
the window of applicability $1.4 <  \sqrt{s} < 1.8$ we obtain a satisfactory
description of the multipole amplitudes. Close to the production
threshold we
do not expect our model to describe the multipole amplitudes well, in
particular
the real parts, because the strong energy dependence implied by the
one-pion exchange
contribution is not treated properly in our present scheme.

It is instructive to discuss the pseudo-scalar meson-baryon scattering
lengths in some detail. A reliable
extraction of a given scattering length from the data set requires that
the model considers all rapid energy
variations induced for example by the opening of inelastic channels or
the presence of close by resonances. For
instance the $\eta N$ threshold is quite close to the $N(1535)$
resonance. Thus a good value for the
$\eta N$ scattering length should be based on a model describing the
$N(1535)$ resonance in a realistic manner.
Similarly we expect that it is important to consider the inelastic $\rho
N $ and $\omega N$ channels when deriving
the $K \Lambda $ and $K \Sigma $ scattering lengths, simply because the
threshold values of those four channels
are quite close to each other. Since we consider all above inelastic
channels, we are convinced
that our model permits reliable extractions of the pseudo-scalar meson
baryon scattering lengths:
\begin{eqnarray}
&& a_{\eta N}^{(\frac{1}{2})} \simeq  (0.43 +i\,0.21 )\,{\rm fm } \;,
\qquad \;\;\,
a_{K \Lambda}^{(\frac{1}{2})} \simeq  (0.26 +i\,0.10)\, {\rm fm }
\;, \qquad
\nonumber\\
&& a_{K \Sigma }^{(\frac{1}{2})} \simeq  (-0.15+i\,0.09)\,{\rm fm }
\,,\qquad
a_{K \Sigma }^{(\frac{3}{2})} \simeq  (-0.13+i\,0.04)\,{\rm fm } \,.
\label{r:etan}
\end{eqnarray}
We observe that our value for the eta-meson nucleon scattering length is
quite consistent
with previous analyses \cite{Sauerman} with $a_{\eta N}\simeq (0.51
+i\,0.21)$ fm,
and \cite{etaN} with $a_{\eta N}\simeq (0.55 +i\,0.30)$. For a detailed
discussion
of the various analyses including a more complete collection of
references we suggest
the recent work by Green and Wychech \cite{Green:Wychech}. Their latest
scattering length, which favors a relatively large real part of about 1 fm, was
obtained by a
simultaneous analysis of the pion and photon
induced eta production data. For the kaon-hyperon scattering lengths the
existing literature
is much more sparse. The value $a_{K \Lambda} \simeq  (0.06 +i\,0.03)$
fm was obtained recently in \cite{Nieves:Arriola} in a coupled channel
approach which does not consider the important $\rho N$, $\omega N$ and $\pi
\Delta$ channels (see also \cite{Feuster:Mosel:1}).
Their value differs from our result in (\ref{r:etan}) by about a factor
four.

\begin{figure}[b]
\begin{center}
\includegraphics[width=12.5cm,clip=true]{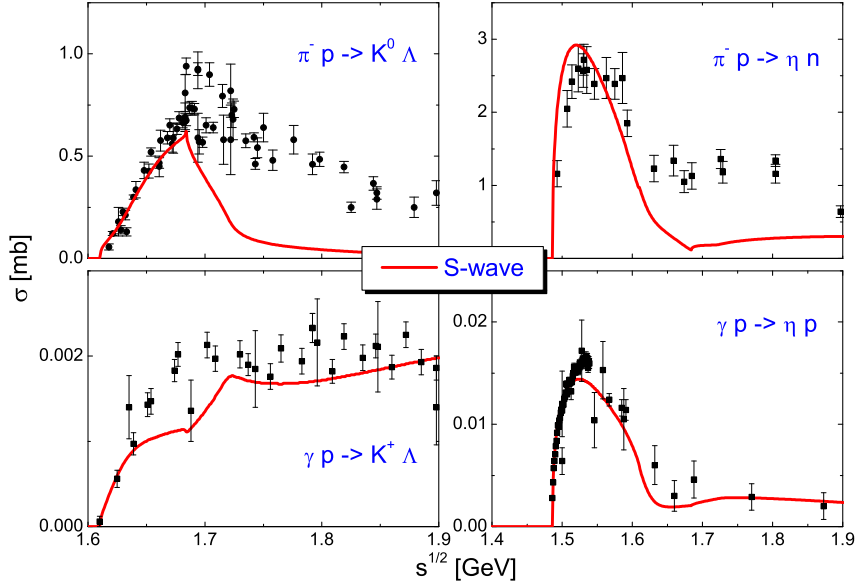}
\end{center}
\caption{Pion- and $\gamma$-induced $\eta N$ and $K \Lambda$ production
cross sections. The data are taken from \cite{Krusche:et:al,Tran:et:al,Landolt:Boernstein}.}
\label{fig:5}
\end{figure}

The values for the scattering lengths derived in our work
are a particular interpretation of the pion- and photon-induced
eta and kaon production data of Fig. \ref{fig:5}. The pion- and $\gamma$-induced
kaon production cross section with an associated $\Sigma$ hyperon will be shown in the
next section when presenting the s-wave dynamics in the isospin three half channel. The
discrepancy of our model
and some of the cross section data in Fig. \ref{fig:5} starting somewhat
above threshold  is
expected since higher partial waves are not included in the analysis.
The s-wave dominance for the considered reactions
is confirmed qualitatively by available differential cross section data
\cite{Krusche:et:al,Brown:et:al,Baker:et:al:2}.
Note that there is no one-pion-exchange contribution
in any of these cross sections. An improved treatment would profit from high quality
differential production
cross sections and polarization data not available for all reactions at
present.

\begin{figure}[b]
\begin{center}
\includegraphics[width=10.5cm,clip=true]{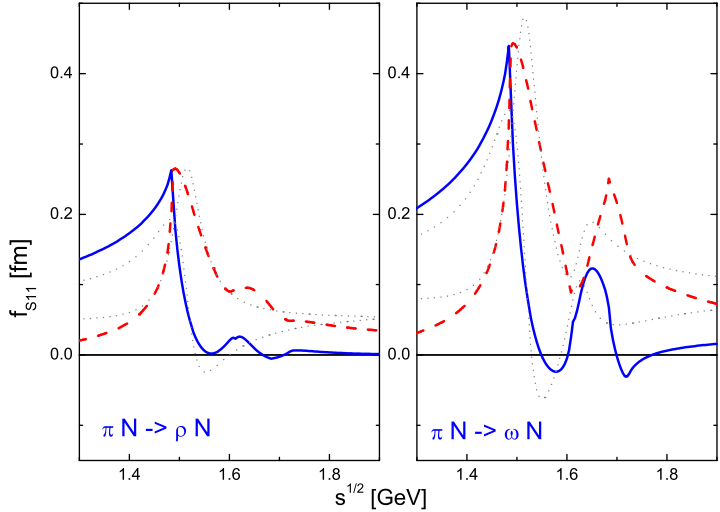}
\end{center}
\caption{Pion-induced vector-meson production amplitudes with
$I=J=\frac{1}{2}$. The solid and dashed lines represent the real and imaginary parts
of the amplitudes. The dotted lines follow from the schematic resonance exchange model
defined in (\ref{rep-S11}).}
\label{fig:6}
\end{figure}

\subsubsection{Scattering amplitudes}

\begin{figure}[b]
\begin{center}
\includegraphics[width=13.5cm,clip=true]{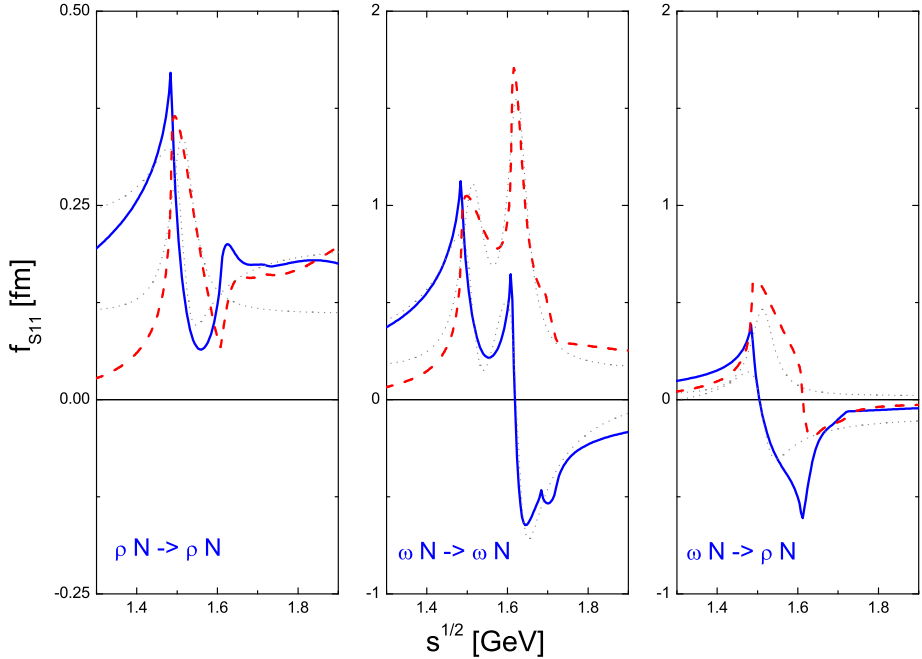}
\end{center}
\caption{Vector-meson nucleon scattering amplitude with
$I=J=\frac{1}{2}$. Solid, dashed and dotted lines as in Fig. \ref{fig:6}.}
\label{fig:7}
\end{figure}

Our model analysis leads to well defined vector-meson production
and vector-meson nucleon scattering amplitudes. These amplitudes
constitute the central result of our work. In Fig. \ref{fig:6} we
present the pion-induced vector-meson production amplitudes
introduced in (\ref{f-vp}). The vector-meson nucleon scattering
amplitudes are displayed in Fig. \ref{fig:7}. The $N(1535)$ and
$N(1650)$ resonances lead to peak structures in the imaginary
parts of the subthreshold amplitudes. Those structures reflect the
subthreshold strength seen in the $\rho$-meson production cross section
of Fig. \ref{fig:1}. It is convenient to represent the result in
terms of the reduced scattering amplitudes $f^{(IJ \pm)}_{VN\to
VN}(\sqrt{s}\,)$,
\begin{eqnarray}
&& f^{(\frac{1}{2}\frac{1}{2}+)}_{\rho N \to \rho N}(\sqrt{s}\,) =
\frac{N_{\rho N}(\sqrt{s}\,)}{8\pi\,\sqrt{s}}\,
M_{33}^{(\frac{1}{2}+)}(\sqrt{s},0)\,,\qquad
\nonumber\\
&& f^{(\frac{1}{2}\frac{1}{2}+)}_{\omega N \to \omega
N}(\sqrt{s}\,) = \frac{N_{\omega
N}(\sqrt{s}\,)}{8\pi\,\sqrt{s}}\,M_{44}^{(\frac{1}{2}+)}(\sqrt{s},0)\,,\qquad
\label{}
\end{eqnarray}
where the normalization factor $N_{\rho N}(\sqrt{s}\,)$ and $N_{\omega
N}(\sqrt{s}\,)$ was
introduced in (\ref{om-spec}). In $N_{\rho N}(\sqrt{s}\,)$ we use
the nominal $\rho$-meson mass $m_\rho= 779$ MeV as introduced
in (\ref{rho-explicit}). The vector-meson nucleon s-wave scattering
lengths
with $I=J=\frac{1}{2}$ can be read of Fig. \ref{fig:7} at $\sqrt{s}
= m_N+m_V$. It is evident from Fig. \ref{fig:7} that the
repulsive $\omega$-meson nucleon scattering length is a direct
consequence of the close by N(1650) resonance just below the
$\omega N$ threshold. The $\rho$-meson nucleon scattering length
is small and attractive in this channel because the N(1650) resonance
couples weakly to the $\rho N$ channel and because the real part of the
amplitude is
strongly affected by an attractive background term effect.
The scattering lengths are an important piece of information required
for the study of vector-meson propagation in a dense nuclear
environment,
\begin{eqnarray}
&& a_{\rho N}^{(\frac{1}{2}\frac{1}{2})} =
f^{(\frac{1}{2}\frac{1}{2}+)}_{\rho N \to \rho
N}(m_N+m_\rho\,)\simeq (0.17+i\,0.16)\,{\rm fm } \,, \quad
\nonumber\\
&& a_{\omega N}^{(\frac{1}{2}\frac{1}{2})} =
f^{(\frac{1}{2}\frac{1}{2}+)}_{\omega N \to \omega
N}(m_N+m_\omega\,) \simeq (-0.45+i\,0.31)\,{\rm fm } \,. \label{a:S11}
\end{eqnarray}

We analyze the scattering amplitudes in more detail within a
schematic resonance
exchange model. The amplitudes can be represented in terms of the
coupling
constants $g^{(R)}_{\pi N}$, $g^{(R)}_{V N}$, the phases
$\phi^{(R)}_{\pi V}$, $\phi^{(R)}_{\omega \rho}$,
and the background parameters $b^{(\frac{1}{2}\frac{1}{2})}_{\pi N\to V
N}$ and
$b^{(\frac{1}{2}\frac{1}{2})}_{V N\to V' N}$,
\begin{eqnarray}
&& f^{(\frac{1}{2}\frac{1}{2}+)}_{V N \to V'N}(\sqrt{s}\,) \simeq
-\,\frac{e^{\,i\,\phi^{(1535)}_{V V'}}\,|g_{V
N}^{(1535)}|\,|g_{V'N}^{(1535)}|}{\sqrt{s}-m_{1535}+\frac{i}{2}\,\Gamma_{1535}}
-\,\frac{e^{\,i\,\phi^{(1650)}_{V V'}}\,|g_{V N}^{(1650)}|\,|g_{V
'N}^{(1650)}|}{\sqrt{s}-m_{1650}+\frac{i}{2}\,\Gamma_{1650}}
\nonumber\\
&& \qquad \qquad \qquad \;+ \,b^{(\frac{1}{2}\frac{1}{2}+)}_{V N
\to V' N}\,,
\nonumber\\
&& f^{(\frac{1}{2}\frac{1}{2}+)}_{\pi N \to V N}(\sqrt{s}\,)
\simeq -\,\frac{e^{\,i\,\phi^{(1535)}_{\pi V}}\,|g_{\pi
N}^{(1535)}|\,|g_{V
N}^{(1535)}|}{\sqrt{s}-m_{1535}+\frac{i}{2}\,\Gamma_{1535}}
-\,\frac{e^{\,i\,\phi^{(1650)}_{\pi V}}\,|g_{\pi N}^{(1650)}|\,
|g_{V N}^{(1650)}|}{\sqrt{s}-m_{1650}+\frac{i}{2}\,\Gamma_{1650}}
\nonumber\\
&& \qquad \qquad \qquad \;+ \,b^{(\frac{1}{2}\frac{1}{2}+)}_{\pi N
\to V N}\,. \label{rep-S11}
\end{eqnarray}
Our amplitudes, as shown in Figs. \ref{fig:6}, \ref{fig:7} by solid
lines for the real parts and dashed lines for the imaginary parts, may be
crudely interpreted in terms of the resonance parameters collected in
Tab. \ref{tab:1b}. These parameters were obtained by fitting
the amplitudes in the resonance region. Thus, the  resonance mass
and width parameters were fixed by fitting the elastic omega-meson nucleon
amplitude. The coupling constants of the N(1535) and N(1650) resonances to the
$\rho N$ channel were then determined from the
$\omega N \to \rho N$ amplitude using the previously extracted
resonance mass and width parameters. Similarly the coupling strength of the
resonances to the $\pi N$ channel were obtained in terms of previously determined
parameters by adjusting $g_{\pi N}^{(R)}$ to the pion-induced $\omega $-production
amplitude of Fig. \ref{fig:6}. We find particularly striking the quite
different phase parameters $\phi_{\pi \omega}^{(R)}$ of $\simeq -9\,^\circ $
and $\simeq -90\,^\circ $ for the $N(1535)$ and $N(1650)$ resonances. Note that
consistency of the resonance exchange picture requires the phase relation,
\begin{eqnarray}
\phi_{\pi \omega}^{(R)} \simeq \phi_{\pi \rho}^{(R)}- \phi_{\omega \,\rho}^{(R)} \,,
\label{phase-relation}
\end{eqnarray}
which is reasonably well satisfied  for the $N(1535)$ resonance.
The background parameters, part of our schematic
resonance exchange model, may be viewed as resonance-subtracted
effective scattering lengths,
\begin{eqnarray}
&& b^{(\frac{1}{2}\frac{1}{2})}_{+,\, \pi N \to \omega N} \simeq
(0.10+i\,0.09)\,{\rm fm} \,, \quad
b^{(\frac{1}{2}\frac{1}{2})}_{+,\, \pi N \to \rho N} \simeq
(0.07+i\,0.05)\,{\rm fm} \,,
\nonumber\\
&& b^{(\frac{1}{2}\frac{1}{2})}_{+,\, \omega N \to \omega N} \simeq
(0.15+i\,0.15)\,{\rm fm} \,, \quad
b^{(\frac{1}{2}\frac{1}{2})}_{+,\, \rho N \to \rho N} \simeq
(0.21+i\,0.11)\,{\rm fm} \,,
\nonumber\\
&& b^{(\frac{1}{2}\frac{1}{2})}_{+,\, \omega N \to \rho N} \simeq
(-0.07+i\,0.02)\,{\rm fm} \,,
\label{b:S11}
\end{eqnarray}
In Fig. \ref{fig:6} and \ref{fig:7} we included the result of the
schematic model (\ref{rep-S11}) for the vector-meson production and
scattering amplitudes shown by thin dotted lines. As is clearly demonstrated the
strong energy dependence induced by the nearby $\eta N$ and $K \Lambda $
thresholds prohibits a quantitative application of the resonance
exchange model (\ref{rep-S11}). In particular there is no unique
interpretation of the N(1650) resonance
contributions in terms of simple coupling strengths to the $\pi N$, $\rho N$ channels.
It is instructive to compare the background parameters $b_{VN \to V N}$
with the scattering lengths $a_{VN\to VN}$ given in (\ref{a:S11}). One
concludes that a substantial
fraction of the scattering length is a background term effect. We
emphasize that such
background term effects are crucial for the proper description of the
in-medium propagation
of vector mesons in nuclear matter. They act like an effective mean
field which shifts the energy of the in-medium vector-meson state.

\begin{table}[t]
\begin{tabular}{|r||c|c|c||c|c|c||c|c|c|} \hline
& $g_{\pi N} $ & $g_{\omega N} $ & $g_{\rho N}$& $\phi_{\pi \omega}$
[$\,^\circ $]&
$\phi_{\pi \rho}$ [$\,^\circ $]&
$\phi_{\omega \rho}$ [$\,^\circ $]&  $ m$[MeV ] & $\Gamma $ [MeV ]\\
\hline \hline

$N(1535)$    &  0.19 & 0.39 & 0.20 & -8.5 &-7.5 & 2.7& 1513 & 69 \\
\hline

$N(1650)$   & 0.04 & 0.45 & 0.00 & -89.3& -& - &  1624 & 60 \\ \hline

\end{tabular}
\vspace*{2mm} \caption{Resonance and background parameters
in the $I J^P =\frac{1}{2}\,\frac{1}{2}^-$ channel .}
\label{tab:1b}
\end{table}

We wish to make three points here. First, the schematic representation
(\ref{rep-S11}) of the amplitudes is respecting the unitarity condition only
approximatively. In the presence of
two resonances or a background term the 'unitarization' of the
amplitudes is not unique. Since we aim at only a qualitative interpretation of our
result we do not
use a more complicated and necessarily less transparent representation
of the amplitudes. In any case, the course of improving the representation
(\ref{rep-S11}) would ultimately lead us back to the original model or some improved
version of it.
Second we point out the importance of the background terms $b$ in
(\ref{rep-S11}).
It is clear that one may view the background terms as a particular model
for the
energy dependence of the resonance self energy. And third,
even the most general off-shell behavior in the meson-baryon resonance vertex
function, defines only a particular and incomplete model for the
background term.
One may always write down further 4-point
interaction terms, in the spirit of effective field theory, which would
alter the background term at will.

In the following subsection we will further study the
schematic resonance coupling constants by relating the parameters of
Tab. \ref{tab:1b}
to the most general and covariant 3-point interaction vertices
describing the coupling
of the on-shell resonances to the meson-baryon pairs. We will compare
our results
with recent values for the resonance coupling constants obtained by
Brown and Riska \cite{Brown:Riska} within the quark model.

\subsubsection{Resonance coupling constants}

The appropriate covariant interaction terms which couple the
resonances $ N(1535) $ and $N(1650)$ to the pion-nucleon channel
are readily constructed,
\begin{eqnarray}
&&{\mathcal L}_{\frac{1}{2}\,\frac{1}{2}^-}^{(\pi N)} =
{\textstyle{1\over \sqrt{3}}}\,f^{(1535)}_{\pi N}\, \bar
N_{1535}\,\vec \pi \cdot \vec \tau \,N +
{\textstyle{1\over\sqrt{ 3}}}\,f^{(1650)}_{\pi N}\, \bar
N_{1650}\,\vec \pi \cdot \vec \tau \,N +{\rm h.c.} \,,
\label{def-R:S11}
\end{eqnarray}
where we identify,
\begin{eqnarray}
&& \Big| f_{\pi N}^{(1535)} \Big|= \sqrt{
\frac{8\,\pi\,m_{1535}}{N_{\pi
N}^{(+)}(m_{1535})}}\,\Big|g_{\pi N}^{(1535)}\Big|\simeq 0.85 \,,
\nonumber\\
&& \Big|f_{\pi N}^{(1650)}\Big| = \sqrt{\frac{8\,\pi\,m_{1650}}{N_{\pi
N}^{(+)}(m_{1650})}}\,\Big|g_{\pi N}^{(1650)}\Big| \simeq 0.19 \,.
\label{g:S11:piN}
\end{eqnarray}
Note that we do not consider further interaction terms in
(\ref{def-R:S11}) which are redundant if the resonance, the nucleon and pion are
strictly on-shell. As discussed above such terms do affect the background terms
in the
pion induced vector-meson production amplitudes of Fig. \ref{fig:6}.
Here
we do not take up this issue since we believe that the reliable
determination of any background parameter is possible only via a coupled
channel
analysis. The coupling constants
$\big|f^{(1535)}_{\pi N, BR}\big| \simeq 3.6 $ and $\big|f^{(1650)}_{\pi N, BR} \big|\simeq
2.5 $
of Brown and Riska \cite{Brown:Riska} differ significantly from our
values given in (\ref{g:S11:piN}). We use a subscript
'BR' to identify the coupling constants of  \cite{Brown:Riska}.
On the other hand, our value for the N(1535) resonance coupling constant
compares favorably with the analysis \cite{Sauerman} where the value
$\big| f^{(1535)}_{\pi N } \big|\simeq 1.2 $, representative for various analyses, was
obtained.
The coupling constant $\big| f^{(1650)}_{\pi N } \big| \simeq 2.0 $ of
\cite{Sauerman} is significantly larger than our result. This may reflect the
fact that the $\pi \Delta $ and $\rho N$
channels, which are expected to be more important for the heavier
resonance,  were not explicitly included in \cite{Sauerman}.
Moreover, the extraction of the N(1650) resonance coupling strength to
the $\pi N$ channel from the $\pi N$ partial-wave amplitude is obscured
by a substantial background, which introduces a large ambiguity in the
resonance width and coupling parameters. Note that we do not consider
a possible phase parameter, $\phi_{\pi N}^{(R)}$, in the resonance coupling constant
\footnote{The phase parameters discussed here are not directly related to the
resonance and background phase parameters, $\phi_R$ and $\phi_B$, frequently
introduced at the level of the $S$-matrix, where one writes
$S = e^{2\,i \,(\phi_B+\phi_{R})}$ near the resonance pole. The background
phase accounts for the contributions of the non-resonant background, which in the
parameterization (\ref{rep-S11}) is absorbed in the background parameters $b_{VN\to V'N}$
and $b_{\pi N\to VN}$.},
\begin{eqnarray}
f_{\pi N}^{(R)} = e^{i\, \phi_{\pi N}^{(R)}}\,| f_{\pi N}^{(R)}| \,.
\label{discuss:phase}
\end{eqnarray}
It is evident that only relative phases, like
$\phi_{\pi \rho}^{(R)} =\phi_{\pi N}^{(R)}- \phi_{\rho N}^{(R)}$, are observable.
That reflects the freedom to absorb for instance the
phase $\phi_{\pi N}^{(R)}$ into the definition of the resonance field. Given that
our model leads to resonance coupling constants which deviate strongly from those
of the quark-model, we do not see much point to discuss the phase parameters in
this section. The values of our relative phase parameters are given in
Table \ref{tab:1b}.

We turn to the coupling constants of the resonances to the
vector-meson nucleon channels. The most general on-shell vertex permits
two
independent interaction vertices parameterized by axial-vector,
$f_{V \!N, A}^{(R)}$, and pseudo-scalar, $f_{V \!N, P}^{(R)}$,
coupling constants,
\begin{eqnarray}
&& {\mathcal L}_{\frac{1}{2}\,\frac{1}{2}^-}^{(V\! N)}=
{\textstyle{1\over 3}}\,f^{(1535)}_{\rho N,A}\, \bar
N_{1535}\,\gamma_5\,\gamma_{\,\mu}\,\vec \rho^\mu \cdot \vec
\tau \, N +{\textstyle{i\over 3}}\,f^{(1535)}_{\rho N,P}\, \bar
N_{1535}\,\gamma_5\,\vec \rho^{\,\mu} \cdot \vec
\tau \,\big( \partial_\mu\,N \big) \nonumber\\
&&\qquad+ \,{\textstyle{1\over \sqrt{3}}}\,f^{(1535)}_{\omega
N,A}\, \bar N_{1535} \,\gamma_5\,\gamma_\mu\,\omega^\mu \, N +
\,{\textstyle{i\over \sqrt{3}}}\,f^{(1535)}_{\omega N,P}\, \bar
N_{1535} \,\gamma_5\,\omega^\mu \, \big( \partial_\mu\,N \big)
\nonumber\\
&&\qquad + \,{\textstyle{1\over 3}}\,f^{(1650)}_{\rho N,A}\, \bar
N_{1650}\,\gamma_5\,\gamma_\mu\,\vec \rho^{\,\mu } \cdot \vec
\tau \, N + \,{\textstyle{i\over 3}}\,f^{(1650)}_{\rho N,P}\, \bar
N_{1650}\,\gamma_5\,\vec \rho^{\,\mu } \cdot \vec
\tau \, \big( \partial_\mu\,N \big)\nonumber\\
&&\qquad+ \,{\textstyle{1\over \sqrt{3}}}\,f^{(1650)}_{\omega
N,A}\,\bar N_{1650} \,\gamma_5\,\gamma_\mu\,\omega^\mu \, N +
{\textstyle{i\over \sqrt{3}}}\,f^{(1650)}_{\omega N, P}\,\bar
N_{1650} \,\gamma_5\,\omega^\mu \, \big( \partial_\mu\,N \big)
+{\rm h.c.}\,.\nonumber\\ \label{def-R:S11:VN}
\end{eqnarray}
The relative normalization factor of the $\omega N$ and $\rho N$
channels in (\ref{def-R:S11}) is motivated by the corresponding
factors in (\ref{r-def}). We illustrate the
different effects of the axial-vector versus the pseudo-scalar term
by evaluating the partial resonance-decay width,
\begin{eqnarray}
&& \Gamma_{R}^{(\rho N)} (\sqrt{s}\,) =
\frac{2}{3}\,\Big(E_N+m_N\Big)\,\Bigg(2\, \left(f^{(R)}_{\rho
N,A}\right)^2
\nonumber\\
&& \qquad \qquad \;\, +\Big(
\frac{\sqrt{s}-m_N}{m_\rho}\,f^{(R)}_{\rho N,A}
-\Big(E_N-m_N\Big)\,\frac{\sqrt{s}}{m_\rho }\,\,f^{(R)}_{\rho N,P}
\Big)^2 \Bigg)\,\frac{p_{\rho N}}{8\,\pi \,\sqrt{s}} \,,
\label{S11-self}
\end{eqnarray}
where $E_N^2= m_N^2+p_{\rho N}^2$. A folding with the $\rho$-meson
spectral function (\ref{rho-explicit}) analogous to
(\ref{rho-loop}) is understood in (\ref{S11-self}). The width of the
resonance in
Breit-Wigner approximation is given with $\Gamma_R = \Gamma_R(m_R)$.
The result (\ref{S11-self}) demonstrates that the
axial-vector coupling term with $f_{\rho N,A}^{(R)}$ dominates the
resonance-decay width. The pseudo-scalar contribution with
$f_{\rho N,P}^{(R)}$ is suppressed by at least two powers of the
phase-space factor $p_{\rho N}$. Note that the presence of two
independent coupling constants in (\ref{S11-self}) reflects the
two helicity states (\ref{j12-trafo}) a vector-meson nucleon pair
with $J={\textstyle{1 \over 2}}$ can occupy. Based on the
leading projectors introduced in (\ref{proj-v1}) we
identify the coupling constants
\begin{eqnarray}
&& g_{V\! N}^{(R)} =
\sqrt{\frac{N_{V\!N}(m_{R})}{8\,\pi\,m_{R}}}\,f_{V\!N,A}^{(R)}\,,
\quad m_R\,f_{V\!N,P}^{(R)} = f_{V\!N,A}^{(R)}
\,,
\nonumber\\
&& f_{\rho N,A}^{(1535)} \simeq 0.97\,,\quad
f_{\omega\!N,A}^{(1535)} \simeq 1.89\,,\quad
f_{\rho N,A}^{(1650)} \simeq 0.00\,,\qquad
f_{\omega\!N,A}^{(1650)} \simeq 2.17\,,
\label{det:S11:coupling}
\end{eqnarray}
where $R = (1535, 1650)$ and $V= (\rho, \omega)$ (see (\ref{rep-S11})).
The reliable determination of the pseudo-scalar coupling constant $f_{\rho
N,P}^{(R)}$ would require the consideration of subleading projectors
not considered in this work. Taking our values for the
resonance coupling constants in (\ref{det:S11:coupling})
we find partial $\rho N$-decay widths of less than 1 MeV  for
the N(1535) and N(1650) resonance.
We compare our resonance coupling constants with values of the
recent work \cite{Brown:Riska} derived within the quark model. We
observe
strong deviations for the vector meson coupling
constants $\big| f^{(1535)}_{\rho N,BR}\big|
\simeq 8.7 $, $\big|f^{(1535)}_{\omega N,BR}\big| \simeq 7.8 $ and
$\big|f^{(1650)}_{\rho N,BR} \big|\simeq 2.2 $, $\big|f^{(1650)}_{\omega N,BR} \big|\simeq
3.8 $ from our values in (\ref{det:S11:coupling}).
We should mention that the comparison of the vector-meson coupling
constants is
subtle. In \cite{Brown:Riska} it is implied $f_{VN,P}^{(R)} = 0$ in
conflict
with our model. The quark model appears to suggest axial-vector type
coupling
constants $f_{VN,A}^{(R)} $ for which we recollected their values above.
If we use the value $\big| f^{(1535)}_{\rho N,BR} \big|
\simeq 8.7 $  together with the $\rho$-meson spectral density
of (\ref{rho-explicit}) in (\ref{S11-self}) we obtain a partial $\rho
N$-decay width
of 26 MeV for the N(1535) resonance. Moreover, we observe that this
values is rather
sensitive to the precise form of the $\rho$-meson spectral function. If
we use a
simple Breit-Wigner model with $m_\rho \simeq 770$ MeV and $\Gamma_\rho
\simeq 150 $ MeV,
as applied in the analysis of Manley and Salesky \cite{Manley}, that
partial decay widths
becomes 143 MeV instead, clearly an unreasonable value.
It was emphasized in \cite{Brown:Riska} that only the ratios of the
resonance coupling constants are expected to be reasonable because
many-body quark
operators not considered in \cite{Brown:Riska} may change their results.
However, even taking
ratios only does not lead to a consistent matching of our values to
those of the
quark model. This is an interesting result which may help to
discriminate the
quark-model picture of resonances from the dynamical scenario in which
all resonances
except the baryon octet and decuplet ground states are considered
to be generated by important coupled channel effects \cite{LK,Krehl}.

We discuss the coupling of the s-wave resonances to the
photon-nucleon state as probed in the multipole amplitude. The
most general gauge invariant interaction vertex as required for on-shell
particles
is readily constructed \cite{Benmerrouche},
\begin{eqnarray}
{\mathcal L}_{\frac{1}{2}\,\frac{1}{2}^-}^{(\gamma N)} &=&
\frac{e}{4\,m_R}\,\bar
R\,\gamma_5\,\sigma_{\mu \nu} \,\Big(f_{\gamma N, S}^{(R)}+f_{\gamma N,
V}^{(R)} \,\tau_3 \Big) \,N\,F^{\mu \nu}  +{\rm h.c.} \,.\label{def:V:A}
\end{eqnarray}
where e is the unit charge ($e^2/(4\pi) \simeq 1/137$).
The isoscalar and isovector coupling constants $f_{\gamma
N,S}^{(R)}$ and $f_{\gamma N,V}^{(R)}$ for a given resonance, $R$,
are predicted by the generalized vector-meson dominance assumption
(\ref{gamma-ansatz:k}) in terms of the hadronic coupling constants
introduced in (\ref{def-R:S11:VN}) and the channel independent
parameters $g_{S(V),i}^{(\pm)}$ quantifying the generalized
vector-meson dominance assumption. The effective interaction
Lagrangian density describing the transition of the resonance into
the $\gamma N$ state is identified\footnote{The vector-meson dominance
assumption is defined with respect to the
covariant projector algebra. That implies a particular off-shell
structure of the hadronic transition vertex. Here we apply the
correct off-shell form as is probed by the leading projector, i.e. we
assume that $m_R\,f^{(R)}_{VN,P} = f^{(R)}_{VN,A} $ holds strictly.},
\begin{eqnarray}
{\mathcal L}_{\frac{1}{2}\,\frac{1}{2}^-}^{(\gamma N)} &=& e\,\bar R
\,\gamma_5\,\gamma^\mu\Big( {\textstyle{i\over \sqrt{3}}}\,f_{\omega N,A}^{(R)}\,
\Gamma_{S,\nu} +{\textstyle{i\over 3}}\,f_{\rho
N,A}^{(R)}\, \Gamma_{V,\nu}\,\tau_3 \Big)\,\Big(N\,F^{\mu \nu}
\Big)
\nonumber\\
&-& e\,\bar R \,\gamma_5\,\Big( {\textstyle{1\over \sqrt{3}}}\,f_{\omega N,P}^{(R)}\, \Gamma_{S,\nu}
+{\textstyle{1\over 3}}\,f_{\rho N,P}^{(R)}\,
\Gamma_{V,\nu}\,\tau_3 \Big)\, \Big((\partial^\mu\,N)\,F^{\mu \nu}
\Big) +{\rm h.c.} \,, \label{S11-gamma:a}
\end{eqnarray}
with the electromagnetic field strength tensor $F_{\mu \nu }=
\partial_\mu\,A_\nu - \partial_\nu\,A_\mu $ and the transition
operator $\Gamma_{S(V)}^\nu$,
\begin{eqnarray}
&& \Gamma_{S(V),\nu}^{}=
\Gamma_{S(V),\nu}^{(+)}+\Gamma_{S(V),\nu}^{(-)}\,,
\nonumber\\
&& \Gamma_{S(V),\nu}^{(\pm)}  =
\frac{m_{R}\pm i\,\gamma \cdot
\partial }{2\,m_{R}}\,\Big( \frac{g_{S(V),2}^{(\pm)}\mp
g_{S(V),1}^{(\pm)}}{m_{R}\,m_\omega }\,i\,\partial_\nu +
\frac{g_{S(V),1}^{(\pm)}}{m_\omega } \,\gamma_\nu \Big)  \,.
\label{S11-gamma}
\end{eqnarray}
In (\ref{S11-gamma}) the transition tensor (\ref{tensor-ansatz}) is cast into a form
that leads to the gauge invariant expressions (\ref{S11-gamma:a}). The linear
dependence on the photon 4-momentum $q_\mu$ in $\Gamma^{\mu \nu}_{(S,V)}(q;w)$
is absorbed into the field strength tensor $F_{\mu \nu}$, explaining why
(\ref{tensor-ansatz}) and (\ref{S11-gamma}) are Lorentz tensors of different degrees.
We obtain the result:
\begin{eqnarray}
f_{\gamma N, S}^{(R)} &=& \frac{1}{\sqrt{3}} \left( 2\,g^{(+)}_{S,1}
\,\frac{m_R-m_N}{m_\omega} +g^{(+)}_{S,2}
\,\frac{m_R+m_N}{m_\omega} \right)\,f_{\omega N,
A}^{(R)} \nonumber\\
&+& \frac{1}{\sqrt{3}}g^{(-)}_{S,1} \,\frac{m_R+m_N}{m_\omega} \, \Big(
f_{\omega N, A}^{(R)} -m_R\,f_{\omega N, P}^{(R)} \Big) \,,
\nonumber\\
f_{\gamma N, V}^{(R)} &=& \frac{1}{3} \left( 2\,g^{(+)}_{V,1}
\,\frac{m_R-m_N}{m_\omega} +g^{(+)}_{V,2}
\,\frac{m_R+m_N}{m_\omega} \right)\,f_{\rho N,
A}^{(R)} \nonumber\\
&+& \frac{1}{3}\,g^{(-)}_{V,1}
\,\frac{m_R+m_N}{m_\omega} \, \Big( f_{\rho N,
A}^{(R)} -m_R\,f_{\rho N, P}^{(R)} \Big) \,.
 \label{}
\end{eqnarray}
Our present scheme with (\ref{det:S11:coupling}) cannot determine the coupling
constants $g_{S(V),i}^{(-)}$. These coupling constants should be derived studying
photon-induced production of p-wave nucleon and isobar resonances. In Tab. \ref{tab:1c}
we compare the resulting electromagnetic resonance coupling constants with values
obtained in a previous analysis \cite{Feuster:Mosel}. We conclude that there is
reasonable agreement given the quite large uncertainties in the
coupling constants.

\begin{table}[t]
\begin{tabular}{|r||c|c||c|c|c|} \hline
& $f^{(1535)}_{\gamma N,S} $ & $f^{(1650)}_{\gamma N,S} $ &
$f^{(1535)}_{\gamma N,V}$&
$f^{(1650)}_{\gamma N,V}$ \\ \hline \hline

this work    &  0.13 & 0.18 & 0.47 & 0.00  \\ \hline

 \cite{Feuster:Mosel}    & 0.04-0.22 & 0.06-0.41 & 1.10-1.25 & 0.01-0.13
\\ \hline

\end{tabular}
\vspace*{2mm} \caption{Electromagnetic resonance coupling constants
in the $I J^P =\frac{1}{2}\,\frac{1}{2}^-$ channel .}
\label{tab:1c}
\end{table}

\subsection{$I \,J^P=\frac{3}{2}\,\frac{1}{2}^-$ channel}

We proceed with the $ I J^P=\frac{3}{2}\,\frac{1}{2}^-$ sector
where we include the four channels $\pi N$, $\pi \Delta $, $\rho
N$, and $K \Sigma$.  Note that here the $\omega N$ channel does not
contribute since we assume exact isospin conservation. The 10
parameters of this sector, determined by our best fit, are given in
Tab. \ref{tab:2}. As may be expected, the parameters, $g_{11},
g_{14}$ and $g_{44}$ describing the coupling strength in the $\pi
N$ and $K \Sigma $ channels do not satisfy the SU(3) constraint,
given in terms of the $h_i$ parameters specified in Tab.
\ref{tab:cs}. In fact we find strong deviations from those values
indicating that there are important SU(3) symmetry breaking effects
in this sector, once part of the SU(3) dynamics, e.g. the $K_\mu
\,\Sigma$ or $K\, \Sigma_\mu $ channels, is integrated out.

\begin{table}[h]
\begin{tabular}{|r||c|c|c|c|} \hline
$g^{(\frac{3}{2},+)}_{0}$\,$\Big[m_\pi^{-\frac{n+m}{2}}\Big]$& $\pi N
[n=1]
$& $\pi \Delta [n=3]$ & $\rho N[n=1] $  & $K \Sigma [n=1]$\\ \hline
\hline

$\pi N[m=1]$  & -29.41  & 0.41 & -25.46  & 11.81\\ \hline

$\pi \Delta [m=3] $  & 0.41 & 12.69  & 14.86 & -53.40\\ \hline

$\rho N [m=1]$  & -25.46 & 14.86 & -26.44  &  -5.77\\ \hline

$ K \Sigma [m=1] $  & 11.81 & -53.40 & -5.77 & 5.00\\ \hline

\end{tabular}
\vspace*{2mm} \caption{Coupling constants in the $I J^P
=\frac{3}{2}\,\frac{1}{2}^-$ channel.
}
\label{tab:2}
\end{table}

\begin{figure}[t]
\begin{center}
\includegraphics[width=10.5cm,clip=true]{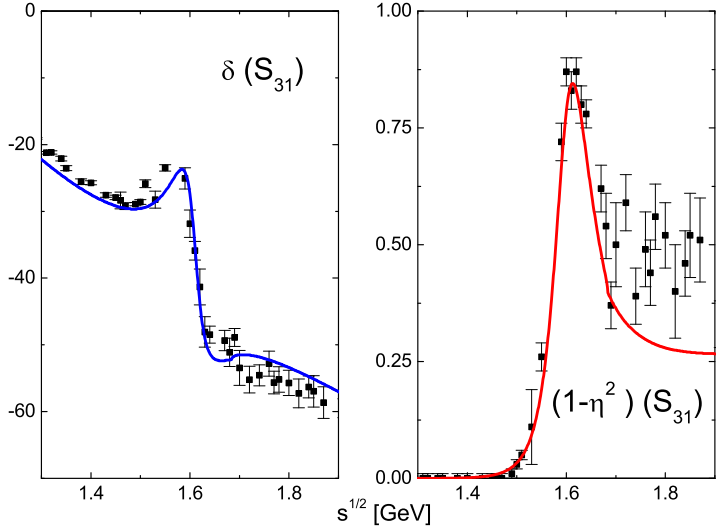}
\end{center}
\caption{Pion-nucleon scattering phase shift $\delta_{S_{31}}$ and
inelasticity parameter $\eta_{S_{31}}$ of the $S_{31}$ channel of our theory and of
the single-energy analysis SP98 \cite{Arndt:piN,SAID}.}
\label{fig:8}
\end{figure}

\begin{figure}[b]
\begin{center}
\includegraphics[width=13.5cm,clip=true]{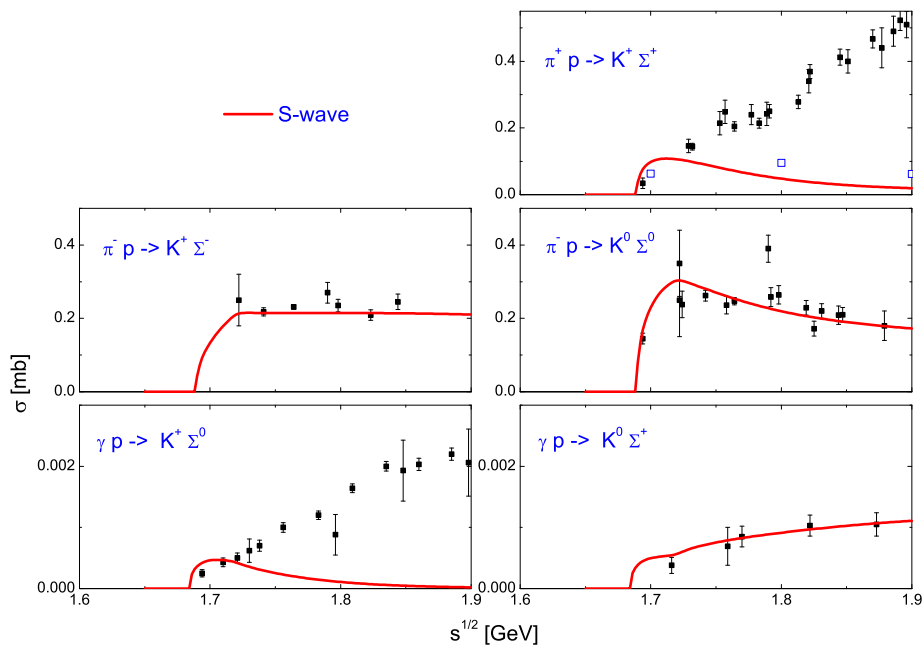}
\end{center}
\caption{Pion- and photon-induced $K \Sigma$ production cross sections. The
data are taken from
\cite{Bockhorst:et:al,Tran:et:al,Goers:et:al,Candlin:et:al,Candlin:PW,Landolt:Boernstein}.}
\label{fig:9}
\end{figure}

Our model reproduces the pion-nucleon phase shift and the
inelasticity parameter nicely, as shown in Fig. \ref{fig:8}. The
rapid energy variation close to $\sqrt{s}\simeq $ 1620 MeV reflects
the presence of the isobar resonance $\Delta(1620)$ with quantum
numbers  $I \,J^P = \frac{3}{2}\,\frac{1}{2}^-$. The fit is also
constrained by the pion-induced $ K^+\Sigma^+, K^+\,\Sigma^-$ and
$K^0\,\Sigma^0$ production cross section shown in Fig. \ref{fig:9}.
For the $\pi^+p\to K^+ \Sigma^+$ reaction data of good quality on
the differential cross section is available. These were used in the
partial-wave analysis of \cite{Candlin:PW}. We utilize this, by fitting our
model to the s-wave part of the production cross section, extracted
in this analysis. In Fig. \ref{fig:9} the s-wave cross section,
which is reasonably well reproduced by our model, is shown as open
squares, while the total cross section is given by the filled
squares. The difference between the total cross section and the
s-wave part, starting at about $\sqrt{s}\simeq 1750$ MeV
illustrates the importance of higher partial waves, not included in
our model. In the case of the reactions $ \pi^- p\to
K^+\,\Sigma^-\,, K^0\,\Sigma^0$ the data set does not permit a
reliable partial wave analysis. The few available data points
\cite{Good:Kofler,Baker:et:al:1} are consistent with s-wave dominance up to
$\sqrt{s}\simeq 1.9$ GeV, as suggested by the fit. Similar
conclusions can be drawn in the $\gamma$-induced $K^+
\Sigma^0$ and $K^0 \Sigma^+$ production
cross sections, also shown in Fig. \ref{fig:9}. We find that our
model provides a reasonable interpretation of the data, in
particular in the $K^0 \Sigma^+$ channel. The importance of higher
partial waves in the $K^+ \Sigma^0$ channel is, at least on a
qualitative level, consistent with available differential cross
sections \cite{Tran:et:al}, which show large non-isotropic
contributions already at $p_{\rm lab}$ = 1.5 GeV.

\begin{figure}[t]
\begin{center}
\includegraphics[width=9.5cm,clip=true]{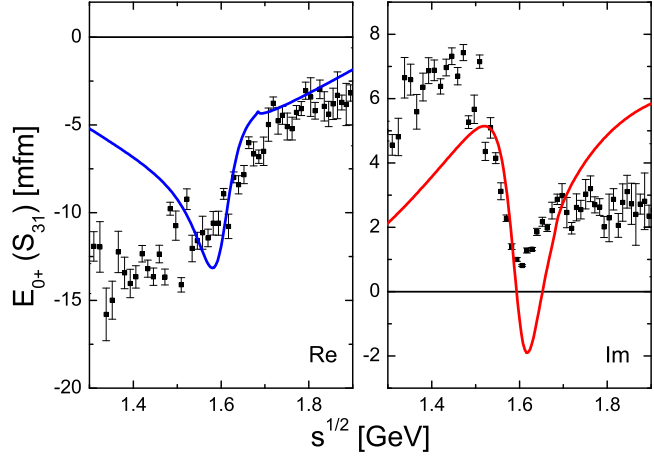}
\end{center}
\caption{Electric multipole amplitudes  $E_{0+}(S_{31})$ of
$\gamma$-induced pion production. We show the results of the
single-energy solution SM00 \cite{Arndt:gamma,SAID}.}
\label{fig:11}
\end{figure}

\begin{figure}[t]
\begin{center}
\includegraphics[width=10.5cm,clip=true]{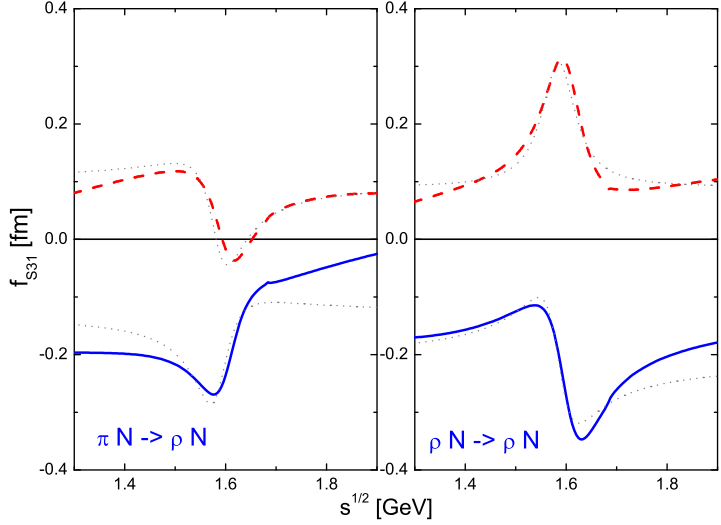}
\end{center}
\caption{Vector-meson production and scattering amplitude with
$I=\frac{3}{2}$ and $J=\frac{1}{2}$. The solid and dashed lines show the real
and imaginary parts of the amplitudes. The dotted lines are the result of the schematic
resonance exchange model introduced in (\ref{rep-S31}).}
\label{fig:12}
\end{figure}

We now turn to the $\gamma$ induced production of pions. In Fig.
\ref{fig:12} we compare the multipole amplitude $E_{0+}(S_{31})$
with the analysis of \cite{Arndt:gamma}. Given the simplicity of our
model we obtain a reasonable description of the real and the
imaginary part of $E_{0+}(S_{31})$ in the energy range of interest
here. The discrepancy
between our model and the empirical analysis may have several
causes. First, the generalized vector-meson dominance assumption
(\ref{gamma-ansatz:k}) for the electromagnetic interactions may be
too restrictive. If other couplings, not related to the
vector-meson amplitudes, play an important role, it is possible
that our model does not have enough freedom to reproduce the data in
the different partial waves and reaction channels simultaneously.
On the other hand, it may be that the deficiencies of our model reflect the
importance of inelastic channels, like the $K \Sigma_\mu$ or $K_\mu
N$, which so far are not included. A further possible source of the
discrepancy is the kinematically suppressed contributions from the
$\rho N$ states (\ref{j12-trafo}), which are not included in the
model. These define further contributions to the multipole
amplitudes $E_{0+}$. However, since these contributions are
suppressed by at least two powers of the phase space factor
$p_{\gamma N}= (\sqrt{s}-m_N^2/\sqrt{s}\,)/2$ they could play a
role only if the corresponding coupling constants are anomalously large.
The discrepancy at low energies is, as noted above, presumably related
to the strong energy dependence of the one-pion exchange contribution, which
is not treated properly in our scheme.

We now discuss the $\rho$-meson scattering and production
amplitudes. In Fig. \ref{fig:12} we present the pion-induced
$\rho$-meson production amplitude introduced in (\ref{f-pn:vn}).
and the $\rho$-meson nucleon scattering amplitude in the $I=\frac{3}{2}$ and
$J=\frac{1}{2}$ sector. The presence of the $\Delta(1620)$
resonance leads to a peaked structure in the imaginary parts of the
subthreshold amplitudes. Note we show the reduced amplitudes, which
for elastic $\rho N$ scattering is given by
\begin{eqnarray}
&& f^{(\frac{3}{2}\,\frac{1}{2}+)}_{\rho N \to \rho N}(\sqrt{s}\,)
= \frac{N_{\rho N}}{8\pi\,\sqrt{s}}\,
M_{33}^{(\frac{3}{2}+)}(\sqrt{s},0)\,.\qquad
\end{eqnarray}
This means that the diagonal amplitude at threshold equals the $\rho$-meson
nucleon s-wave scattering length
\begin{eqnarray}
 && a_{\rho N}^{(\frac{3}{2}\frac{1}{2})} =
f^{(\frac{3}{2}\frac{1}{2}+)}_{\rho N \to \rho
N}(m_N+m_\rho\,)\simeq (-0.25+i\,0.09)\,{\rm fm} \,, \label{a:S31}
\label{}
\end{eqnarray}
We find a repulsive s-wave scattering length in this channel, while
in the isospin one-half channel we found an attractive one. The
isospin averaged scattering length of the $J={\textstyle{1\over 2}
}$ channel,
\begin{eqnarray}
\bar a_{\rho N}^{(\frac{1}{2})}=
{\textstyle{1\over 3}}\,a_{\rho N}^{(\frac{1}{2}\frac{1}{2})}
+{\textstyle{2\over 3}}\,a_{\rho N}^{(\frac{3}{2}\frac{1}{2})}
\simeq (-0.11+i\,0.11) \,{\rm fm},
\label{rhoN1}
\end{eqnarray}
is therefore somewhat reduced.

We also analyze the production and scattering amplitudes in terms
of a resonance model
\begin{eqnarray}
&& f^{(\frac{3}{2}\frac{1}{2})}_{+,\,\rho N \to \rho N}(\sqrt{s}\,)
\simeq -\,
\frac{|g_{\rho
N}^{(1620)}|^2}{\sqrt{s}-m_{1620}+\frac{i}{2}\,\Gamma_{1620}}
+ b^{(\frac{3}{2}\frac{1}{2})}_{+,\,\rho N \to \rho N}\,,
\nonumber\\
&& f^{(\frac{3}{2}\frac{1}{2})}_{+,\,\pi N \to \rho N}(\sqrt{s}\,)
\simeq -\,
\frac{e^{\,i\,\phi^{(1620)}_{\pi \rho }}\,|g_{\pi N}^{(1620)}|\,|g_{\rho
N}^{(1620)}|}{\sqrt{s}-m_{1620}+\frac{i}{2}\,\Gamma_{1620}}
 + b^{(\frac{3}{2}\frac{1}{2})}_{+,\,\pi N \to \rho N}\,,
\label{rep-S31}
\end{eqnarray}

with the coupling constants  $g^{(1620)}_{\pi N}$,
$g^{(1620)}_{\rho N}$, the relative phase $\phi_{\pi
\rho}^{(1620)}$ and channel-dependent background terms
$b^{(\frac{3}{2}\frac{1}{2})}_{\pi N \to \rho N}$ and
$b^{(\frac{3}{2}\frac{1}{2})}_{\rho N \to \rho N}$. The resulting
resonance parameters are given in Tab. \ref{tab:2b}. We obtain a
fairly small decay width for the $\Delta (1620)$ resonance. Our
value is about a factor two smaller than that given by the Particle
Data Group \cite{PDB}, $\Gamma_{1620} \simeq 150$ MeV. This is
possibly a consequence of the dynamic generation of the isobar
resonance which typically introduces a pronounced energy dependence
in the resonance self energy.

\begin{table}[b]
\begin{tabular}{|r||c|c||c||c|c|} \hline
& $g_{\pi N} $ &  $g_{\rho N}$ & $\phi_{\pi \rho}$ [$\,^\circ $] &  $
m$[MeV ] & $\Gamma $ [MeV ]\\ \hline \hline

$\Delta(1620)$  &  0.17 &  0.21 & 130 & 1585 & 80 \\ \hline

\end{tabular}
\vspace*{2mm} \caption{Resonance parameters in the $I J^P
=\frac{3}{2}\,\frac{1}{2}^-$ channel .}
\label{tab:2b}
\end{table}

In Fig. \ref{fig:12} the amplitudes of the resonance-exchange model
is shown by the dotted lines. Both amplitudes are fairly well
represented by this simple model (\ref{rep-S31}). We note, however,
that it is crucial to allow for a strong, repulsive background
contribution,
\begin{eqnarray}
&& b^{(\frac{3}{2}\frac{1}{2})}_{-,\, \pi N \to \rho N} \simeq
(-0.13+i\,0.10)\,{\rm fm} \,,\quad
b^{(\frac{3}{2}\frac{1}{2})}_{-,\, \rho N \to \rho N} \simeq
(-0.21+i\,0.09)\,{\rm fm} \,.
\label{b:S31}
\end{eqnarray}
The sizeable imaginary part of the background parameters in (\ref{b:S31})
may be a reflection of our 'small' decay width of the resonance.
By forcing all of the imaginary part of the amplitude into an effective width parameter,
the total decay width is enhanced and approaches the value given by the Particle
Data Group. In the subsequent section we compare these results with quark-model
predictions for the coupling constants.

\subsubsection{Resonance coupling constants}

The interaction term which couples the resonances $ \Delta(1620) $
to the pion-nucleon channel is readily constructed,
\begin{eqnarray}
&& {\mathcal L}_{\frac{3}{2}\,\frac{1}{2}^-}^{(\pi N)} =
f^{(1620)}_{\pi N}\, \bar \Delta_{1620}\,\vec \pi \cdot  \vec
T \,N  +{\rm h.c.} \,
\nonumber\\
&& \Big| f_{\pi N}^{(1620)}\Big| = \sqrt{\frac{8\,\pi\,m_{1620}}{N_{\pi
N}^{(+)}(m_{1620})}}\, \Big|g_{\pi N}^{(1620)} \Big|\simeq 0.19 \,,
\label{}
\end{eqnarray}
where we used the results presented in the previous section and assumed on-shell kinematics.
The corresponding value for the resonance coupling constant obtained from the
quark model \cite{Brown:Riska} is $\big| f_{\pi N,BR}^{(1620)} \big| \simeq 1.67$. Again
we a find large deviation.

We turn to the coupling constants of the resonances to the
vector-meson nucleon channels. The most general on-shell vertex
permits two independent interaction terms parameterized by
axial-vector, $f_{\rho N, A}^{(1620)}$, and pseudo-scalar, $f_{\rho
N, P}^{(1620)}$, coupling constants,
\begin{eqnarray}
&& {\mathcal L}_{\frac{3}{2}\,\frac{1}{2}^-}^{(\rho N)}=
{\textstyle{1\over \sqrt{3}}}\,f^{(1620)}_{\rho N,A}\, \bar
\Delta_{1620}\,\gamma_5\,\gamma_{\,\mu}\,\vec \rho^\mu \cdot
\vec T\, N
\nonumber\\
&& \qquad \;\; +\,{\textstyle{i\over \sqrt{3}}}\,f^{(1620)}_{\rho
N,P}\, \bar \Delta_{1620}\,\gamma_5\,\vec \rho^{\,\mu} \cdot
\vec T\,\big(
\partial_\mu\,N \big) +{\rm h.c.} \,.  \label{def-R:S31:VN}
\end{eqnarray}
We illustrate the different effects of the axial-vector versus the
pseudo-scalar term by evaluating the partial decay width of the resonance:
\begin{eqnarray}
&& \Gamma_{1620}^{(\rho N)} (\sqrt{s}\,) =
\frac{2}{3}\,\Big(E_N+m_N\Big)\,\Bigg(2\, \left(f^{(1620)}_{\rho
N,A}\right)^2
\nonumber\\
&& \qquad \qquad \;\, +\Big(
\frac{\sqrt{s}-m_N}{m_\rho}\,f^{(1620)}_{\rho N,A}
-\Big(E_N-m_N\Big)\,\frac{\sqrt{s}}{m_\rho }\,\,f^{(1620)}_{\rho
N,P} \Big)^2 \Bigg)\,\frac{p_{\rho N}}{8\,\pi \,\sqrt{s}} \,,
\label{S31-self}
\end{eqnarray}
where $E_N^2= m_N^2+p_{\rho N}^2$. A folding with the $\rho$-meson
spectral function (\ref{rho-explicit}) analogous to
(\ref{rho-loop}) is understood. The result (\ref{S31-self}) demonstrates
that the pseudo-scalar
contribution with $f_{\rho N,P}^{(1620)}$ is suppressed by at
least two powers of the phase-space factor $p_{\rho N}$.  We identify
the coupling constants implied by the leading projectors introduced in
(\ref{proj-v1}),
\begin{eqnarray}
&& \Big| f_{\rho N,A}^{(1620)}\Big| = \sqrt{\frac{8\,\pi\,m_{1620}}{N_{\rho
N}(m_{1620})}}\,\Big| g_{\rho N}^{(1620)} \Big|\simeq 1.04\,, \quad
m_{1620}\,f_{\rho N,P}^{(1620)} = f_{V\!N,A}^{(1620)}
\,.\label{det:S31:coupling}
\end{eqnarray}
As in the $S_{11}$ sector, a reliable determination of the pseudo-scalar
coupling constant $f_{\rho N,P}^{(1620)}$ would require subleading
projectors, which are not considered in this work. The axial-vector
resonance coupling constant in (\ref{det:S31:coupling}) agrees
reasonably well with the quark-model result $\big| f_{\rho N,BR}^{(1620)} \big|\simeq 1.52$
of \cite{Brown:Riska}. Our value for this coupling constant implies
a partial $\rho N$-decay width for the $\Delta(1620)$ resonance of about
1 MeV when an energy dependent $\rho$-meson self energy (\ref{example:rho}) is employed.

We now discuss the electromagnetic interactions of the $\Delta (1620)$.
The most general on-shell interaction vertex for the coupling of
this resonance to the photon-nucleon state is of the form
\begin{eqnarray}
{\mathcal L}_{\frac{3}{2}\,\frac{1}{2}^-}^{(\gamma N)}
&=&\frac{e}{4\,m_R}\,
f_{\gamma N, V}^{(1620)} \,\bar
\Delta_{1620}\,T_3\,\gamma_5\,\sigma_{\mu \nu} \, \,N\,F^{\mu \nu}
 +{\rm h.c.} \,.\label{}
\end{eqnarray}
The coupling constant can be determined by analyzing the corresponding multipole
amplitudes. By using the generalized vector-meson dominance assumption
(\ref{gamma-ansatz:k}), we can relate the isovector coupling constant
$f_{\gamma N,V}^{(1620)}$ to the hadronic coupling constants
introduced in (\ref{def-R:S31:VN}) and the channel independent
parameters $g_{S(V),i}^{(\pm)}$, which quantifies the strength of the photon-vector meson
conversion matrix element. It is then straightforward to construct the
effective interaction Lagrangian density, which
describes the transition of the resonance into the $\gamma N$
state
\begin{eqnarray}
{\mathcal L}_{\frac{3}{2}\,\frac{1}{2}^-}^{(\gamma N)} &=&
i\,e\,{\textstyle{1\over \sqrt{3}}}\,f_{\rho N,A}^{(1620)}\, \bar
\Delta_{1620} \,T_3\,\gamma_5\,\gamma^\mu \,
\Gamma_{V,\nu} \,\Big(N\,F^{\mu \nu} \Big)
\nonumber\\
&-&e\,{\textstyle{1\over \sqrt{3}}}\,f_{\rho N,P}^{(1620)}\, \bar
\Delta_{1620} \,T_3\,\gamma_5\, \Gamma_{V,\nu}\,
\Big((\partial^\mu\,N)\,F^{\mu \nu} \Big) +{\rm h.c.} \,,
\label{S31-gamma}
\end{eqnarray}
with the electromagnetic field strength tensor $F_{\mu \nu }=
\partial_\mu\,A_\nu - \partial_\nu\,A_\mu $ and the transition
operator $\Gamma_{V}^\nu$ given in (\ref{S11-gamma}). We then find:
\begin{eqnarray}
f_{\gamma N, V}^{(1620)} &=& \frac{1}{\sqrt{3}}\left(
2\,g^{(+)}_{V,1} \,\frac{m_{1620}-m_N}{m_\omega}
+g^{(+)}_{V,2} \,\frac{m_{1620}+m_N}{m_\omega}
\right)\,f_{\rho N,
A}^{(1620)} \simeq 0.93\,,
 \label{gamma:1620}
\end{eqnarray}
where (\ref{det:S31:coupling}) is strictly applied. Our value in
(\ref{gamma:1620}) does not agree too well with the range of values
$|f_{\gamma N, V}^{(1620)}| \simeq 0.01-0.45 $
obtained in \cite{Feuster:Mosel}. As was emphasized in \cite{Feuster:Mosel}
for the resonance $\Delta(1620)$ the coupling constants are not determined reliably.

\subsection{$I \,J^P=\frac{1}{2}\,\frac{3}{2}^-$ channel}

We continue with the $ I J^P=\frac{1}{2}\,\frac{3}{2}^-$ sector
where 10 parameters are required to account for the four channels
$\pi N$, $\pi \Delta $, $\rho N$, and $\omega N$. The parameters of our
best fit are given in Tab. \ref{tab:3}. The inelastic channels
are of almost the same importance in the coupled channel matrix,
with roughly equal coupling strength to the $\pi N$ channel.
In particular there is no indication that it would be legitimate to
integrate out the $\omega N$ channel.
\begin{table}[h]
\begin{tabular}{|r||c|c|c|c|} \hline
$g^{(\frac{1}{2},-)}_{1}$\,$\Big[m_\pi^{-\frac{n+m}{2}}\Big]$& $\pi
N[n=3] $
&$\pi \Delta [n=1]$ & $\rho N[n=1] $ &$\omega N [n=1]$ \\ \hline \hline

$\pi N [m=3]$  & 4.50 & 5.19 & 6.70 & 9.94  \\ \hline

$\pi \Delta [m=1]$  & 5.19 & -11.90 & -7.37 & 24.72 \\ \hline

$\rho N [m=1]$   & 6.70 & -7.37 & 5.26 & 23.88 \\ \hline

$\omega N [m=1]$  & 9.94 & 24.72 & 23.88 & 24.65 \\ \hline

\end{tabular}
\vspace*{2mm} \caption{Coupling constants in the $I J^P
=\frac{1}{2}\,\frac{3}{2}^-$ channel.}
\label{tab:3}
\end{table}

\begin{figure}[b]
\begin{center}
\includegraphics[width=10.5cm,clip=true]{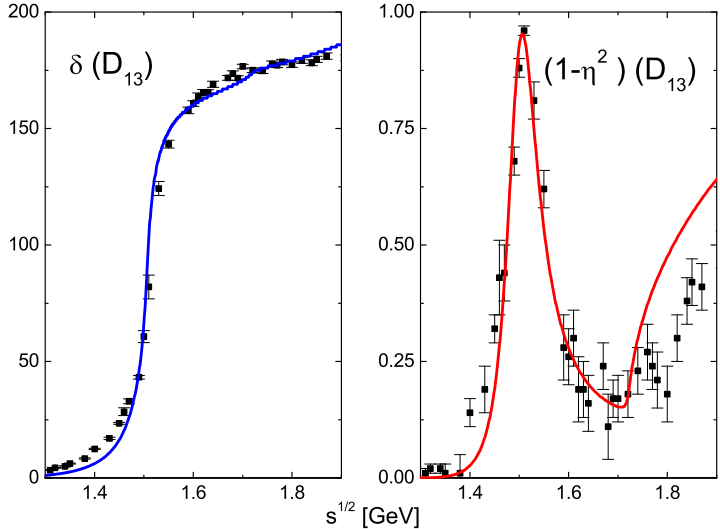}
\end{center}
\caption{Pion-nucleon scattering phase shift $\delta_{D_{13}}$ and
inelasticity parameter $\eta_{D_{13}}$ of the $D_{13}$ channel.
We compare to the single-energy solution SP98 of \cite{Arndt:piN,SAID}}
\label{fig:13}
\end{figure}

The pion-nucleon phase shift and its inelasticity parameter, presented
in Fig. \ref{fig:8},
are well reproduced by our model. The phase shift and
inelasticity parameters clearly show the  presence of the nucleon
resonance $N(1520)$ with $I=\frac{1}{2}$ and $J=\frac{3}{2}$. The phase
shift passes
through $90$ degree close to
$\sqrt{s}\simeq $ 1520 MeV. It is crucial to realize that the fit is
constrained also
by the vector-meson production cross section shown already in Fig.
\ref{fig:1} and Fig. \ref{fig:2}. Note
that at present the pion-induced $\pi \Delta (1232)$ production does not
constrain the
model since differential production cross sections are not available.
\begin{figure}[t]
\begin{center}
\includegraphics[width=13.5cm,clip=true]{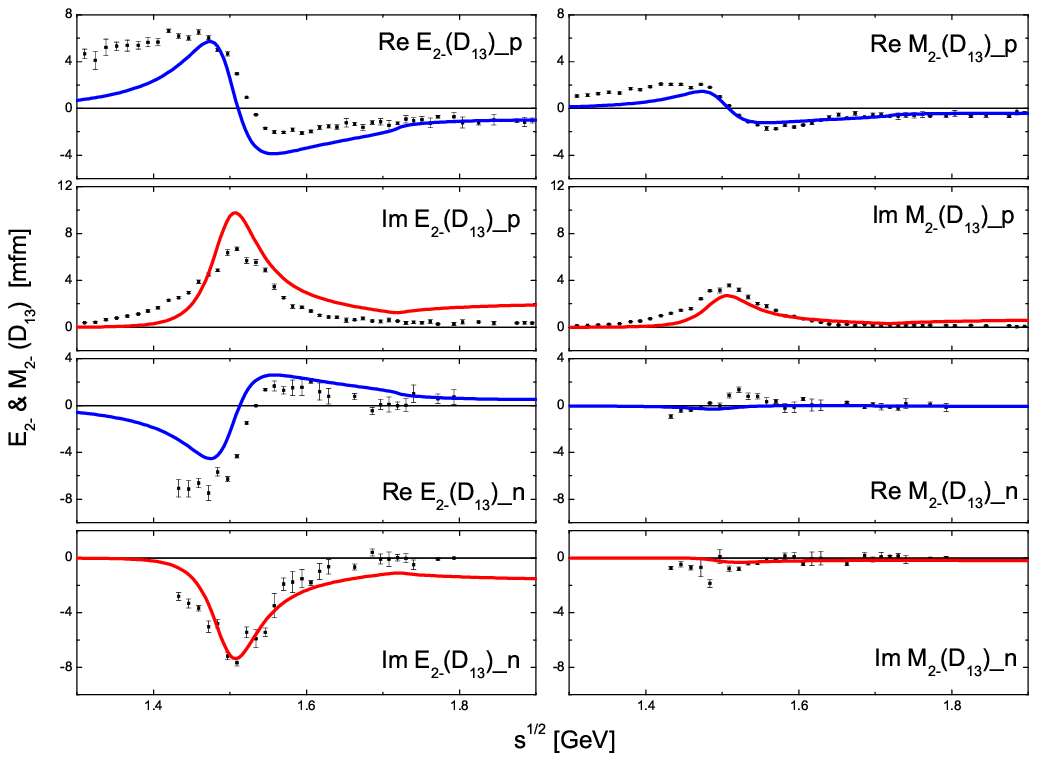}
\end{center}
\caption{Electric and magnetic multipole amplitudes
$E_{2-}(D_{13})_{p,n}$ and
$M_{2-}(D_{13})_{p,n}$ of photon-induced pion production (see
(\ref{match-multipole})). It is shown the single energy solution SM00
\cite{Arndt:gamma,SAID}}
\label{fig:14}
\end{figure}
Of particular importance for the determination of the parameter set in
Tab. \ref{tab:3} are the
electric and magnetic multipole amplitudes $E_{2-}^{(p,n)}(D_{13})$ and
$M_{2-}^{(p,n)}(D_{13})$
of the $\gamma$ induced pion production process. One important remark is
here
in order. The neglected $\rho N$ and $\omega N$ states in
(\ref{j33-trafo}) define
further contributions to the multipole amplitudes $E_{2-}(D_{13})$ and
$M_{2-}(D_{13})$. However,
any contribution of those additional states is suppressed by at least
two
powers of the phase space factor $p_{\gamma N}=
(\sqrt{s}-m_N^2/\sqrt{s}\,)/2$. In view of this
uncertainty we believe that Fig. \ref{fig:14} demonstrates a fair
description of the resonance
structures in the multipole amplitudes.

\begin{figure}[t]
\begin{center}
\includegraphics[width=10.5cm,clip=true]{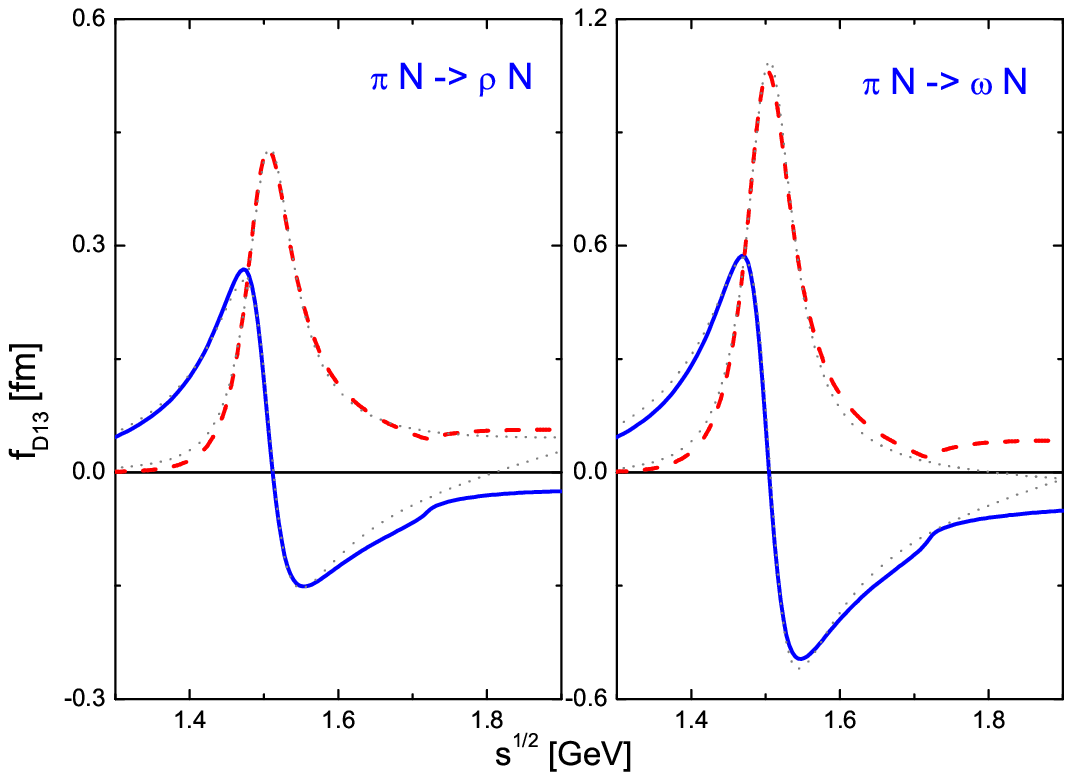}
\end{center}
\caption{Pion-induced vector-meson production amplitudes with
$I=\frac{1}{2}$ and $J=\frac{3}{2}$.
The real and imaginary parts of the amplitudes are shown
by solid and dashed lines. The results of the schematic resonance
exchange model (\ref{rep-D13}) are represented by dotted lines.}
\label{fig:15}
\end{figure}
\begin{figure}[t]
\begin{center}
\includegraphics[width=13.5cm,clip=true]{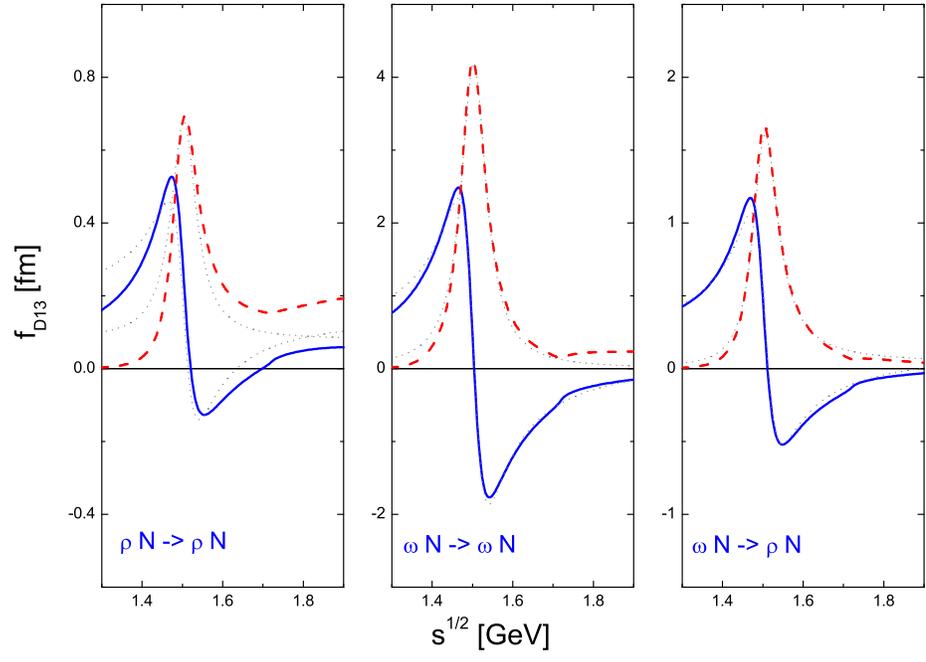}
\end{center}
\caption{Vector-meson nucleon scattering amplitude with $I=\frac{1}{2}$
and $J=\frac{3}{2}$. Solid, dashed and dotted lines as in Fig. \ref{fig:15}.}
\label{fig:16}
\end{figure}

We now turn to the vector-meson scattering and production amplitudes.
In Fig. \ref{fig:15} we present the pion-induced
vector-meson production amplitudes $I=\frac{1}{2}$ and
$J=\frac{3}{2}$ sector
while the corresponding vector-meson nucleon scattering amplitudes
are shown in Fig. \ref{fig:16}. The presence of the $N(1520)$ resonance
below threshold is clearly reflected in the reduced scattering amplitudes,
\begin{eqnarray}
&& f^{(\frac{1}{2}\frac{3}{2}-)}_{\rho N \to \rho N}(\sqrt{s}\,) =
\frac{N_{\rho N}}{8\pi\,\sqrt{s}}\,
M_{33}^{(\frac{1}{2}-)}(\sqrt{s},1)\,,
\nonumber\\
&& f^{(\frac{1}{2}\frac{3}{2}-)}_{\omega N \to \omega
N}(\sqrt{s}\,) = \frac{N_{\omega
N}}{8\pi\,\sqrt{s}}\,M_{44}^{(\frac{1}{2}-)}(\sqrt{s},1)\,.
\label{}
\end{eqnarray}

The vector-meson nucleon s-wave scattering length in this sector
corresponds to the value of the elastic  amplitudes at threshold, i.e.
$\sqrt{s}=m_N+m_V$
\begin{eqnarray}
&& a_{\rho N}^{(\frac{1}{2}\frac{3}{2})} =
f^{(\frac{1}{2}\frac{3}{2})}_{+,\,\rho N \to \rho N}(m_N+m_\rho\,)
\simeq (0.02+i\,0.15)\,{\rm fm }\,,
\quad
\nonumber\\
&& a_{\omega N}^{(\frac{1}{2}\frac{3}{2})} =
f^{(\frac{1}{2}\frac{3}{2})}_{+,\,\omega N \to \omega N}(m_N+m_\omega\,)
\simeq (-0.43+i\,0.15)\,{\rm fm }\;.
\label{a:D13}
\end{eqnarray}
The $\omega$-meson nucleon scattering
length is repulsive
reflecting a strong coupling of the subthreshold resonance N(1520) to the $\omega N$
channel. The small imaginary part results from the various inelastic
channels that are open at
the $\omega N$ threshold. For the $\rho$-meson we find an imaginary part
of similar size, but a very small, attractive, real part.
Using the scattering lengths obtained in the
$S_{11}$ sector we can now compute the spin averaged $\omega$-meson
nucleon scattering length,
\begin{eqnarray}
\bar a_{\omega N} =
{\textstyle{1\over 3}}\,a_{\omega N}^{(\frac{1}{2}\frac{1}{2})}
+{\textstyle{2\over 3}}\,a_{\omega N}^{(\frac{1}{2}\frac{3}{2})}
\simeq (-0.44+i\,0.20)\, {\rm fm}.
\label{}
\end{eqnarray}
This implies that an $\omega$ meson in low-density nuclear matter experiences
a repulsive mass shift. This would rule out the existence of
$\omega$-mesic
bound states at the nucleus surface. This is in contrast to the
results of Klingl {\em et al.}
\cite{Klingl} who find an attractive scattering length for the
$\omega$-meson, which would be conducive for $\omega$-mesic atom states.
However, it is possible that there are
strong non-linearities in the nuclear optical potential that lead to
sufficient
attraction at somewhat higher density and thus to deeply bound
$\omega$-mesic atom states.
We return to this issue in section 6 when discussing the vector-meson
spectral densities
in nuclear matter.

It is useful to explore to what extend the production and scattering
amplitudes may be
represented in terms of a schematic resonance exchange model. We
introduce a set of
resonance coupling constants $g^{(1520)}_{\pi N}$ and $g^{(1520)}_{V
N}$,
phase parameters $\phi_{\pi V}$ and $\phi_{V V'}$, in terms of which we
fit
the resonance structure of the amplitudes,
\begin{eqnarray}
&& f^{(\frac{1}{2}\frac{3}{2})}_{-,\,\pi N \to V N}(\sqrt{s}\,) \simeq
-\frac{e^{\,i\,\phi_{\pi V}}\,|g_{\pi N}^{(1520)}|\,|g_{V
N}^{(1520)}|\,p^2_{\pi N}}
{m_{1520}^2\,(\sqrt{s}-m_{1520}+\frac{i}{2}\,\Gamma_{1520})}
+\,\frac{p^2_{\pi N}}{m_{1520}^2}\,b^{(\frac{1}{2}\frac{3}{2})}_{-,\,\pi
N \to  V N} \,.
\nonumber\\
&& f^{(\frac{1}{2}\frac{3}{2})}_{-,\,V N \to V' N}(\sqrt{s}\,) \simeq
-\frac{e^{\,i\,\phi_{V V'}}\,|g_{V N}^{(1520)}|\,|g_{V'
N}^{(1520)}|}{\sqrt{s}-m_{1520}+\frac{i}{2}\,\Gamma_{1520}}
 + b^{(\frac{1}{2}\frac{3}{2})}_{-,\,V N \to V' N}\,.
\label{rep-D13}
\end{eqnarray}
Important background effects in the amplitudes are parameterized by
the channel-dependent parameters
$b^{(\frac{1}{2}\frac{3}{2})}_{X\to V N}$.
In Fig. \ref{fig:15} and  \ref{fig:16} we confront the schematic
expressions (\ref{rep-D13})
with our vector-meson production and scattering amplitudes. The dotted
lines show
real and imaginary parts of the amplitudes as given by the schematic
resonance exchange
model (\ref{rep-D13}). The resonance parameters
are collected in Tab. \ref{tab:3b}. The resonance exchange model leads
to a fair
reproduction of all but the elastic $\rho N$ amplitude provided the
background terms,
\begin{eqnarray}
&& b^{(\frac{1}{2}\frac{3}{2})}_{-,\, \pi N \to \omega N} \simeq
(1.12-i\,0.18)\,{\rm fm} \,, \quad
b^{(\frac{1}{2}\frac{3}{2})}_{-,\, \pi N \to \rho N} \simeq
(0.57+i\,0.11)\,{\rm fm} \,,
\nonumber\\
&& b^{(\frac{1}{2}\frac{3}{2})}_{-,\, \omega N \to \omega N} \simeq
(0.24+i\,0.00)\,{\rm fm} \,, \quad
b^{(\frac{1}{2}\frac{3}{2})}_{-,\, \rho N \to \rho N} \simeq
(0.16+i\,0.08)\,{\rm fm} \,,
\nonumber\\
&&
b^{(\frac{1}{2}\frac{3}{2})}_{-,\, \rho N \to \omega N} \simeq
(0.15+i\,0.03)\,{\rm fm}\,,
\label{b:D13}
\end{eqnarray}
are incorporated. Note that the coupling strength of the N(1520) to the
$\rho N$ state
was extracted from the off diagonal $\omega N \to \rho N$ amplitude
using the values of the
resonance mass and width as obtained from the elastic $\omega N$
amplitude.
The elastic $\rho N$ amplitude was then fitted in terms of only the
background
parameters. The discrepancy found for that amplitude
reflects intrinsic limitations of the schematic resonance model which
does not
properly account for the energy dependence of the resonance self energy.

\begin{table}[t]
\begin{tabular}{|r||c|c|c||c|c|c||c|c|} \hline
& $g_{\pi N} $ & $g_{\omega N} $  & $g_{\rho N}$ &
$\phi_{\pi \omega}$ [$\,^\circ $]& $\phi_{\pi \rho}$ [$\,^\circ $]&
$\phi_{\omega \rho}$ [$\,^\circ $] &  $ m$[MeV ] & $\Gamma $ [MeV ]\\
\hline \hline

$N(1520)$  &  2.71 & 0.90 & 0.34 & 0.6 & -7.0 & -9.2 & 1503 & 76 \\
\hline

\end{tabular}
\vspace*{2mm} \caption{Resonance parameters in the $I J^P
=\frac{1}{2}\,\frac{3}{2}^-$ channel .}
\label{tab:3b}
\end{table}

\subsubsection{Discussion of N(1520) resonance coupling constants}

We compare the resonance coupling constants with the quark-model values found in
\cite{Brown:Riska}. Since that comparison has subtle aspects we do
this in some detail.  The d-wave nucleon resonance N(1520) with
$I\,J^P=\frac{1}{2}\,\frac{3}{2}^-$ couples to the
pion-nucleon  channel in the following form:
\begin{eqnarray}
&&{\mathcal L}_{\frac{1}{2}\,\frac{3}{2}^-}^{(\pi N)} =
\frac{f^{(1520)}_{\pi N}}{m_\pi}\, \bar R_{\mu} \,(\partial^\mu
\,\vec \pi )\cdot \vec \tau \, i\,\gamma_5\, N +{\rm h.c.} \,,
\label{R:D13}
\end{eqnarray}
with $R_\mu =N_{1520,\,\mu}$, the nucleon field, $N$, and the
pion field $\pi$.  We identify
\begin{eqnarray}
\Big| f_{\pi N}^{(1520)} \Big| =
\frac {\sqrt{8\,\pi\,N_{\pi N}^{(+)}(m_{1520})}}
{m_{1520}^{3/2}/m_\pi} \,\Big| g_{\pi N}^{(1520)} \Big|\simeq 1.44
\,, \label{}
\end{eqnarray}
and conclude that our value compares favorably with the value
$\big| f_{\pi N,BR}^{(1520)} \big| \simeq 1.71$, which was obtained in the quark
model
\cite{Brown:Riska}.

In contrast to the simple one-parameter structure of the
pion-nucleon vertex the vector-meson nucleon vertices permit a
much richer structure leading altogether to six coupling
constants. We construct the most general on-shell interaction vertex,
\begin{eqnarray}
{\mathcal L}_{\frac{1}{2}\,\frac{3}{2}^-}^{(V\! N)} &=&
f^{(1520)}_{\omega N,S} \, \bar R_\mu \,N\,\omega^\mu
+{\textstyle{1\over \sqrt{3}}}\,f^{(1520)}_{\rho N,S} \, \bar
R_\mu \,\vec \tau\,N \,\vec \rho^\mu +i\,f^{(1520)}_{\omega N,V}
\, \bar R_\mu \,\gamma_\alpha \,(\partial_\mu  N )\,\omega^\alpha
\nonumber\\
&+&{\textstyle{i\over \sqrt{3}}}\,f^{(1520)}_{\rho N,V} \, \bar
R_\mu\,\vec \tau \,\gamma_\alpha \,(\partial_\mu  N )\,\vec
\rho^\alpha +i\,f^{(1520)}_{\omega N,T}\,  \bar R_\mu
\,\sigma_{\alpha \beta}\, (\partial_\mu
\partial_\alpha N)\, \omega^\beta
\nonumber\\
&+& {\textstyle{i\over \sqrt{3}}}\,f^{(1520)}_{\rho N,T}\,  \bar
R_\mu\,\vec \tau \,\sigma_{\alpha \beta}\, (\partial_\mu
\partial_\alpha N)\, \rho^\beta +{\rm h.c.}\,, \label{D13-general}
\end{eqnarray}
describing the interaction of the vector meson fields $\omega_\mu
$ and $\rho_\mu$. The three coupling constants $f_S, f_V$ and
$f_T$ are independent quantities and need to be determined
separately. In our work we evaluate only the scalar term,
$f_{S}$,
\begin{eqnarray}
&&g_{VN}^{(1520)}
=\sqrt{\frac{N_{VN}(m_{1520})}{8\,\pi\,m_{1520}}}\,
f_{VN,S}^{(1520)}\,, \quad f_{VN,V}^{(1520)}=f_{VN,T}^{(1520)}=0\,,
\nonumber\\
&& \Big| f_{\rho N,S}^{(1520)} \Big| \simeq 1.65\,,\qquad
\Big|f_{\omega N,S}^{(1520)} \Big| \simeq 4.35 \,.
\label{}
\end{eqnarray}
We will demonstrate explicitly that the determination of $f_V$ and
$f_T$ requires the control of terms suppressed by the phase space
factor $p_{\rho N}^2$ or $p_{\omega N}^2$.

In our previous analyses
\cite{Hirschegg}, where we did not include constraints from
the $\gamma$ induced production data we obtained a significantly
larger ratio $f^{(1520)}_{\rho N}/ f^{(1535)}_{\omega N} \simeq 0.9
$. We point out, however, that once the $\gamma$ induced production
data, in particular the electric and magnetic multipole amplitudes
$E_{2-} $ and $M_{2-}$ describing the $\gamma$-induced pion
production data, are considered (see Fig. \ref{fig:14}), the
coupling of the $N(1520)$ resonance to the $\rho N$ channel is
reduced. Here we observe that the extraction of the coupling
constant $f^{(1520)}_{\rho N}$ from the hadronic sector only
appears rather model dependent. In fact this observation was in
part our motivation for extending the analysis of \cite{Hirschegg}.
Data on the dilepton production process
$\pi^- p\to n \,e^+ \,e^-$ are expected to provide further
constraints on the vector-meson coupling strengths to the
subthreshold baryon resonances. In this reaction the subthreshold
vector-meson nucleon production amplitudes, required for the
determination of $f^{(1520)}_{\rho N}$ and $f^{(1520)}_{\omega N}$
\cite{SLF} are probed. We emphasize that any microscopic theory of
the in-medium properties of vector mesons requires the information
encoded in these coupling constants as input. This will be
discussed in detail in section 6.

In order to illustrate the effect of
the different coupling constants we evaluate the imaginary part of
the reduced resonance self energy,
\begin{eqnarray}
&& \Gamma_{1520}^{(\rho N)} (\sqrt{s}\,) =2\,\Bigg(
\left(f^{(1520)}_{\rho
N,S}\right)^2 +(E_N-m_N)^2\,\left(f^{(1520)}_{\rho N,V}\right)^2 +
\Bigg( \frac{E_\rho}{m_\rho}\,f^{(1520)}_{\rho N,S}
\nonumber\\
&& \qquad \qquad \;\, -\Big(E_N-m_N\Big)\,\Big(
\frac{\sqrt{s}+m_N}{m_\rho}\,f^{(1520)}_{\rho N,V}
+\frac{s-m_\rho^2-m_N^2}{2\,m_\rho }\,f^{(1520)}_{\rho N,T} \Big)
\Bigg)^2
\nonumber\\
&& \qquad \qquad \;\, + \Big(f^{(1520)}_{\rho N,S}-
(E_N-m_N)\,f^{(1520)}_{\rho N,V}\Big)^2 \Bigg)\,\big(E_N+m_N
\big)\,\frac{p_{\rho N}}{8\,\pi \,\sqrt{s}} \,, \label{D13-self}
\end{eqnarray}
where a folding with the $\rho$-meson spectral function
(\ref{rho-explicit}) analogous to (\ref{rho-loop}) is understood.
Also $\sqrt{s}= E_\rho +E_N $ and $E_N^2= m_N^2+p_{\rho N}^2$. A
technical remark is in order here. The expression for the partial
decay width in (\ref{D13-self}) is obtained by projecting the imaginary
part of the polarization tensor for the d-wave $N(1520)$ resonance onto
$J^P ={\textstyle{3 \over 2}}^-$. Clearly the contributions of the vector
and tensor coupling
constants $f_{\rho N,V}$ and $f_{\rho N,T}$ to the width are
kinematically suppressed by the factor $p_{\rho N}^2$, compared to
the contribution of the scalar coupling constant $f_{\rho N,S}$.
This observation supports the reasoning in section 3.1, where we
argued that the additional projectors for the $\omega N$ and $\rho
N$ channels are kinematically suppressed. Therefore they are not
considered in this work. Using our value for the resonance coupling
constant we obtain a $\rho N$-partial decay width of 2 MeV for a
$\rho$-meson spectral function with an energy dependent, and 13 MeV
with an energy-independent $\rho$-meson width. Thus, for a given
value of the $\rho N N(1520)$ coupling constant, the $\rho N$ width
of the $N(1520)$ depends crucially on the model for the
$\rho$-meson spectral function.

As noted at the beginning of this section, the extraction of the
$\rho$-production cross section and in particular the coupling of
the $N(1520)$ to the $\rho N$ channel from hadronic reactions alone
is model dependent. We attempt to remove this ambiguity by
considering the additional constraints provided by the multipole
amplitudes for the reaction $\gamma N \rightarrow \pi N$ within a
generalized vector meson dominance model. Within this framework, we
find a consistent description of the $\rho$ production cross
section of \cite{Brody:et:al} and of the multipole amplitudes in all four
channels.

The quantitative comparison with the result of \cite{Brown:Riska}
is more subtle since the quark model appears to favor vector- and
tensor-type couplings of the $N(1520)$ to the $\rho N$ channel

\begin{eqnarray}
{\mathcal L}_{BR} &=& i\,\frac{g^{(\omega)}_{BR}}{m_\omega^2 } \,
 \bar N \,\sigma_{\mu \nu}\,R_\kappa
\,(\partial^\nu\,\partial^\kappa\,\omega^\mu )
+i\,\frac{g^{(\rho )}_{BR}}{m_\rho^2 } \,\bar N \,\sigma_{\mu \nu}\,\vec
\tau\,R_\kappa \,
(\partial^\nu\,\partial^\kappa\,
\vec \rho^{\,\mu} )
+{\rm h.c.}  \,,
\label{BR-vertex}
\end{eqnarray}
while in our model the scalar coupling dominates. It is
straightforward to identify the on-shell equivalent coupling
constants,
\begin{eqnarray}
&& f_{\rho N, T}^{(1520)}\Big|_{BR} = 2\,\sqrt{3}\,\frac{g^{(\rho
)}_{BR}}{m_\rho^2 }  \,,\qquad
f_{\omega N, T}^{(1520)}\Big|_{BR} =
2\,\frac{g^{(\omega)}_{BR}}{m_\omega^2 }
\nonumber\\
&& f_{V N, S}^{(1520)}\Big|_{BR} = 0 \,, \qquad
f_{V N, V}^{(1520)}\Big|_{BR} =
-{\textstyle{1\over2}}\,\big(m_{1520}-m_N\big)\,f_{V N,
T}^{(1520)}\Big|_{BR} \,,
\label{1520:BR}
\end{eqnarray}
introduced in (\ref{D13-general}). Using the coupling constants obtained by
Brown and Riska \cite{Brown:Riska}, $g_{BR}^{(\rho)} \simeq 4.5$ and
$g_{BR}^{(\omega)} \simeq 7.7$, in (\ref{1520:BR}), we find a $\rho N$-decay
width for the N(1520) resonance of 0.04 MeV and 0.38 MeV for an energy dependent and energy
independent $\rho$-meson width, respectively. Thus the partial
width implied by the quark model results is much smaller than the
partial widths, 2 MeV and 13 MeV, obtained in our model. This is by and large
a consequence of the phase-space suppression, with
$\Gamma^{(\rho N)}_{1520} \sim p_{\rho N}^3$ implied by the vertex
(\ref{BR-vertex}).

It is instructive to express the coupling constants $f_{\omega
N,S}$, $f_{\omega N, V}$ and $f_{\omega N,T}$ in terms of helicity
matrix elements $h^{(L,T)}_{\omega N,h} $ where the upper index
$(L,T)$ refers to whether the matrix element involves a transverse
or longitudinal vector meson,
\begin{eqnarray}
f_{\omega N,S}^{(1520)}&=& -h^{(T)}_{\omega N, 3/2} \,, \nonumber\\
f_{\omega N,V}^{(1520)} &=& \frac{\sqrt{3}\,E_N}{2\, q^2}\,
h^{(T)}_{\omega N,1/2}
-\frac{\sqrt{3}\,m^2_N}{2\, q^2\,m_{1520}}\,h^{(T)}_{\omega N,1/2}
-\frac{m_N+E_N}{2\, q^2}\,h^{(T)}_{\omega N,3/2}\nonumber\\
&+& m_N\,\frac{E_N-E_\omega+m_N}{2\,
q^2\,m_{1520}}\,h^{(T)}_{\omega N,3/2} +
\frac{\sqrt{6}\,m_\omega\,m_N}{2\,m_{1520}\, q^2}\,h^{(L)}_{\omega
N,1/2} \,,
\nonumber\\
f_{\omega N,T}^{(1520)} &=& -\frac{\sqrt{3}\,(m_{1520}+m_N)}{2\,
q^2\,m_{1520}}\,h^{(T)}_{\omega N,1/2} +
\frac{E_N-E_\omega+m_N}{2\,
q^2\,m_{1520}}\,h^{(T)}_{\omega N,3/2} \nonumber\\
&+& \frac{\sqrt{6}\,m_\omega}{2\,m_{1520}\, q^2}\,h^{(L)}_{\omega
N,1/2}\,, \label{g-hel}
\end{eqnarray}
where $E_N= (m_N^2+q^2)^{1/2}$, $E_\omega= (m_\omega^2+q^2)^{1/2}$
and $E_N+E_\omega=m_{1520}$. Analogous results hold for the
coupling constants of the $\rho$ meson. The expressions in
(\ref{g-hel}) clearly demonstrate that the leading moments in
$q^2$ of the helicity amplitudes are correlated. Only if
\begin{eqnarray}
h^{(T)}_{\omega N,3/2} = \sqrt{3}\,h^{(T)}_{\omega N,1/2} =
\sqrt{3}\,h^{(L)}_{\omega N,1/2}/\sqrt{2} \;,\quad  {\rm at} \quad
q^2=0\,, \label{hel-rel}
\end{eqnarray}
holds  one obtains a finite result for $f_{\omega N,S}$,
$f_{\omega N,V}$ and $f_{\omega N, T}$ in the limit of $q^2 \to
0$. Note that the various factors in (\ref{hel-rel}) are in one to
one correspondence to the coefficients given in (\ref{j33-trafo})
when defining the state $|J\!=\!{\textstyle{3\over 2}};1\rangle $.
This phenomenon illustrates a well known effect, namely that
helicity amplitudes are subject to further constraint equations
required as to avoid kinematical singularities. Nevertheless the
helicity matrix elements are useful since they permit a more
direct relation  to corresponding matrix elements involving real
photon states.

For the case of photons gauge invariance requires a
transverse transition tensor. That leads to a
correlation of the parameters in (\ref{D13-general}),
\begin{eqnarray}
f^{(1520)}_{\omega N,S}= (m_{1520}-m_N)\,f^{(1520)}_{\omega N,V}
+m_{1520}\,(E_N-m_N)\,f^{(1520)}_{\omega N,T} \,,\label{}
\end{eqnarray}
if applied to real photons. The fact that there are only two
coupling constants left is readily understood because those two
terms reflect the two
helicity matrix elements $ h^{(T)}_{\omega N,3/2}$ and
$h^{(T)}_{\omega N, 1/2}$. A longitudinal photon state does not
exist implying $h^{(L)}_{\omega N,1/2}=0$ in the case of real
photons. The most general on-shell vertex may be represented by \cite{Benmerrouche},
\begin{eqnarray}
{\mathcal L}_{\frac{1}{2}\,\frac{3}{2}^-}^{(\gamma N)} &=&
\frac{e}{2\,m_R}\,\bar
R_\mu \,\gamma_\nu\,\Big( i\,f_{\gamma N,S_1}^{(1520)}+i\,f_{\gamma
N,V_1}^{(1520)}\,\tau_3 \,
\Big) \,N \,F^{\mu \nu}
\nonumber\\
&+&\frac{e}{4\,m_R^2}\,\bar
R_\mu \,\Big( f_{\gamma N,S_2}^{(1520)}+f_{\gamma
N,V_2}^{(1520)}\,\tau_3 \,
\Big) \,\big(\partial_\nu\,N\big) \,F^{\mu \nu}
+{\rm h.c.}
\,. \label{D13-gamma:def}
\end{eqnarray}

We derive the consequences of the generalized vector-meson dominance
assumption
(\ref{gamma-ansatz:k},\ref{tensor-ansatz}). The effective
interaction Lagrangian density describing the transition of the
resonance into the $\gamma N$ state is:
\begin{eqnarray}
&& {\mathcal L}_{\frac{1}{2}\,\frac{3}{2}^-}^{(\gamma N)} = e\,\bar
R_\mu \,\Big( i\,f_{\omega N,S}^{(1520)}\, \Gamma^{(+)}_{S,\nu}
+{\textstyle{i\over \sqrt{3}}}\,f_{\rho N,S}^{(1520)}\,
\Gamma^{(+)}_{V,\nu}\,\tau_3 \Big)\,\Big(N\,F^{\mu \nu} \Big)
+{\rm
h.c.} \,,
\nonumber\\
&& \Gamma_{S(V),\nu}^{(+)}  = \frac{m_{1520}+ i\,\gamma \cdot
\partial }{2\,m_{1520}}\,\Big( \frac{g_{S(V),2}^{(+)}-
g_{S(V),1}^{(+)}}{m_{1520}\,m_\omega }\,i\,\partial_\nu +
\frac{g_{S(V),1}^{(+)}}{m_\omega } \,\gamma_\nu \Big)
 \,,
\label{D13-gamma}
\end{eqnarray}
where we do not consider effects from the vector and tensor coupling
strength
$f_{\rho N, V(T)}^{(1520)}$ and $f_{\omega N, V(T)}^{(1520)}$. The
proper treatment of these terms requires a careful analysis of the suppressed
$\rho N$ and $\omega N$ states in (\ref{j33-trafo}), in particular the
off-shell structure of the associated projection operators.
It is straightforward to match the on-shell parts of
(\ref{D13-gamma:def}) and
(\ref{D13-gamma}),
\begin{eqnarray}
f_{\gamma N,V_1}^{(1520)} &=&
\frac{2}{\sqrt{3}}\,\frac{m_{1520}}{m_\omega}\,g_{V,1}^{(+)}\,f_{\rho
N,S }^{(1520)}\;, \quad
f_{\gamma N,V_2}^{(1520)} =
\frac{4}{\sqrt{3}}\,\frac{m_{1520}}{m_\omega}\,\Big(g_{V,1}^{(+)}-g_{V,2}^{(+)}\Big)\,
f_{\rho N,S }^{(1520)}
\nonumber\\
f_{\gamma N,S_1}^{(1520)} &=&
2\,\frac{m_{1520}}{m_\omega}\,g_{S,1}^{(+)}\,f_{\omega N,S }^{(1520)}\;,
\quad
f_{\gamma N,S_2}^{(1520)} =
4\,\frac{m_{1520}}{m_\omega}\,\Big(g_{S,1}^{(+)}-g_{S,2}^{(+)}\Big)\,f_{\omega
N,S }^{(1520)} \,.
\label{}
\end{eqnarray}

Finally we derive the isoscalar and isovector helicity matrix
elements $h^{(T)}_{S(V),h}$ commonly used to parameterize the
electromagnetic resonance decay $R\to \gamma N$,
\begin{eqnarray}
h^{(T)}_{S,3/2}&=& -e\,f_{\omega N,S}^{(1520)}
\,\frac{m_{1520}^2-m^2_N}{2\,m_{1520}\,m_\omega}\,\Big(
\frac{m_{1520}-m_N}{m_{1520}+m_N}\, g^{(+ )}_{S,1} +g^{(+
)}_{S,2}\Big) \,,
\nonumber\\
h^{(T)}_{S,1/2}&=& \frac{e}{\sqrt{3}}\,f_{\omega N,S}^{(1520)}
\,\frac{m_{1520}^2-m^2_N}{2\,m_{1520}\,m_\omega }\,\Big(
\frac{m_{1520}-m_N}{m_{1520}+m_N}\, g^{(+ )}_{S,1} - g^{(+
)}_{S,2}\Big) \,,
\nonumber\\
h^{(L)}_{S,1/2} &=&  0 \,.\label{}
\end{eqnarray}
Results for the isovector helicity matrix elements $h^{(T)}_{V,h}$
follow by analogy in terms of ${\textstyle{1\over
\sqrt{3}}}\,f_{\rho N}^{(1520)}$ and $g^{(\pm)}_{V,i}$.
In Table \ref{tab:3c} we compare the electromagnetic
resonance coupling constants with the estimates of \cite{Feuster:Mosel}.
We find deviations by almost a factor of two
in some cases. Furthermore our estimates of the isoscalar coupling constants
are systematically larger than the values of \cite{Feuster:Mosel}. One may expect
that the inclusion of the suppressed $\rho N$ and
$\omega N$ channels (see \ref{j33-trafo}) will lead to a more quantitative
matching of the different approaches.

\begin{table}[t]
\begin{tabular}{|r||c|c||c|c|c|} \hline
& $f^{(1520)}_{\gamma N,S_1} $ & $f^{(1520)}_{\gamma N,S_2} $ &
$f^{(1520)}_{\gamma N,V_1}$&
$f^{(1520)}_{\gamma N,V_2}$ \\ \hline \hline

this work  &  7.84 & 7.63 & 1.72 & 1.67  \\ \hline

\cite{Feuster:Mosel}  & 2.9-4.5 & 4.3-5.8 & 3.1-5.3 & 1.1-4.4 \\ \hline

\end{tabular}
\vspace*{2mm} \caption{Electromagnetic resonance coupling constants
in the $I J^P =\frac{1}{2}\,\frac{3}{2}^-$ channel.}
\label{tab:3c}
\end{table}

\subsection{$I \,J^P=\frac{3}{2}\,\frac{3}{2}^-$ channel}

We now turn to the remaining $ I J^P=\frac{3}{2}\,\frac{3}{2}^-$
sector. Here we include the four channels $\pi N$, $\pi \Delta $,
$\rho N$ and $\eta \Delta$. This leads all together to 10
parameters in this sector. The parameters of our best fit are given
in Tab. \ref{tab:4}.

\begin{table}[h]
\begin{tabular}{|r||c|c|c|c|} \hline
$g^{(\frac{3}{2},-)}_{1}$\,$\Big[m_\pi^{-\frac{n+m}{2}}\Big]$& $\pi N
[n=3] $
&$\pi \Delta [n=1]$ & $\rho N [n=1]$ &$\eta \Delta [n=1]$  \\ \hline
\hline

$\pi N[m=3]$  & 0.36 & -2.47 & -3.68 & 2.24\\ \hline

$\pi \Delta [m=1]$  & -2.47 & 12.34 & 7.61 &-4.47  \\ \hline

$\rho N [m=1]$   & -3.68 & 7.61 & 7.87 & -11.22 \\ \hline

$\eta \Delta [m=1]$  & 2.24 & -4.47 & -11.22  &12.02\\ \hline

\end{tabular}
\vspace*{2mm} \caption{Coupling constants in the $I J^P
=\frac{3}{2}\,\frac{3}{2}^-$ channel.}
\label{tab:4}
\end{table}

\begin{figure}[t]
\begin{center}
\includegraphics[width=10.5cm,clip=true]{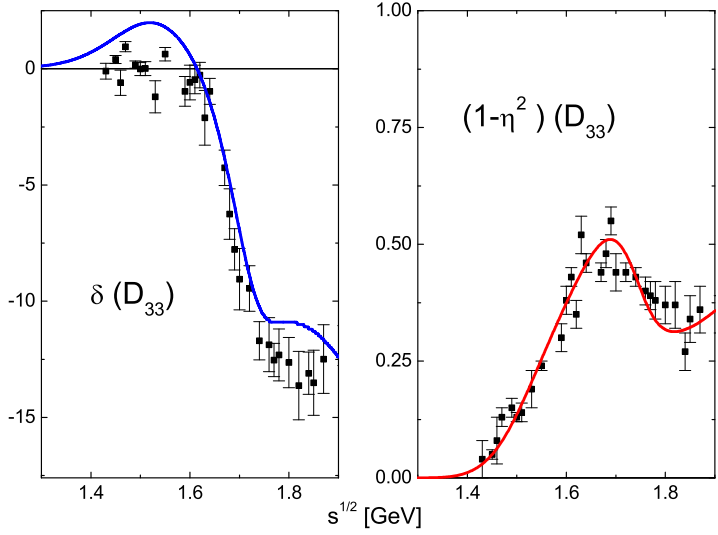}
\end{center}
\caption{Pion-nucleon scattering phase shift $\delta_{D_{33}}$ and
inelasticity parameter $\eta_{D_{33}}$ of the $D_{33}$ channel. We
compare to the single-energy solution SP98 of \cite{Arndt:piN,SAID}.}
\label{fig:17}
\end{figure}

The pion-nucleon phase shift and the inelasticity parameter are
well reproduced by our model as demonstrated in Fig. \ref{fig:17}.
The phase shift and inelasticity parameter exhibit rapid changes at
$\sqrt{s}\simeq $ 1700 MeV signalling the presence of the baryon
resonance $\Delta(1700)$  with quantum numbers  $I \,J^P
=\frac{3}{2}\,\frac{3}{2}^-$.The fit is constrained also by the
vector-meson production cross section shown already in Fig.
\ref{fig:1} and Fig. \ref{fig:2}.

\begin{figure}[b]
\begin{center}
\includegraphics[width=10.5cm,clip=true]{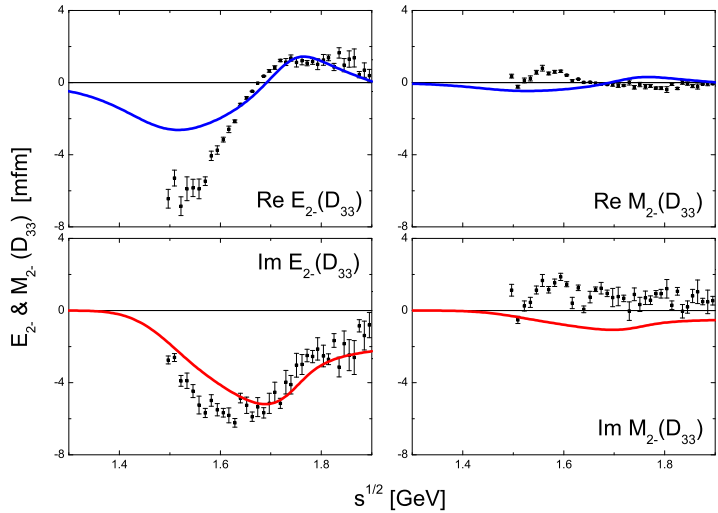}
\end{center}
\caption{Multipole amplitudes $E_{2-}(D_{33})$ and $M_{2-}(D_{33})$
of $\gamma $ induced pion production (see (\ref{match-multipole})).
Our results shown by solid lines are compared to those of the single-energy
solution SM00 of \cite{Arndt:gamma,SAID}.}
\label{fig:18}
\end{figure}

Further constraints are given by the electric and magnetic
multipole amplitudes $E_{2-}(D_{33})$ and $M_{2-}(D_{33})$. In Fig.
\ref{fig:18} we compare the appropriate electric and magnetic
multipole amplitudes with the empirical SM00 analysis of
\cite{Arndt:gamma,SAID}. We obtain a satisfactory description of the real
and imaginary part, which indicates that the contributions we
include are indeed the leading ones.

\begin{figure}[b]
\begin{center}
\includegraphics[width=10.5cm,clip=true]{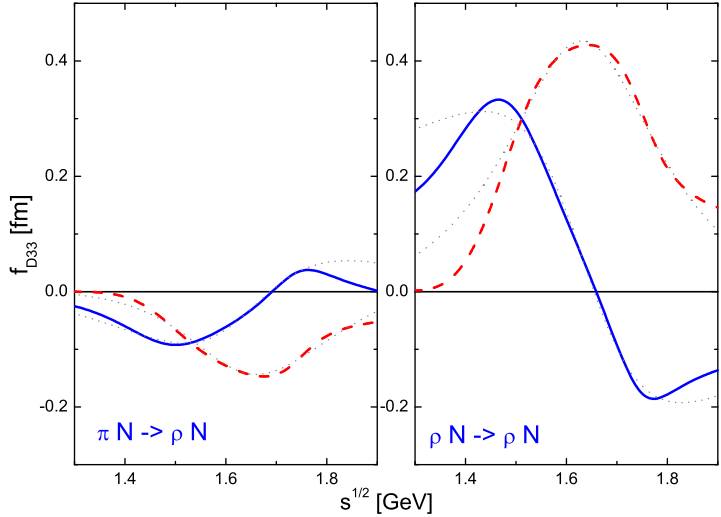}
\end{center}
\caption{Vector-meson production and scattering amplitude with
$I=J=\frac{3}{2}$. The solid and dashed lines represent the real and imaginary parts
of the amplitudes. The dotted lines follow from the schematic resonance exchange model
defined in (\ref{rep-D33}).}
\label{fig:19}
\end{figure}

We turn to the $\rho$-meson scattering and production amplitudes.
In Fig. \ref{fig:19} we present the pion-induced $\rho$-meson
production amplitudes introduced in (\ref{f-vp}). The $\rho $-meson
nucleon scattering amplitude with $I\!=\!\frac{3}{2}$ and
$J\!=\!\frac{3}{2}$ is also shown in Fig. \ref{fig:19}. The isobar
$\Delta(1700)$ resonance leads to a peak structure in the imaginary
parts of the subthreshold amplitudes. Again it is useful to
introduce a reduced scattering amplitude,
\begin{eqnarray}
&& f^{(\frac{3}{2}\frac{3}{2}-)}_{\rho N \to \rho N}(\sqrt{s}\,) =
\frac{N_{\rho
N}}{8\pi\,\sqrt{s}}\,M_{33}^{(\frac{3}{2}-)}(\sqrt{s},1)\,,
\nonumber\\
&& a_{\rho N}^{(\frac{3}{2}\frac{3}{2})} =
f^{(\frac{3}{2}\frac{3}{2}-)}_{\rho N \to \rho
N}(m_N+m_\rho\,)\simeq (-0.13+i\,0.38)\,{\rm fm} \,, \label{a:D33}
\end{eqnarray}
which at threshold equals the scattering length. We find a
repulsive scattering length consistent with the dominance of the
$\Delta (1700)$ resonance in the $I\, J^P =
\frac{3}{2} \frac{3}{2}^-$ amplitude. With (\ref{a:D33}) the
collection of s-wave $\rho$-meson nucleon scattering lengths is
complete, and we are ready to compute the spin and isospin averaged
scattering length $\bar a_{\rho N}$,
\begin{eqnarray}
 \bar a_{\rho N} =
{\textstyle{1\over 3}}\,a_{\rho N}^{(\frac{1}{2})}
+{\textstyle{2\over 3}}\,a_{\rho N}^{(\frac{3}{2})}
\simeq (-0.09+i\,0.24) \,{\rm fm},
\label{rhoNav}
\end{eqnarray}
where $a_{\rho N}^{(\frac{1}{2})}$ is the isospin averaged
$J\!=\!{\textstyle{1\over 2}}$ scattering length given in
(\ref{rhoN1}) and $a_{\rho N}^{(\frac{3}{2})}$ the isospin averaged
$J\!=\!{\textstyle{3\over 2}}$ scattering length,
\begin{eqnarray}
&& \bar a_{\rho N}^{(\frac{3}{2})}=
{\textstyle{1\over 3}}\,a_{\rho N}^{(\frac{1}{2}\frac{3}{2})}
+{\textstyle{2\over 3}}\,a_{\rho N}^{(\frac{3}{2}\frac{3}{2})}
\simeq (-0.08+i\,0.30) \,{\rm fm}.
\label{rhoN3}
\end{eqnarray}
It is obvious that the scattering length $a_{\rho
N}^{(\frac{3}{2}\frac{3}{2})}$, presented in this section, is of
major importance for the spin and isospin averaged scattering
length because it contributes with the relatively large weight
factor $4/9$. Our result for the averaged scattering length is in
qualitative agrement with previous model calculations of the
$\rho$-meson self energy \cite{HFN,CS,KKW} in nuclear matter if
interpreted in terms of an effective scattering length. For dilute
nuclear matter the isospin and spin averaged scattering $\bar
a_{\rho N} $ of (\ref{rhoNav}) leads to a considerable broadening
but only a small mass shift for the $\rho$-meson. We return to the
discussion of $\rho$-meson propagation in nuclear matter again in
the next section where the influence of the nucleon and isobar
resonances is discussed.

In Fig. \ref{fig:19} we show the vector-meson production and
scattering amplitudes obtained in our model as well as those of a
schematic resonance exchange model, defined by,
\begin{eqnarray}
&& f^{(\frac{3}{2}\frac{3}{2})}_{+,\,\pi N \to \rho N}(\sqrt{s}\,)
\simeq
-\frac{e^{\,i\,\phi^{(1700)}_{\pi \rho}}|g_{\pi N}^{(1700)}|\,|g_{\rho
N}^{(1700)}|\,p_{\pi
N}^2}{m_{1700}^2\,(\sqrt{s}-m_{1700}+\frac{i}{2}\,\Gamma_{1700})}
 + \frac{p^2_{\pi N}}{m_{1700}^2}\,b^{(\frac{3}{2}\frac{3}{2})}_{+,\,\pi
N \to \rho N}\,,
\nonumber\\
&& f^{(\frac{3}{2}\frac{3}{2})}_{+,\,\rho N \to \rho N}(\sqrt{s}\,)
\simeq
-\frac{|g_{\rho
N}^{(1700)}|^2}{\sqrt{s}-m_{1700}+\frac{i}{2}\,\Gamma_{1700}}
 + b^{(\frac{3}{2}\frac{3}{2})}_{+,\,\rho N \to \rho N}\,,
\label{rep-D33}
\end{eqnarray}
with the coupling constants $g^{(1700)}_{\pi N}$, $g^{(1700)}_{\rho
N}$, a phase parameter $\phi_{\pi
\rho}^{(1700)}$, and the background parameters
$b^{(\frac{3}{2}\frac{3}{2})}_{\pi N\to \rho N}$,
$b^{(\frac{3}{2}\frac{3}{2})}_{\rho N\to \rho N}$. The results of
the schematic model are shown by the dotted lines. This simple
model yields a reasonable description of the amplitudes. Note that
in the model for the production amplitude in (\ref{rep-D33}) we
include a phase-space factor $p_{\pi N}^2$ which stems from the
fact that the initial state is in a d-wave. The resonance
parameters of our fit to the vector-meson production and scattering
amplitudes are collected in Tab. \ref{tab:4b}. The background
parameters are:
\begin{eqnarray}
&& b^{(\frac{3}{2}\frac{3}{2})}_{-,\, \pi N \to \rho N} \simeq
(-0.56+i\,0.32)\,{\rm fm} \,,\quad
b^{(\frac{3}{2}\frac{3}{2})}_{-,\, \rho N \to \rho N} \simeq
(0.06-i\,0.07)\,{\rm fm} \,.
\label{b:D33}
\end{eqnarray}

\begin{table}[t]
\begin{tabular}{|r||c|c||c||c|c|c|} \hline
& $g_{\pi N} $ & $g_{\rho N}$ & $\phi_{\pi \rho }$ [$\,^\circ $]&  $
m$[MeV ] & $\Gamma $ [MeV ]\\ \hline \hline

$\Delta (1700)$  &  2.29  & 0.71 & -175 & 1631 & 394 \\ \hline

\end{tabular}
\vspace*{2mm} \caption{Resonance parameters in the $I J^P
=\frac{3}{2}\,\frac{3}{2}^-$ channel .}
\label{tab:4b}
\end{table}

\subsubsection{Discussion of $\Delta$(1700) resonance coupling
constants}

We compare the resulting resonance coupling constants with values
of \cite{Brown:Riska}.  The d-wave isobar resonance
$\Delta(1700)$ with $(I\,J)^P=(\frac{3}{2}\,\frac{3}{2})^-$ couples
to the pion-nucleon  and $\rho$-meson nucleon channels in the
following form:
\begin{eqnarray}
{\mathcal L}_{\frac{3}{2}\,\frac{3}{2}^-}^{(\pi N)} &=&
\Big(f^{(1700)}_{\pi N}/m_\pi\Big) \, \bar
\Delta_{1700,\mu } \,(\partial^\mu \,\vec \pi )\cdot \vec
T \, i\,\gamma_5\, N +{\rm h.c.} \,,
\nonumber\\
{\mathcal L}_{\frac{3}{2}\,\frac{3}{2}^-}^{(\rho N)}
&=&f^{(1700)}_{\rho N,S} \, \bar
\Delta_\mu \,\vec T\,N \,\vec \rho^{\,\mu}
+i\,f^{(1700)}_{\rho N,V} \, \bar
\Delta_\mu\,\vec T \,\gamma_\alpha \,(\partial_\mu  N
)\,\vec \rho^{\,\alpha}
\nonumber\\
&+& i\,f^{(1700)}_{\rho N,T}\,  \bar
\Delta_\mu\,\vec T\,\sigma_{\alpha \beta}\,
(\partial_\mu
\partial_\alpha N)\, \vec \rho^{\,\beta} +{\rm h.c.} \,.
\label{R:D33}
\end{eqnarray}
The value  $f^{(1700)}_{\pi N,BR}\simeq $ 2.66 of \cite{Brown:Riska}
describing the strength with which the isobar resonance couples to the
$\pi N$ state,
compares reasonably well with our result,
\begin{eqnarray}
\Big| f_{\pi N}^{(1700)} \Big| =\sqrt{3}\,m_\pi\,
\frac {\sqrt{8\,\pi\,N_{\pi N}^{(+)}(m_{1700})}}
{m_{1700}^{3/2}} \,\Big| g_{\pi N}^{(1700)}\Big| \simeq 1.89 \,.
\label{}
\end{eqnarray}
The three coupling constants $f_S, f_V$ and $f_T$ for the $\rho N$
channel, introduced in (\ref{R:D33}), are independent quantities.
In our work we determined the
scalar term only, which dominates the vertex, unless the vector and
tensor coupling strengths are anomalously large,
\begin{eqnarray}
\Big| f_{\rho N, S}^{(1700)} \Big| =\sqrt{\frac{8\,\pi\,m_{1700}}{N_{ \rho
N}(m_{1700})}}\, \Big|g_{\rho N}^{(1700)}\Big| \simeq 3.42\,,
\quad f_{\rho N,V}^{(1700)} = f_{\rho N,T}^{(1700)}=0\,.
 \label{}
\end{eqnarray}
This should be compared with the quark-model results which imply a
vanishing scalar coupling strength but finite values for the
kinematically suppressed vector- and tensor-type vertices (see
(\ref{1520:BR})). In order to facilitate the comparison with
various models we evaluate the partial $\rho N$-decay width of the
isobar resonance,
\begin{eqnarray}
&& \Gamma_{1700}^{(\rho N)} (\sqrt{s}\,) =2\,\Bigg(
\left(f^{(1700)}_{\rho N,S}\right)^2
+(E_N-m_N)^2\,\left(f^{(1700)}_{\rho N,V}\right)^2 + \Bigg(
\frac{E_\rho}{m_\rho}\,f^{(1700)}_{\rho N,S}
\nonumber\\
&& \qquad \qquad \;\, -\Big(E_N-m_N\Big)\,\Big(
\frac{\sqrt{s}+m_N}{m_\rho}\,f^{(1700)}_{\rho N,V}
+\frac{s-m_\rho^2-m_N^2}{2\,m_\rho }\,f^{(1700)}_{\rho N,T} \Big)
\Bigg)^2
\nonumber\\
&& \qquad \qquad \;\, + \Big(f^{(1700)}_{\rho N,S}-
(E_N-m_N)\,f^{(1700)}_{\rho N,V}\Big)^2 \Bigg)\,\big(E_N+m_N
\big)\,\frac{p_{\rho N}}{8\,\pi \,\sqrt{s}} \,, \label{D33-self}
\end{eqnarray}
in terms of the three coupling constants, $f_{S,V,T}^{(1700)}$,
$\sqrt{s}= E_\rho +E_N $ and $E_N^2= m_N^2+p_{\rho N}^2$. In
(\ref{D33-self}) a folding with the $\rho$-meson spectral function
(\ref{rho-explicit}) analogous to (\ref{rho-loop}) is understood.
With our value for the resonance coupling constant we find a $\rho
N$-decay width of 71 MeV for an energy dependent, and 175 MeV for
an energy independent width of the $\rho$-meson. Using the quark-model
coupling constants instead \cite{Brown:Riska}, we obtain 0.03
MeV and 0.12 MeV, respectively, for the $\rho$N partial widths of
the $\Delta(1700)$.

Using the generalized vector-meson dominance assumption
(\ref{gamma-ansatz:k},\ref{tensor-ansatz}), the on-shell vertex may
be represented by,
\begin{eqnarray}
{\mathcal L}_{\frac{3}{2}\,\frac{3}{2}^-}^{(\gamma N)} &=&
\frac{e}{2\,m_R}\,\bar
R_\mu \,\gamma_\nu\,i\,f_{\gamma N,V_1}^{(1520)}\,T_3 \,N \,F^{\mu \nu}
+\frac{e}{4\,m_R^2}\,\bar
R_\mu \,f_{\gamma N,V_2}^{(1520)}\,T_3 \,
\big(\partial_\nu\,N\big) \,F^{\mu \nu}
+{\rm h.c.}
\nonumber\\
&=& i\,e\,f_{\rho N,S}^{(1700)}\, \bar
\Delta_\mu \,T_3\, \Gamma_{V,\nu}^{(+)} \,\big(N\,F^{\mu
\nu} \big) +{\rm h.c.} \,, \label{D33-gamma}
\end{eqnarray}
with the transition operator $\Gamma_{S(V)}^\nu $ specified in
(\ref{D13-gamma}). We obtain values,
\begin{eqnarray}
\Big| f_{\gamma N,V_1}^{(1700)} \Big|
\simeq 3.23\;, \qquad \Big| f_{\gamma N,V_2}^{(1700)} \Big|  \simeq 3.14 \;,
\label{gam:D33}
\end{eqnarray}
that are comparable with the values $f_{\gamma N,V_1}^{(1700)}
\simeq 1.6-2.3$ and $f_{\gamma N,V_2}^{(1700)} \simeq 2.7-4.8$
suggested in \cite{Feuster:Mosel}.

\section{Implications for vector-meson propagation in nuclear matter}

\begin{figure}[t]
\begin{center}
\includegraphics[width=11.5cm,clip=true]{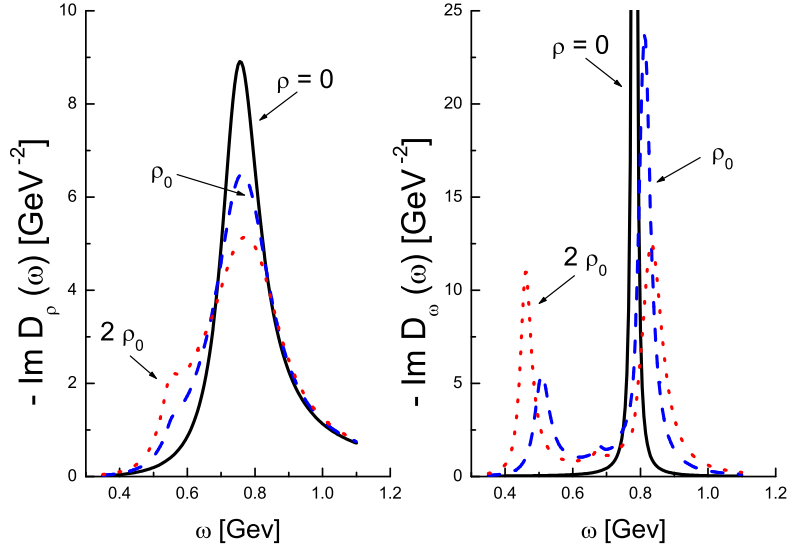}
\end{center}
\caption{Imaginary parts of the
$\rho$-meson and $\omega$-meson propagators in nuclear matter at $\rho
= \rho_0$ and $\rho = 2\, \rho_0$, compared to those in the vacuum.}
\label{fig:20}
\end{figure}

In this section we present results for the propagators of the
$\rho$ and $\omega$ mesons at rest in nuclear matter, obtained
with the scattering amplitudes presented in section 5, to leading
order in density. The low-density theorem states that the self
energy, $\Delta m_V^2(\omega )$, of a vector meson in nuclear
matter is given by~\cite{LDT}
\begin{equation}
\Delta m_V^2(\omega ) = -\frac{8\,\pi\,\sqrt{s}}{N_{VN }(\sqrt{s}\,)}\,
f_{V N}\,(\sqrt{s}\,)\,\Bigg|_{\sqrt{s}=\omega +m_N} \,\,\rho_N + \dots,
\label{LDT}
\end{equation}
where $\omega$ is the energy of the vector meson and $\rho_N$ the
nucleon density.
The normalization factor $N_{VN} (\sqrt{s}\,)
\simeq 2\,m_N$, was introduced
in (\ref{om-spec}). In spin and isospin saturated nuclear matter,
the shifts of the mass and width, given by $\Delta m_V^2(\omega )$,
are in this approximation proportional to the spin and isospin
averaged s-wave scattering amplitude,
\begin{eqnarray}
&& f_{\rho N}(\sqrt{s}\,) = \frac{1}{9}\,
f^{(\frac{1}{2}\frac{1}{2})}_{+,\,\rho N \to \rho N}(\sqrt{s}\,)
+\frac{2}{9}\,
f^{(\frac{3}{2}\frac{1}{2})}_{+,\,\rho N \to \rho N}(\sqrt{s}\,)
\nonumber\\
&& \qquad \quad +\frac{2}{9}\,
f^{(\frac{1}{2}\frac{3}{2})}_{-,\,\rho N \to \rho N}(\sqrt{s}\,)
+\frac{4}{9}\,
f^{(\frac{3}{2}\frac{3}{2})}_{-,\,\rho N \to \rho N}(\sqrt{s}\,) \,,
\nonumber\\
&& f_{\omega N}(\sqrt{s}\,) = \frac{1}{3}\,
f^{(\frac{1}{2}\frac{1}{2})}_{+,\,\omega N \to \omega
N}(\sqrt{s}\,) +\frac{2}{3}\,
f^{(\frac{1}{2}\frac{3}{2})}_{-,\,\omega N \to \omega
N}(\sqrt{s}\,)\,. \label{}
\end{eqnarray}
In Fig.~\ref{fig:20} we show the resulting propagators at the
saturation density of nuclear matter, $\rho_0 = 0.17
\mbox{~fm}^{-3}$ and at $\rho = 2\, \rho_0$. For the $\rho$ meson
we note an enhancement of the width, and a downward shift in
energy, due to the mixing with the baryon resonances at $\sqrt{s} =
1.5 - 1.6$ GeV.  As the density is increased, the width of the
$\rho$-like peak is amplified and more strength is shifted down to
the resonance-hole region at lower energies. At nuclear saturation
density the center of gravity of the energy-weighted sum rule is
shifted down by about 3 \% only. Compared to our previous
preliminary result \cite{Hirschegg} we find much less attraction in
the $\rho$-meson spectral function. This is a consequence of a significantly
reduced coupling of the d-wave N(1520) resonance to the $\rho N$
state. In the detailed discussions of the previous sections we
linked this result to the fact that we now include the data on
photon-induced reactions in a systematic fashion \cite{RW}. In previous analyses
the photon-induced multipole amplitudes were not considered systematically in this
context. In particular our spectral function for the $\rho$ meson
shows considerably less strength in the low-mass region as compared to
previous works \cite{Post:Leupold:Mosel,PLPM,Kim:Rapp:Brown:Rho}.
The in-medium propagator of the $\omega$ meson exhibits two distinct quasi-particles, an
$\omega$-meson like mode, which is shifted up somewhat in energy,
and a resonance-hole like mode at low energies. This confirms our previous
result \cite{Hirschegg} almost quantitatively. The low-lying mode
carry about 15 \% on the energy-weighted sum rule. The
center-of-gravity is shifted down by $\simeq 4 $~\% at nuclear
saturation density. However, we stress that the structure of the
in-medium $\omega$-meson spectral function clearly cannot be
characterized by this number alone. Note that as compared to the recent
work \cite{Post:Mosel} we obtain a significantly stronger coupling of
the $\omega $ meson to the N(1520) nucleon-hole state leading to a much more
pronounced effect of the resonance-hole  state in the spectral function.

A simple estimate on the accuracy of the leading order result
follows by investigating the size of the subleading term, which is
due to the Pauli blocking of intermediate states. In order to
compute the leading and subleading mass shifts of the $\omega$
meson we need only the s-wave scattering lengths $a_{\omega
N}^{1/2}\simeq (-0.45+i\,0.31)$ fm and $a_{\omega N}^{3/2} \simeq
(-0.43+i\,0.15)$ fm. We obtain a model independent result, in
complete analogy with the corresponding expression for the
kaon~\cite{Lutz:SK,Lutz:Korea},
\begin{eqnarray}
\Delta\,m_\omega^2 &=&-4\,\pi \left(1+\frac{m_\omega}{m_N}\right)
\Bigg[
 \frac{1}{3}\,a_{\omega N}^{(\frac{1}{2})}
+\frac{2}{3}\,a_{\omega N}^{(\frac{3}{2})}
\Bigg] \rho
\nonumber\\
&+&\frac{4}{\pi^2}\,\frac{1-x^2+x^2 \log x^2 }{(1-x)^2}\,
\Bigg[  \frac{1}{3}\,\left(a_{\omega N}^{(\frac{1}{2})} \right)^2
+\frac{2}{3}\,\left(a_{\omega N}^{(\frac{3}{2})} \right)^2
\Bigg] k_F^4+{\mathcal O}\left(k_F^5 \right) \,,
\label{del-om}
\end{eqnarray}
where $x= m_\omega/m_N $.  At nuclear saturation density the
correction term of order $k_F^4$ in (\ref{del-om}) implies a
further repulsive mass shift of about $1$ MeV and increase of the decay
width of $2$ MeV for the $\omega $ meson. This should be compared to the leading
order repulsive mass and width shift of $43$ MeV and $39$ MeV respectively. We
expect that the results obtained with only the leading term in the low-density
expansion are qualitatively correct at normal nuclear matter
density. However, on a quantitative level, the spectral functions
may change when higher order terms in the density expansion are
included. For instance, we expect that the in-medium properties of
the baryon resonances depend sensitively on the meson spectral
functions. If this is the case, a self consistent calculation, which
corresponds to a partial summation of terms in the density expansion, would have to be
performed~\cite{Lutz:SK,Lutz:Korpa,PLPM}.

\section{Summary and conclusion}

In this work we have computed the scattering amplitudes that describe
the s-wave scattering of the light vector mesons off nucleons. To
leading order in a density expansion these amplitudes, in
particular the subthreshold parts, determine the spectral
functions of the $\rho$- and $\omega$-meson in nuclear matter.
Since there are no data on vector-meson nucleon scattering we
constrained our analysis by all relevant elastic and inelastic
$\gamma N$ and $\pi N$ data. The coupled channel unitarity
condition together with covariance, the causality property of local quantum
field theory and the data set then lead to fairly robust predictions for the vector-meson
nucleon scattering amplitudes. We explored the
possibility that the nucleon and isobar resonances, which do not belong to the
large-$N_c$ ground state of QCD, are generated by coupled channel dynamics \cite{LK}.
The merit of this conjecture, which admittedly is rather extreme, consists
in a significant reduction of the number of parameters,
since such a scheme does not require any parameters describing the
properties of bare resonances. Based on our analysis, where all relevant resonances in the
channels considered were successfully generated by coupled channel
dynamics, we see no indication that would disprove this
conjecture~\cite{LK}.

In contrast to previous works which typically apply the K-matrix formalism, our
coupled channel study is based on the covariant projector
technique introduced in \cite{LK}. This leads to loop functions and scattering amplitudes
which are consistent with a dispersion-integral representation. The known
drawback of the K-matrix formalism, the violation of analyticity and causality,
is avoided. We developed a generalization of the vector meson dominance assumption
that is appropriate
for an effective field theory formulated in terms of quasi-local two-body interaction terms only.
Moreover, this work is the first study of the vector-meson nucleon scattering processes
where the $\rho N$, $\omega N$ and $\gamma N$ states were considered simultaneously.
Altogether, in our approach we incorporate the
$\gamma N $, $\pi N$, $\pi \Delta $, $\rho N$, $\omega N$, $\eta N$, $K \Lambda$ and $K \Sigma
$ states. We allow for s- and d-wave states in the  $\pi N $ and $\pi \Delta$ channels
since only those states couple to the s-wave vector-meson nucleon states.
In the remaining hadronic channels we determine the strength of the leading s-wave states only.
We approximate the three-body final
state, $\pi \pi N$, by including the $\rho N$ and $ \pi \Delta $ channels where we
use energy dependent decay widths of the $\rho$ meson and $\Delta$ isobar.
The two-pion total cross section was not fitted directly, since our
model is restricted to low angular momentum states. Furthermore, we
did not include the $\sigma N$ channel explicitly. It would anyway
be difficult to discriminate its effects from the $\pi \Delta$
channel, which in our scheme is only indirectly constrained by
data. Hence, we consider the latter as an effective channel, which
subsumes also the residual effect of the $\sigma N$ channel.

Since there is no
model-independent partial-wave decomposition of the two-pion production data available,
the $\sigma N$ channel was not considered. It would be extremely difficult to discriminate its
effects from the $\pi \Delta$ channel in our present scheme. The latter channel, which
is constrained in our scheme by data only rather indirectly, is considered as an effective
channel in the sense that it is supposed to describe some residual effects of the $\sigma N$
channel also.

Due to the presence of nucleon and isobar resonances, the vector-meson nucleon
scattering amplitudes show
rapid energy variations. In particular we find that the s-wave resonances $N(1535)$ and
$N(1650)$ show a strong $\omega N$ component, which are required to simultaneously generate
both resonances.
On the other hand, only the $N(1535)$ but not the $N(1650)$ resonance
appears to couple significantly to the $\rho N$ channel. Similarly striking
is our result that the d-wave $N(1520)$ resonance couples strongly to the
$\omega N$ channel but with only much reduced strength to the $\rho N$ channel.
This result is a consequence of our systematic inclusion of the photon-induced
scattering data not done previously. We extracted coupling constants of the nucleon
and isobar resonances to the vector-meson nucleon states and compared those to predictions
of the quark model. In some cases we find a strong disagreement.

It is gratifying to obtain a simultaneous description of all considered photon- and
pion-induced production data.
Production cross sections, for which a
t-channel exchange of a single pion is dominantly contributing, are typically described well only at
energies rather close to the production threshold. The long-ranged one-pion exchange
contribution is expected to yield significant strength in higher partial waves,
not considered in this work.
The remaining empirical pion- and photon-induced production
cross sections are fairly well described up to significantly higher energies, thus
demonstrating s-wave dominance for those reactions.

An improved description of the considered electromagnetic multipole
amplitudes as well as the
production cross sections is expected once $\rho N$ and $\omega N$ states with subleading
angular momentum characteristics are included in the analysis.

In nuclear matter our scattering amplitudes imply an $\omega$-meson spectral
functions with considerable support at energies smaller than the free-space mass
representing resonance nucleon-hole type excitations. On the other hand,
as an immediate consequence of the moderate coupling of the $\rho N$ channel
to the $N(1520)$ d-wave resonance, the in-medium effects for the $\rho$ meson
are found to be significantly smaller as compared to previous calculations.
We emphasize that the rigorous evaluation of the vector-meson spectral functions
requires an approach in which the transformation of the
scattering amplitudes from the center of mass system into the laboratory frame is well defined.
This transformation can be quite non-trivial for states with high spin and angular momentum.
Clearly, the covariant projector technique applied in this work fulfils this requirement.
We anticipate that a realistic evaluation of the in-medium spectral functions of the light vector
mesons requires a self consistent many-body approach, in particular when the nucleon and isobar
resonances show important in-medium modifications \cite{Lutz:Korpa}.

The results presented in this paper are relevant for the experimental program at
GSI. The HADES detector will help to further explore the properties of
the light vector mesons in nuclear matter by measuring their
dilepton final state with high accuracy. Complementary experimental
programs are pursued at Jefferson Lab \cite{CEBAF}, MAMI \cite{MAMI}
and KEK \cite{KEK}
with photon and nucleon induced reactions off nuclei. To further
substantiate the structure of the
vector-meson nucleon scattering amplitudes it would be desirable to
establish a more
microscopic understanding of the effective interaction vertices employed
in our work. We expect that a
significant parameter reduction is feasible by a proper extension of the
$\chi$-BS(3) approach (chiral Bethe-Salpeter approach for the SU(3) flavor group)
in \cite{LK} to include additional inelastic channels like $\pi \Delta $, $\omega N$
and $\rho N$.

\section*{Acknowledgments}

M.F.M.L. and B.F. acknowledge encouragement and fruitful discussions with M. Soyeur.
Gy.W. was supported by the Hungarian research Foundation (OTKA) grants
T 32038 and T 30855. We thank V. Koch for a critical reading of the manuscript.

\newpage

\section{Appendix A: Projector approach for $\Delta $-isobars}

We elaborate on the inclusion of the $\pi \Delta$ channel. The
$J\!=\!{\textstyle{1\over 2}}$ and $J\!=\!{\textstyle{3\over 2}}$
projectors are
\begin{eqnarray}
&& \big[Y_{0,\,\mu }^{(+)}(\bar q, q;w) \big]_{21} = -3\,P_{\mu
\alpha}^{(-)}(w)\,\bar q^\alpha  \,i\,\gamma_5 \,,
\nonumber\\
&& \big[Y_{0,\,\mu  }^{(+)}(\bar q, q;w) \big]_{12} =
+i\,\gamma_5\,q^\alpha  \,3\,P_{\alpha \mu}^{(-)}(w)\,,
\nonumber\\
&& \big[Y_{0,\,\mu \nu}^{(+)}(\bar q, q;w) \big]_{32} =
\sqrt{3}\,P_{\mu \alpha}^{(-)}(w)\,\bar q^\alpha \left(\gamma_\nu
-\frac{w_\nu}{w^2}\,\wslash \right) \,, \qquad
\nonumber\\
&&\big[Y_{0,\,\mu \nu }^{(+)}(\bar q, q;w) \big]_{23} =
\sqrt{3}\left(\gamma_\mu -\frac{w_\mu}{w^2}\,\wslash \right)
q^\alpha  \,P_{\alpha \nu}^{(-)}(w)\, ,
\nonumber\\
&& \big[Y_{0,\,\mu \nu }^{(+)}(\bar q, q;w) \big]_{22}  =
9\,P_{\mu \alpha}^{(-)}(w)\,\bar q^\alpha \,q^\beta  \,P_{\beta
\nu}^{(-)}(w) \,,
\label{proj-v1-d}
\end{eqnarray}
and
\begin{eqnarray}
&& \big[Y_{1,\,\mu }^{(-)}(\bar q, q;w) \big]_{21} =
-\sqrt{3}\,P_{\mu \alpha}^{(+)}(w)\,q^\alpha  \,i\,\gamma_5 \,,
\qquad \big[Y_{1,\,\mu \nu}^{(-)}(\bar q, q;w) \big]_{32} = P_{\mu
\nu}^{(+)}(w) \,,
\nonumber\\
&& \big[Y_{0,\,\mu  }^{(+)}(\bar q, q;w) \big]_{12} =
+\sqrt{3}\,i\,\gamma_5\,\bar q^\alpha  \,P_{\alpha \mu}^{(+)}(w)
 \,, \qquad
\big[Y_{1,\,\mu \nu }^{(-)}(\bar q, q;w) \big]_{23} = P_{\mu
\nu}^{(+)}(w)\, ,
\nonumber\\
&& \big[Y_{1,\,\mu \nu }^{(-)}(\bar q, q;w) \big]_{22}  =
P_{\mu \nu}^{(+)}(w) \,,
\label{proj-v1-d}
\end{eqnarray}
where we introduced the auxiliary objects
\begin{eqnarray}
P_{\mu \nu}^{(\pm)}(w) &=& \frac{1}{2}\left( g_{\mu \nu}-\frac{w_\mu
\,w_\nu}{w^2}\right)
\left(1\pm \frac{\wslash }{\sqrt{w^2}}\right)
\nonumber\\
&-& \frac{1}{6}
\left(\gamma_\mu -\frac{w_\mu}{w^2}\,\wslash \right)
\left(1 \mp \frac{\wslash }{\sqrt{w^2}}\right)
\left(\gamma_\nu -\frac{w_\nu}{w^2}\,\wslash \right) \;.
\label{}
\end{eqnarray}
Note that in the $J\!=\!\frac{3}{2}$ channel the projectors couple
to further projectors not considered in this work. This amounts to
neglecting contributions from the $\pi \Delta $ states with
$J=\frac{3}{2}$ but $L=2$. Finally we provide the loop functions
of the $\pi \Delta $ channels. The loop functions are specified
first in the zero-width approximation,
\begin{eqnarray}
J_{22}^{(I+)} (\sqrt{s},0) &=& N_{\pi \Delta}^{(+)}(\sqrt{s})\,I_{\pi
\Delta}(\sqrt{s}\,)
\,, \quad
J_{22}^{(I-)} (\sqrt{s},1) =N_{\pi \Delta}^{(-)}(\sqrt{s})\, I_{\pi
\Delta}(\sqrt{s}\,) \;,
\nonumber\\ \nonumber\\
N_{\pi \Delta}^{(+)}(\sqrt{s}\,)&=&
\Big( E_\Delta -m_\Delta \Big)\,\frac{2}{3}
\left( 2\,\frac{E_\Delta}{m_\Delta}-1\right)^2 \,p_{\pi \Delta}^2 \;,
\nonumber\\
N_{\pi \Delta}^{(-)}(\sqrt{s}\,)&=& \Big( E_\Delta +m_\Delta \Big)
\left(
\frac{5}{9}+\frac{2}{9}\,\frac{E_\Delta}{m_\Delta}+\frac{2}{9}\,\frac{E_\Delta^2}{m_\Delta^2}
\right) \;,
\label{delta-loop}
\end{eqnarray}
with $\sqrt{s}
=\sqrt{m_\Delta^2+p_{\pi \Delta}^2}+\sqrt{m_\pi^2+p_{\pi \Delta}^2}$ and
$E_\Delta =\sqrt{m_\Delta^2+p_{\pi \Delta}^2}$.
The threshold behavior of the loop functions (\ref{delta-loop}) confirms
that in the $J=\frac{1}{2}$ channel the
considered $\pi \Delta $ state has $L=2$ but in the $J=\frac{3}{2}$
channel $L=0$.
The finite decay width is included by analogy with our treatment of the
vector-meson loop functions in (\ref{rho-loop}). For the isobar we use
the
spectral density
\begin{eqnarray}
\rho_\Delta(q^2 ) &=& -\Im \left( \zeta_\Delta \,(q^2-m_\Delta^2 )
- \Pi_\Delta (q^2)  \right)^{-1} \,,
\nonumber\\
\Pi_\Delta (q^2) &=& -\frac{C^2}{3\,f^2}\,
 \,\Big( m_N+\sqrt{m_N^2+p_{\pi N}^2}\,\Big)\,p_{\pi N}^2\, \Big( I_{\pi
N} (\sqrt{q^2}\,)-\Re I_{\pi N}(m_\Delta) \Big)\,
\nonumber\\
&& \quad \quad \times
\left( \frac{\lambda_\Delta^2+\bar p_{\pi N}^2}{\lambda_\Delta^2+p_{\pi
N}^2} \right)^2 \,,
\label{del-spec}
\end{eqnarray}
with $\sqrt{q^2} =\sqrt{m_N^2+p_{\pi N}^2}+\sqrt{m_\pi^2+p_{\pi N}^2} $
and
$m_\Delta =\sqrt{m_N^2+\bar p_{\pi N}^2}+\sqrt{m_\pi^2+\bar p_{\pi
N}^2}$. The parameters are
adjusted as to reproduce the pion-nucleon $P_{33}$ phase shift. We use
$f= 90$ MeV, $C \simeq 2.05$,
$m_\Delta \simeq 1232 $ MeV, $\zeta_\Delta \simeq 0.9$ and
$\lambda_\Delta \simeq 316$ MeV. The incorporation of the
dipole form factor in (\ref{del-spec}) leads to a quantitative
description of the $P_{33}$ phase up
to about $\sqrt{s} \simeq 1.5$ GeV \cite{delta:sp}. Finally the isobar
spectral function is normalized by the cutoff
$\Lambda_\Delta \simeq 1.78$ GeV where $\Lambda_\Delta $ is introduced
by analogy with (\ref{def-rho-norm}).

The pion-induced isobar production cross sections are:
\begin{eqnarray}
\sigma_{\pi^-p\,\to\,\pi^- \,\Delta^+ \;} \;&=& 4\,\pi
\,\frac{p_{\pi \Delta }}{p_{\pi N}}\,\frac{4}{9}\, \Bigg( \Big|
\frac{1}{2}\,f^{(\frac{1}{2}\frac{1}{2}+ )}_{\pi N \rightarrow \pi
\Delta}-\sqrt{\frac{2}{5}}\,f^{(\frac{3}{2}\frac{1}{2}+ )}_{\pi N
\rightarrow \pi  \Delta}\Big|^2
\nonumber\\
&&\qquad \quad \;+2\,\Big|
\frac{1}{2}\,f^{(\frac{1}{2}\frac{3}{2}- )}_{\pi N \rightarrow \pi
\Delta}-\sqrt{\frac{2}{5}}\, f^{(\frac{3}{2}\frac{3}{2}- )}_{\pi N
\rightarrow \pi \Delta}\Big|^2 \Bigg) \;,
\nonumber\\
\sigma_{\pi^-p\,\to\,\pi^+ \,\Delta^- \;} \;&=& 4\,\pi
\,\frac{p_{\pi \Delta }}{p_{\pi N}}\,\frac{1}{3}\, \Bigg( \Big|
f^{(\frac{1}{2}\frac{1}{2}+)}_{\pi N \rightarrow \pi
\Delta}+\sqrt{\frac{2}{5}}\,f^{(\frac{3}{2}\frac{1}{2}+ )}_{\pi N
\rightarrow \pi  \Delta}\Big|^2
\nonumber\\
&&\qquad \quad \;+2\,\Big| f^{(\frac{1}{2}\frac{3}{2}- )}_{\pi N
\rightarrow \pi \Delta}+\sqrt{\frac{2}{5}}\,
f^{(\frac{3}{2}\frac{3}{2}- )}_{\pi N \rightarrow \pi
\Delta}\Big|^2 \Bigg) \;,
\nonumber\\
\sigma_{\pi^+p\,\to\,\pi^+ \,\Delta^+ \;} \;&=& 4\,\pi
\,\frac{p_{\pi \Delta }}{p_{\pi N}}\,\frac{2}{5}\, \Bigg(
\Big|f^{(\frac{3}{2}\frac{1}{2}+ )}_{\pi N \rightarrow \pi
\Delta}\Big|^2 +2\, \Big| f^{(\frac{3}{2}\frac{3}{2}- )}_{\pi N
\rightarrow \pi \Delta}\Big|^2 \Bigg) \;,
\nonumber\\
\sigma_{\pi^+p\,\to\,\pi^0 \,\Delta^{++} \;} \;&=&
\frac{3}{2}\,\sigma_{\pi^+p\,\to\,\pi^+ \,\Delta^+ \;} \;,
\label{del-prod}
\end{eqnarray}
where we introduced the partial-wave amplitudes
$f^{(IJ\pm)}(\sqrt{s}\,)$. For example in the isospin
${\textstyle{1\over 2}}$ channel we write:
\begin{eqnarray}
&& f^{({\textstyle{1\over 2}} {\textstyle{1\over 2}}+)}_{\pi
N\,\to \pi \Delta}(\sqrt{s}\,) = \frac{\sqrt{N_{\pi
N}^{(+)}\,N_{\pi \Delta}^{(+)} }}{8\,\pi\,\sqrt{s}}\,
\,M^{({\textstyle{1\over 2}}+)}_{21}(\sqrt{s},0)\;,
\nonumber\\
&& f^{({\textstyle{1\over 2}} {\textstyle{3\over 2}}-)}_{\pi
N\,\to \pi \Delta}(\sqrt{s}\,) = \frac{\sqrt{N_{\pi
N}^{(-)}\,N_{\pi \Delta}^{(-)}}}{8\,\pi\,\sqrt{s}}\,p_{\pi N}
\,M^{({\textstyle{1\over 2}} -)}_{21}(\sqrt{s},1)\;. \label{f-vp}
\end{eqnarray}
The energy dependent width of isobar resonance is taken into account by
folding the
cross sections (\ref{del-prod}) with the spectral function
(\ref{del-spec}).

\end{document}